\documentclass[aps,prb,showpacs,twocolumn,floats,epsfig, nofootbib]{revtex4}
\usepackage{epsfig}
\usepackage{amsmath}
\usepackage{amsfonts}
\usepackage{graphicx}
\usepackage{amssymb}
\usepackage{color}
\usepackage{xcolor}
\usepackage{braket}
\usepackage{bm}
\usepackage[plainpages=false,pdfpagelabels,colorlinks=true,linkcolor=red,urlcolor=blue,citecolor=blue,pdftitle={},pdfauthor={},pdfdisplaydoctitle={},pdfduplex=DuplexFlipLongEdge]{hyperref}
\usepackage{amsbsy}
\usepackage{array}
\usepackage{color,ulem,verbatim,float}

\newcommand{\la}{\langle}
\newcommand{\ra}{\rangle}
\begin{document}

\title{Disorder-induced Enhancement of Entanglement Growth in One Dimension: Information Leakage at the scale of localization length}
%\title {Entanglement Dynamics within the Localization Length in an Inhomogeneous Quench}
%\title {Enhancement in Entanglement Growth due to the Breaking of Translational Symmetry}

\author{Roopayan Ghosh and Arnab Das}

\affiliation{School of Physical Sciences, Indian
Association for the Cultivation of Science, Jadavpur,
Kolkata-700032, India.}

\date{\today}

\begin{abstract}
When a group of compactly packed free fermions is allowed to spread over an empty one dimensional lattice, 
	the spreading particles can create entanglement between different parts of the lattice. We show, 
	though breaking of translational invariance (TI) of the lattice by disorder slows down the spreading of 
	local observables, the entanglement entropy of a subsystem can nonetheless receive a remarkable 
	enhancement as long as the subsystem lies within the single-particle localization length. 
	We show, the main mechanism behind this enhancement is the re-entrant exchange 
	of particles between the subparts due to transport of mutual information due to back scattering. 
	We discuss the length and time scales relevant to the phenomenon. We study the phenomenon
	for breaking of TI by both quasi-periodic and random potentials. We further explore the effect
	of randomness only in the initial state. This also exhibits a similar enhancement effect even in 
	a TI lattice. We also touch upon the special case of periodic potential, 
	where qualitatively similar phenomenology emerges, though the coherence in the back scattering 
	in this case leads to effects not captured by our simple yet generic picture. 
	%a particle that returns due to back scattering
	%to a subpart (the system) after entering the other part (the environment), 
	%though does not contribute to the spreading of 
	%particle density from the system to the environment, 
	%can bring back information from the system to the environment, and hence contribute to 
	%the enhancement of entanglement between them. 
	%{\color{green} Please add a few lines on the robustness with initial state as you had already intended}
	%However, in this case the mechanism is more complex because of coherence in 
	%scattering due to equally spaced scatterers. 
\end{abstract}
\maketitle

\section{Introduction} 
%%%%%%%%%%%%%%%%%%%%%%%%%%%%%%%%%%%%%%%%%%%%%%%%%%%%%%%%%%%%%%%%%%%%%%%%%%%%%%%%%%%%%%%%%%%%%%%%%%%%%%%%%%%%
\iffalse
%An inhomogeneous quench refers to evolving a closed quantum system under a 
%local Hamiltonian $H$, 
%starting from an initial state (other than an eigenstate of $H$) that breaks the translational 
%invariance (TI). 
%Since the 
%paradigm includes evolution of a bipartite system whose two different partitions are initialized to 
%very different states, it is a natural setting for describing
%the evolution of a quantum information storage system initialized to a desired state, under the 
%influence of its environment whose state can be quite generic. 
%An elementary setup that mimics the scenario consists of free fermions on a one dimensional lattice, 
%with a section of the lattice, i.e, a set of consecutive lattice  points (the system), 
%is entirely occupied by the fermions, while the rest (the environment) is completely empty --
%the domain wall states~\cite{alba, Mossel_2010,XXmodel,PhysRevE.78.061115,Eisler,PhysRevE.81.061134,
%PhysRevB.88.245114,SciPostPhys.3.3.020,PhysRevE.98.032105,Klobas_2018,Viti_2016}.
%In general these studies are mostly concentrated around cases where the underlying lattice has translational
%invariance (see however, \cite{PhysRevLett.102.100502,PhysRevB.85.035409,PhysRevB.96.195117,Eisler_2012,Tonni_2018,PhysRevB.100.125139}).\\
\fi
%%%%%%%%%%%%%%%%%%%%%%%%%%%%%%%%%%%%%%%%%%%%%%%%%%%%%%%%%%%%%%%%%%%%%%%%%%%%%%%%%%%%%%%%%%%%%%%%%%%%%%%%%%%%

In this work we study effect of disorder
on the dynamics of entanglement formation 
between two non-overlapping parts of a 
one-dimensional lattice due to spreading 
of non-interacting 
fermions on the lattice over time. 

Dynamics of spreading of particles and entanglement in such 
setups have widely been studied under the general protocol ``inhomogeneous quench" for translationally
invariant (TI) lattices,~\cite{alba, Mossel_2010,XXmodel,PhysRevE.78.061115,Eisler,PhysRevE.81.061134,
PhysRevB.88.245114,SciPostPhys.3.3.020,PhysRevE.98.032105,Klobas_2018,Viti_2016} as well as in some 
special systems where translational invariance is broken 
locally (local quenches)~\cite{PhysRevLett.102.100502,PhysRevB.85.035409,PhysRevB.96.195117,Eisler_2012}, or 
in critical systems where conformal symmetry dictates the dynamics~\cite{Tonni_2018,PhysRevB.100.125139,PhysRevLett.93.260602,Refael_2009,PhysRevB.72.140408,PhysRevB.83.045110,SciPostPhys.3.3.019,PhysRevB.94.035152,PhysRevB.96.174301,doi:10.1143/JPSJ.78.014001,Vitagliano_2010,Ram_rez_2014,Ram_rez_2015,Rodr_guez_Laguna_2016,Rodr_guez_Laguna_2017}.
Here we study the effect of extensive (i.e., spread over the entire lattice) disorder in form of random and
quasi-periodic potentials on the growth of entanglement 
in absence of any simplifying symmetry. 

A suitable initial state for studying such spreading dynamics 
is the so-called ``domain wall state", where one 
part of the lattice is completely filled while the rest of the lattice 
is empty (domain wall refers to the boundary between the completely filled and the 
completely empty parts). This initial state has been used
for most of the above mentioned works, as well as for illustrating most of our results here.

Already in the TI case, the dynamics of the entanglement entropy of a subsystem 
of the lattice exhibits a curious feature if one starts with a domain wall 
like initial state: though the spreading of the particle density is linear in time, 
the entanglement entropy of the subsystem grows logarithmically with time~\cite{alba,Eisler_2009}. 
Such slow growth of entanglement is a hall-mark of
many-body localization~\cite{PhysRev.109.1492,BASKO20061126, 
Bardarson_Pollmann_Moore_EE_Grwth,Serbyn_Papic_Abanin_MBL_EE_Grwth}, 
where the disorder suppresses the growth of entanglement. 
In this case, however, the slow entropy growth 
happens in presence of total translational invariance. 

What we show here is, introduction of disorder actually has an {\it opposite effect}
in this case - it actually {\it enhances} the entanglement growth substantially
as long as the subsystem in question is within the localization length from the initial domain wall. 
We also show that within the relevant length-scale, the entanglement growth is linear in 
time. The most spectacular manifestation of the phenomenon is in the weak disorder limit, 
when the localization length is large.

For disordered (quasi-periodic or random) systems, we identify incoherent back scattering as 
the main mechanisms behind this enhancement in the entanglement entropy. We also study the effect of the 
strength of the TI-breaking potential in spreading of the entanglement.
Since for weak disorder the localization length is
significantly large, the understanding of the phenomenon might play a role in designing systems for 
efficient storage and transfer of information in presence of disorder in
quantum devices. 

%\iffalse

Finally, we explore the effect of randomness put solely in the initial state, and demonstrate that
the well-known result of logarithmic growth of entanglement in the TI case for a single domain-wall
initial state~\cite{alba, Mossel_2010,XXmodel,calabrese3}  is a rather fine-tuned result -- small disorder in the occupied domain can
turn this logarithmic growth to a linear one. Our study also shows the phenomenon of enhancement of
entanglement due to breaking of TI of the lattice is not fine-tuned, and is robust to random variation
of initial states within a certain form.

%\fi

We also provide a glimpse of the phenomenology in a rather special case of TI breaking,
namely, the one due to periodic potentials, where a qualitatively similar enhancement 
of entanglement is observed. However, coherence in scattering due to the regularity of the 
lattice structure also plays a dominant role, making the scenario more complex. 
We find entanglement entropy growth shows a $a\log t +b$ scaling 
behaviour for the lattice periodicity $p \ll L$ with the coefficient $a \propto p.$ 
We also recover the linear scaling behaviour found in the TI case~\cite{Eisler_2012} 
for $p\sim L/2$. We then discuss the respective regimes of periodicity where one expects 
the two different behaviours. Curiously, though the back scattering is still the dominant mechanism
of enhancement of entanglement even in the periodic case, we do not see
greater enhancement with increased number of scatterer/unit length here, rather,
the above scaling relations imply a higher growth rate with larger $p,$ i.e., 
smaller number of scatterers per unit length. Coherence in the scattering from the periodically
placed scatterers forbids simple addition of contributions from different scatterers (a detailed
analysis of this is outside the remit of this work).

 The plan of the rest of the paper is as follows. in Sec. \ref{setup} we discuss 
the model Hamiltonians and give a brief summary of the various cases we consider and quantities we calculate. 
Then in Sec. \ref{disordered} we discuss in details the cases of a random disordered potential and a 
quasiperiodic (Fibonacci) potential, showing the growth of entanglement entropy in various scenarios, 
discuss the scalings and explore various factors that control the spreading. 
Then in Sec.\ref{periodic} we touch upon the behaviour of periodic lattices, 
and finally in Sec. \ref{discussion} we discuss our observed results and conclude.

\section{The setup}   
\label{setup} 
\begin{figure}
\centering{
\includegraphics[width=0.8 \columnwidth]{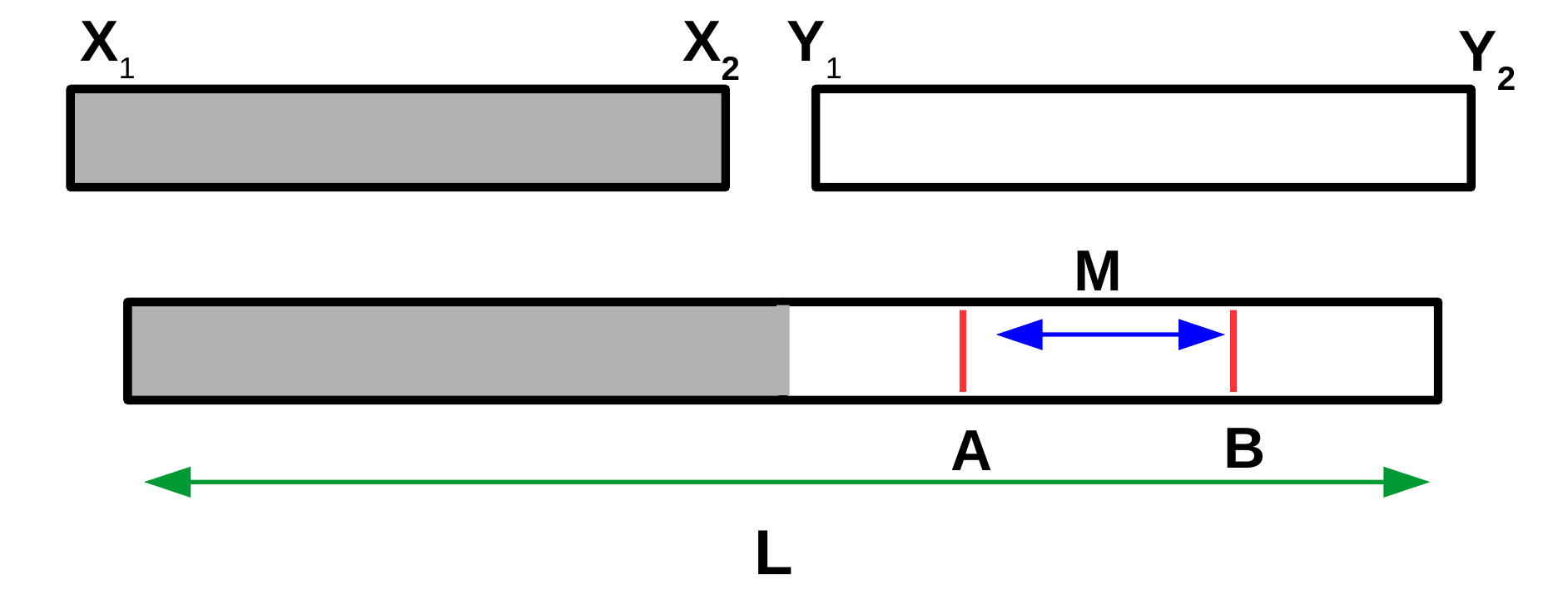}}
\caption{A schematic diagram showing the case of inhomogeneous quench. In an inhomogeneous quench we take two systems $X_1$ $X_2$ and $Y_1$ $Y_2$ and glue $X_2$ and $Y_1$ together at $t=0$. The system $X_1 X_2$ has a different filling fraction(denoted by the grey shade) than $Y_1 Y_2$.  $AB$ denotes the subsystem considered in the work. The system size is denoted as $L$ and the subsystem size as $M$, which represents the number of sites. See text for details. }
\label{model}
\end{figure}
Fig.~\ref{model} describes our quench in a schematic diagram. We take one system $X_1 X_2$ with a certain particle density and another system $Y_1 Y_2$ with a different particle density and then at $t=0$ glue $X_2$ and $Y_1$ together and let the system evolve. This difference in particle densities will induce a flow of particle current which in turn creates entanglement between the two halves of the system.  Equivalently, we can consider the full system $X_1 Y_2$ and start with a domain wall initial state with a domain wall at $i=L/2$ separating the different density sectors. This picture may be more useful for qubit systems for specific initial states where using Jordan Wigner transformation, one can go from the spin $1/2$ language to Fermionic language, and would end up with a domain wall initial state with densities $0$ and $1$ on two sides.  Additionally, domain wall picture is easier to use for analysis and we will refer to it in rest of our work. 
\iffalse
It is expected that the results from inhomogeneous quench will show features intermediate to global and local quenches. In this work however we show that the translation symmetry broken cases generate results qualitatively different from the well known results for global and local quenches. \\
\fi
%%%%%%%%%%%%%%%%%%%%%%%%%%%%%%%%%%%%%%%%%%%%%%%%%%%%%%%%%%%%%%%%%%%%%%%%%%%%%%%%%%%%%%%%%%%%
%%%%%%%%%%%%%%%%%%%%%%%%%%%%%%%%%%%%%%%%%%%%%%%%%%%%%%%%%%%%%%%%%%%%%%%%%%%%%%%%%%%%%%%%%%%%
\noindent
Throughout the paper, the model Hamiltonian used for calculations is.
\begin{equation}
\mathcal{H}=-\frac{J}{2}\sum_m(c_m^{\dagger} c_{m+1}+c_{m+1}^{\dagger} c_m)+J\sum_m \mu_m c_m^{\dagger} c_m
\label{hammaster}
\end{equation}
\noindent where $c_m[c^{\dagger}_m]$ are fermion annihilation[creation] operators and  
$\mu_{m}$ is the form of the onsite potential which we will take to be either random, quasi-periodic or periodic.
The Hamiltonian is now scaled in units of $J$ which is the hopping strength.  
The specifics of the setups are as follows.

\begin{enumerate}
%\item \textbf{Constant:} In this scenario $\delta \mu=0$, thus we have a Translationally Invariant system. A summary of exact analytical calculations \cite{XXmodel, Eisler}in this case is provided in the supplementary material.

	\item \textbf{Quasi-periodic:} 
			We will consider the quasi-periodic potentials
	which can be viewed as a superposition of multiple periodic potentials with incommensurate
		periods.
We will focus on an example of such a sequence, namely, the Fibonacci word sequence 
in the main text. A Fibonacci sequence is generated by the following recursion relation,
\begin{equation*}
F_n=F_{n-1}+F_{n-2}
\end{equation*} 
With $F_0=0$ and $F_1=1$. Thus the well known Fibonacci sequence looks 
		like $0,1,1,2,3,4,8,13, \hdots $. Later, Chuan~\cite{chuan, Chuan1996} introduced a concept of Fibonacci words, defined on the alphabet 
set $\{0,1\}$ in which the length of the $n^{th}$ word in the sequence is given by $F_n$. 
		These words are generated by the concatenation of the previous two words. 
		Formally , $S_n=S_{n-1} S_{n-2}$ where $S_n$ is the $n^{th}$ Fibonacci word. 
		$S_0$ is taken to be $0$ and $S_1=01$. Thus the first few terms of the words are, 
\begin{eqnarray*}
S_0 &=&0 \\
S_1 &=&01 \\
S_2 &=&010 \\
S_3 &=&01001 \\
S_4 &=&01001010 \\
\vdots
\end{eqnarray*}

Even at $N \rightarrow \infty$ it can be shown that $S_N$ 
		has no periodicity and the word is unique. 
		However, it is clear that the letters (digits) in the word are correlated. 
		%The type of sequence in $S_N$ has been labelled as a quasi-periodic 
		%sequence in literature.
		For a system of size $L$ , where $L$ is chosen to be a number in the 
		Fibonacci sequence, we generate the Fibonacci word sequence and then define,
\begin{eqnarray}
\mu_i=\mu_0-\delta\mu , \hspace{0.3 in} S_L^i=0 \nonumber \\
\mu_i=\mu_0+\delta\mu , \hspace{0.3 in} S_L^i=1,
	\label{Fibonacci_Latt}
\end{eqnarray}
\noindent
where we have labelled the $i^{th}$ letter (digit) in $S_n$ as $S_n^i$. This can be shown to result in
a quasi-periodic lattice\cite{1990PhRvB..41.7491Y}.
Results and details for two more such lattices, generated respectively from the Thue-Morse and Rudin-Shapiro 
sequences are relegated to the Supplementary material.
		%Quasiperiodic sequence is an automatic sequence which displays irregular periodicity. We will focus on an example of such sequence, Fibonacci word sequence in the main text. Details of this sequence and a few others are given in the supplementary material. 

	\item \textbf{Random:} The potential on site $\mu_m$ is chosen randomly 
		between $\mu-\delta\mu$ and $\mu+\delta\mu$. 
		For numerical calculations averaging over several realizations of the random numbers is performed.

	\item \textbf{Periodic:} We choose to work with two kinds of periodic potential.\\
(a) $\mu_m=\mu_0+(-1)^{\sum_{n=1}^{L}\delta_{m,np}}\delta \mu$. This represents a periodically varying potential with period $p$, in which every $p^{th}$ site has a potential $\mu-\delta\mu$ and every other site has a potential $\mu+\delta\mu$. \\
(b) 
\begin{eqnarray}
\mu_m&=&\mu_0+\delta\mu, \hspace{0.2 in} m=1 \hdots q/2 \nonumber \\
\mu_m&=&\mu_0-\delta\mu, \hspace{0.2 in} m=q/2+1 \hdots q \nonumber
\end{eqnarray}
		repeated over all the length of the lattice i.e. a square pulse potential 
		varying between $\mu_0+\delta\mu$ and $\mu-\delta\mu$ with period $q$. 
		These two sequences yield qualitatively similar results as will be shown in Sec. \ref{periodic}.
\end{enumerate}
		We consider a system with $L$-sites with open boundary conditions occupied by $L/2$ 
		spinless fermions, with the initial condition,
\begin{align}
\la c_m^{\dagger} c_n \ra&=\delta_{mn} \hspace{0.2 in} m \le L/2 \nonumber\\
	&=0 ~ {\rm otherwise} \label{init}
\end{align}

Starting from $t=0$, our aim is to study the evolution of entanglement between the  
subsystem $AB$ and the rest of the environment. We would calculate two quantities, 
Von Neumann entropy and mutual information for this purpose.\\
To calculate Von Neumann entropy, we first choose a subsystem of $M$ sites. 
For most cases we would deal with a subsystem of $i=L/2+1$ to $i=L/2+M$ 
which gives the same $S$ value as subsystem chosen  between $i=L/2$ to $i=L/2-M+1$. 
We will mostly focus on the case when $M=L/2$, i.e. the bipartite system.\\

Since our Hamiltonian is bilinear in fermions, all its many-body eigenfunctions can be written as 
Slater determinants of one-body eigenfunctions. Hence,one can write the Von Neumann entropy 
for any instant of time $t$ as,\cite{Peschel,Casini}
 \begin{equation}
S(t)=\sum_{i=1}^M\lambda_i(t) \log \lambda_i(t)+(1-\lambda_i(t))\log[1-\lambda_i(t)] 
\end{equation}
where $\lambda(t)$ are the eigenvalues of $C^{res}_{mn}(t)$. $C_{mn}(t)= \la c_m^{\dagger}(t) c_n(t) \ra$ and $res.$ denotes indices restricted to the subsystem under consideration.  \\
$\la c_m^{\dagger}(t) c_n(t) \ra$ can be exactly calculated for any value of time, 
using the Heisenberg picture. For our system the 
expression can be written in terms of single particle eigenfunctions and eigenvalues as,
\begin{eqnarray}
	\la c_m^{\dagger}(t) c_n(t) \ra ~~~~~~~~~~~~~~~~~~~~~~~~~~~~~~~~~~~~~~~~~~~~~~~~~~~~~~~~ && \nonumber\\
	= \sum_{k,l,i,j}R_{km} R_{ln} R_{ki} R_{lj}e^{i(E_k t-E_l t)}\la c^{\dagger}_i (0) c_j(0)\ra && 
\end{eqnarray}
where $E_k$ are the one particle eigenvectors and $R$ is the Unitary matrix diagonalizing 
the one-particle sector of the Hamiltonian. 
We also measure the mutual information between two subsystems labelled by 
$\alpha$ and $\beta$ is defined as follows,
\begin{equation}
\mathcal{M}^{\alpha\beta}=S^{\alpha}+S^{\beta}-S^{\alpha \cup \beta}
\end{equation}

\section{The Disordered Case}  
\label{disordered}
%%%%%%%%%%%%%%%%%%%%%%%%%%%%%%%%%%%%%%%%%%%%%%%%%%%%%%%%%%%%%%%%%%%%%%%%%%%%%%%%
%%                                   Fig - 2
%%%%%%%%%%%%%%%%%%%%%%%%%%%%%%%%%%%%%%%%%%%%%%%%%%%%%%%%%%%%%%%%%%%%%%%%%%%%%%%%
\begin{figure*}[t!]
\includegraphics[width=0.32\linewidth]{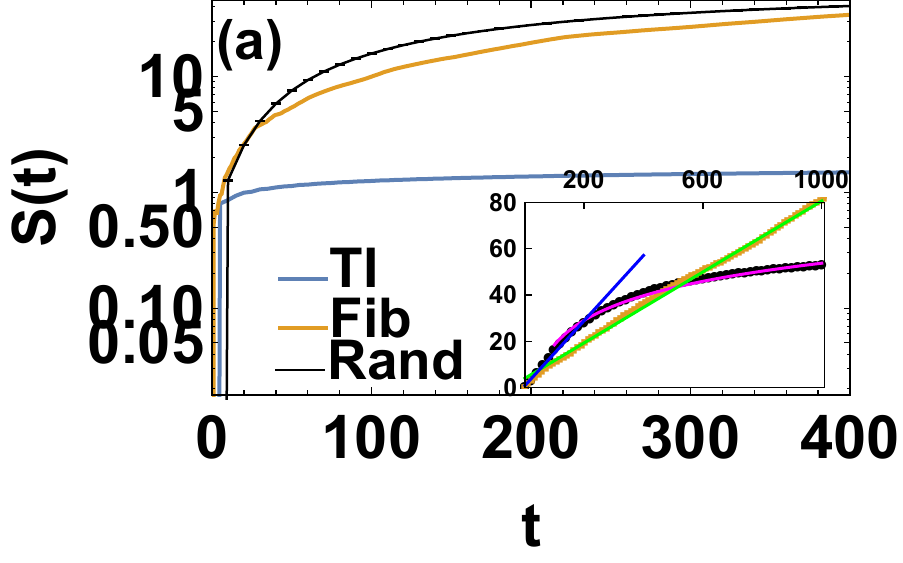}
\includegraphics[width=0.315\linewidth]{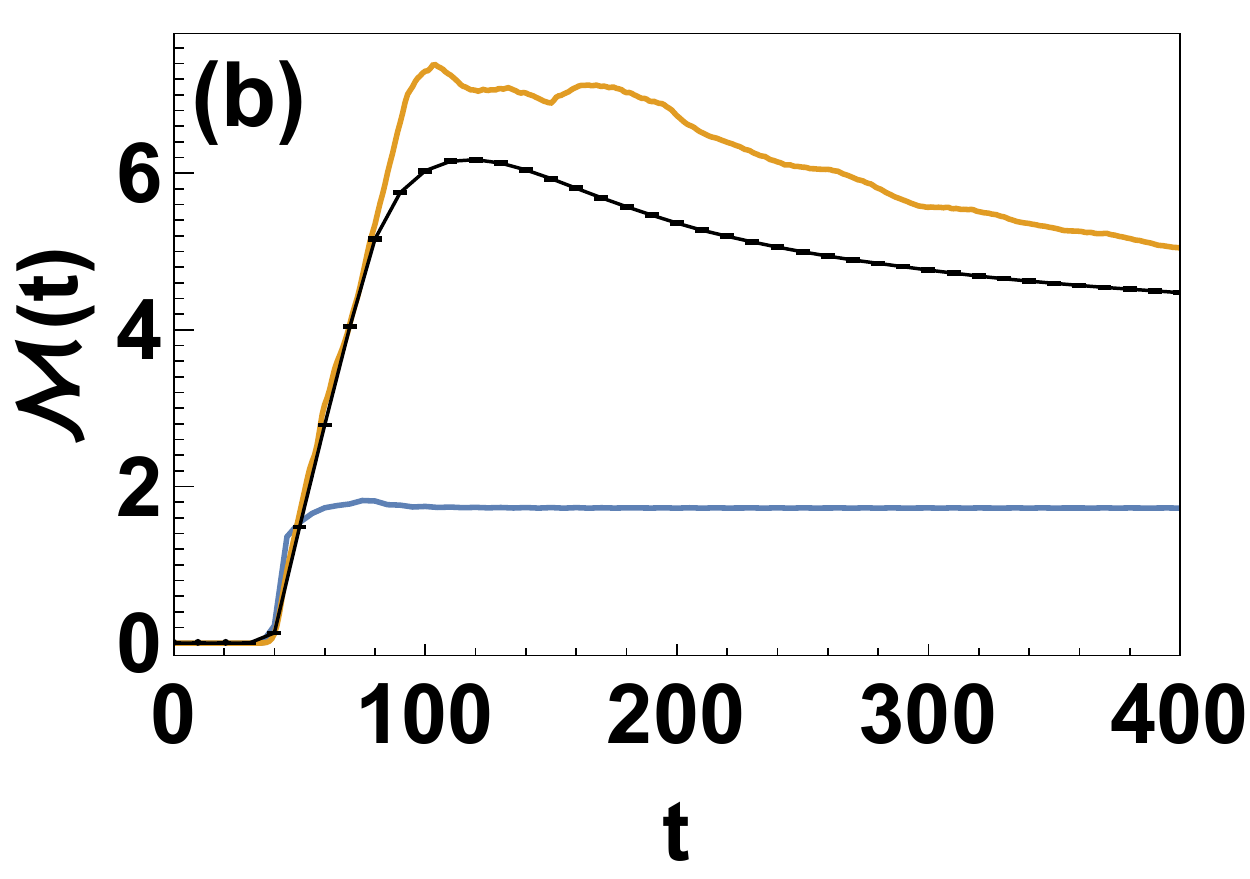}
\includegraphics[width=0.34\linewidth]{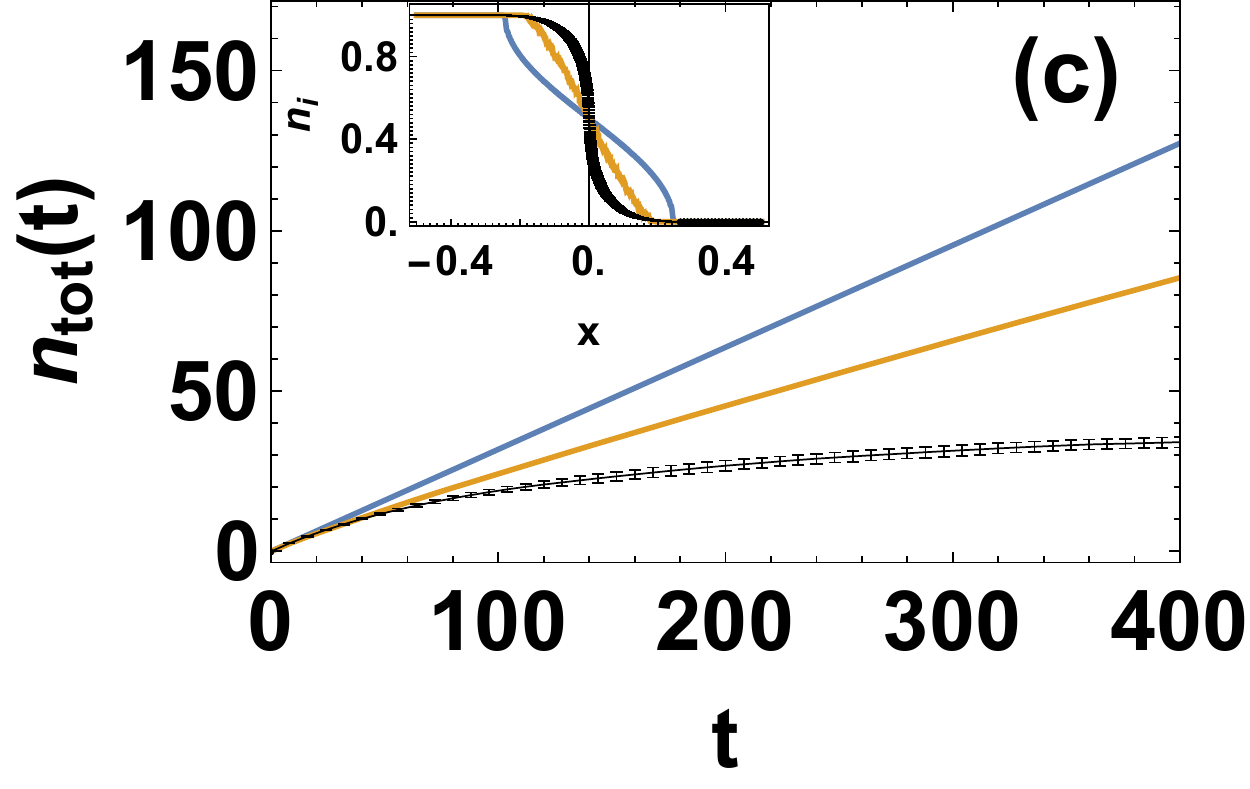}
	\caption{(Colour Online) 
		Growth of entanglement entropy, mutual information and local densities and correlations
		for various lattices.
		We have taken $L=2048$ for the TI and the random case, and $L=2584$ for the Fibonacci case.
		{\bf (a)} 
		Plot of the entanglement entropy $S$ vs time (in units of $J^{-1}$) for a subsystem 
		of size $M=L/2$. 
		The site $A$(see Fig.~\ref{model}) is chosen at $i=L/2+1$, $\mu=\delta \mu=0.1$. 
		 We have averaged over $2000$ disorder realization for the random potential cases. 
		The enhancement of entanglement due to the breaking of translational invariance is clearly visible. 
		This is quite in contrast with the spreading of the density of the particles, which
		is seen to have a slower spreading in TI-broken systems as expected  (shown in (c)). 
		The inset shows the fit of entanglement entropy vs t for the disordered cases. 
		The blue line denotes linear fit to $0.747 + 0.14 t $ while the magenta line denotes a 
		logarithmic fit to$-50.43 + 15.03\log t $ in the two time regimes, where the particles move 
		within and beyond the localization length of the system. The green line is the fit for 
		entanglement entropy growth for Fibonacci systems, which is linear and is $3.97+0.07 t$ 
		{\bf (b)}.
		Plot of mutual information vs time for subsystem $\alpha$ 
		spanning sites $i=L/2+1$ to $i=L/2+40$ and $\beta$ spanning sites $i=L/2+41$ to $i=L/2+80.$
		{\bf (c)} 
		This frame shows the growth of total density within the initially empty subsystem with time: 
		$n_{tot}=\sum_{i=L/2+1}^L \la c_i^{\dagger} c_i \ra $ vs $t,$. 
		The inset is showing $n_i= \la c_i^{\dagger} c_i \ra$ 
		at time $t=500$ versus the normalized lattice coordinate $x$ ($x = (i-L/2)/L$), giving us the
		idea of spreading/localization in space till the given time.}
\label{masterfig}
\end{figure*}
%%%%%%%%%%%%%%%%%%%%%%%%%%%%%%%%%%%%%%%%%%%%%%%%%%%%%%%%%%%%%%%%%%%%%%%%%%%%%%%%%%%%%%%%%%%%
%%%%%%%%%%%%%%%%%%%%%%%%%%%%%%%%%%%%%%%%%%%%%%%%%%%%%%%%%%%%%%%%%%%%%%%%%%%%%%%%%%%%%%%%%%%%
%fig5a, fig5c, fig3a

%%%%%%%%%%%%%%%%%%%%%%%%%%%%%%%%%%%%%%%%%%%%%%%%%%%%%%%%%%%%%%%%%%%%%%%%%%%%%%%%
\subsection{Enhancement of Entanglement}
The central result is summarized in Fig.~\ref{masterfig}. It shows the growth of half-chain entanglement
entropy $S$ of the lattice, starting from a domain wall like initial state where half of the system is full and the
rest is empty. The comparison has been made between lattices with TI, quasi-periodic Fibonacci, and a 
random disordered potential respectively. 

The left frame ( Fig.~\ref{masterfig}~(a)) shows that while the growth of half-chain entanglement is
extremely slow for the TI case, a remarkable enhancement in the growth occurs when the TI is broken 
by introduction of disorder via the Fibonacci and random potentials respectively. The middle frame
shows similar enhancement of growth dynamics for the mutual information ${\cal M}(t)$ due to disorder.

The middle frame compares between growth of mutual information for two adjacent subsystems $\alpha$ 
spanning sites $i=L/2+1$ to $i=L/2+40,$ and $\beta$ spanning sites $i=L/2+41$ to $i=L/2+80$ in a
TI lattice, a Fibonacci lattice and a random lattice respectively. For both cases of disorder, a the
growth of mutual information is much higher compared to the TI case.

\begin{figure}[h!]
\includegraphics[width=0.85\columnwidth]{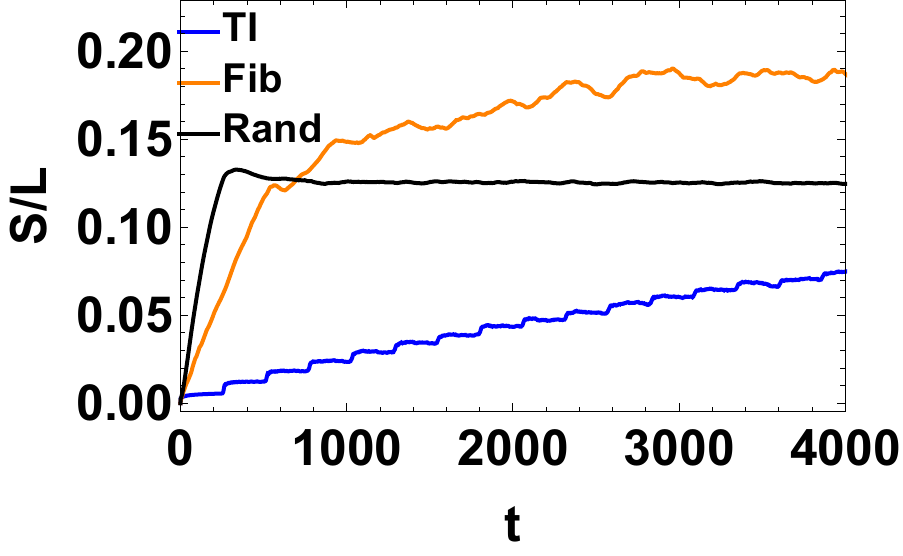}
	\caption{The figure illustrates the mechanism of increase of entanglement entropy
	$S$ of the subsystem of size $M = L/2 $ for a system of length $L = 256.$.  Averaging has been done over $80$ disorder realizations 
	In the TI case,
	the slow (logarithmic) growth is boosted by a steep jump periodically in time, with
	roughly a time-period $\sim 256$. This is approximately the time taken by a particle to
	exit the subsystem and re-enter it after getting reflected from the (open) boundary of the system
	 as particle propagation is ballistic in TI systems. See supplementary material and in Ref. \onlinecite{alba} .
	This illustrates that the faster growth of $S$ on top of the logarithmic one is due to
	re-entrance of the particles to the subsystem after getting back-scattered (in the
	TI case, from the boundary). For the disordered (quasi-periodic/random) cases, 
	there are random scatterers through out the entire lattice, and hence the enhancement of $S$ 
	is extensive, and the steps are smeared out. This results in a super-logarithmic (linear)
	growth of $S$ 
	}
\label{StepMech}
\end{figure}

In contrast, the right frame (Fig.~\ref{masterfig}~(c)) shows the particle density in the initially 
empty subsystem grows much faster in the TI case compared to the two disordered cases as expected. 
The inset of Fig.~\ref{masterfig}~(c) shows the spatial distribution of the particle at $t=500.$ Till this time, 
the spreading in the TI case is appreciably larger compared to the disordered cases. The inset of 
Fig.~\ref{masterfig}~(a) shows the growth of entanglement entropy in the random and Fibonacci potential 
cases respectively. 
It clearly shows two qualitatively different behavours for two different timescales for the random disordered case. 
When the propagating wavefront is within the maximum Thouless localization length\cite{Thouless_1972} ($\sim 250$ 
for the system considered), the entropy increases linearly, but once the wavefront goes beyond that, 
further particle movement gets exponentially suppresed and the growth of EE also gets suppressed 
logarithmically before reaching a steady value. The linear increase can also be 
gleaned from Fig.~\ref{StepMech}. For quasiperiodic Fibonacci system since there 
is no localization in wavefunction, the linear regime continues for a much longer time before 
reaching steady state value based on system size. The linear regime is presented in the inset. 
This is in contrast to the behaviour expected in interacting integrable and non integrable 
systems where one expects a power law $t^{\beta}$, $\beta <1$ rise in entanglement entropy 
for short timescales for the propagating domain wall before a linear regime arrives at 
large time scales and finally saturates \cite{PhysRevB.100.125139, Ljubotina2017, PhysRevB.99.121410}  .
%%%%%%%%%%%%%%%%%%%%%%%%%%%%%%%%%%%%%%%%%%%%%%%%%%%%%%%%%%%%%%%%%%%%%%%%%%%%%%%%
%%                                   Fig - 3
%%%%%%%%%%%%%%%%%%%%%%%%%%%%%%%%%%%%%%%%%%%%%%%%%%%%%%%%%%%%%%%%%%%%%%%%%%%%%%%%
%%%%%%%%%%%%%%%%%%%%%%%%%%%%%%%%%%%%%%%%%%%%%%%%%%%%%%%%%%%%%%%%%%%%%%%%%%%%%%%%
%%%%%%%%%%%%%%%%%%%%%%%%%%%%%%%%%%%%%%%%%%%%%%%%%%%%%%%%%%%%%%%%%%%%%%%%%%%%%%%%
%%                                   Fig - Scale 
%%%%%%%%%%%%%%%%%%%%%%%%%%%%%%%%%%%%%%%%%%%%%%%%%%%%%%%%%%%%%%%%%%%%%%%%%%%%%%%%
\begin{figure*}[t!]
\includegraphics[width=0.48\linewidth]{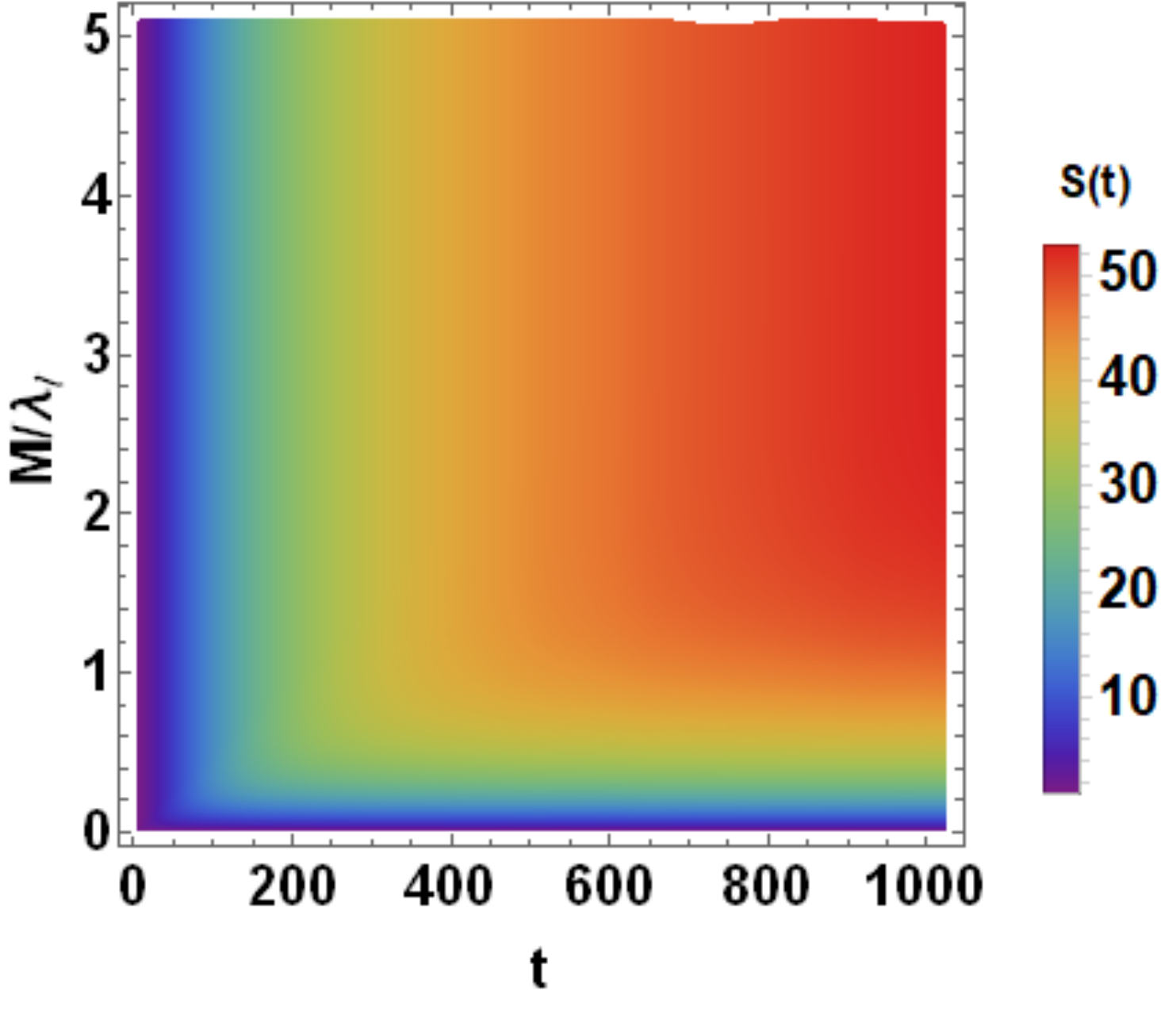}
\includegraphics[width=0.48\linewidth]{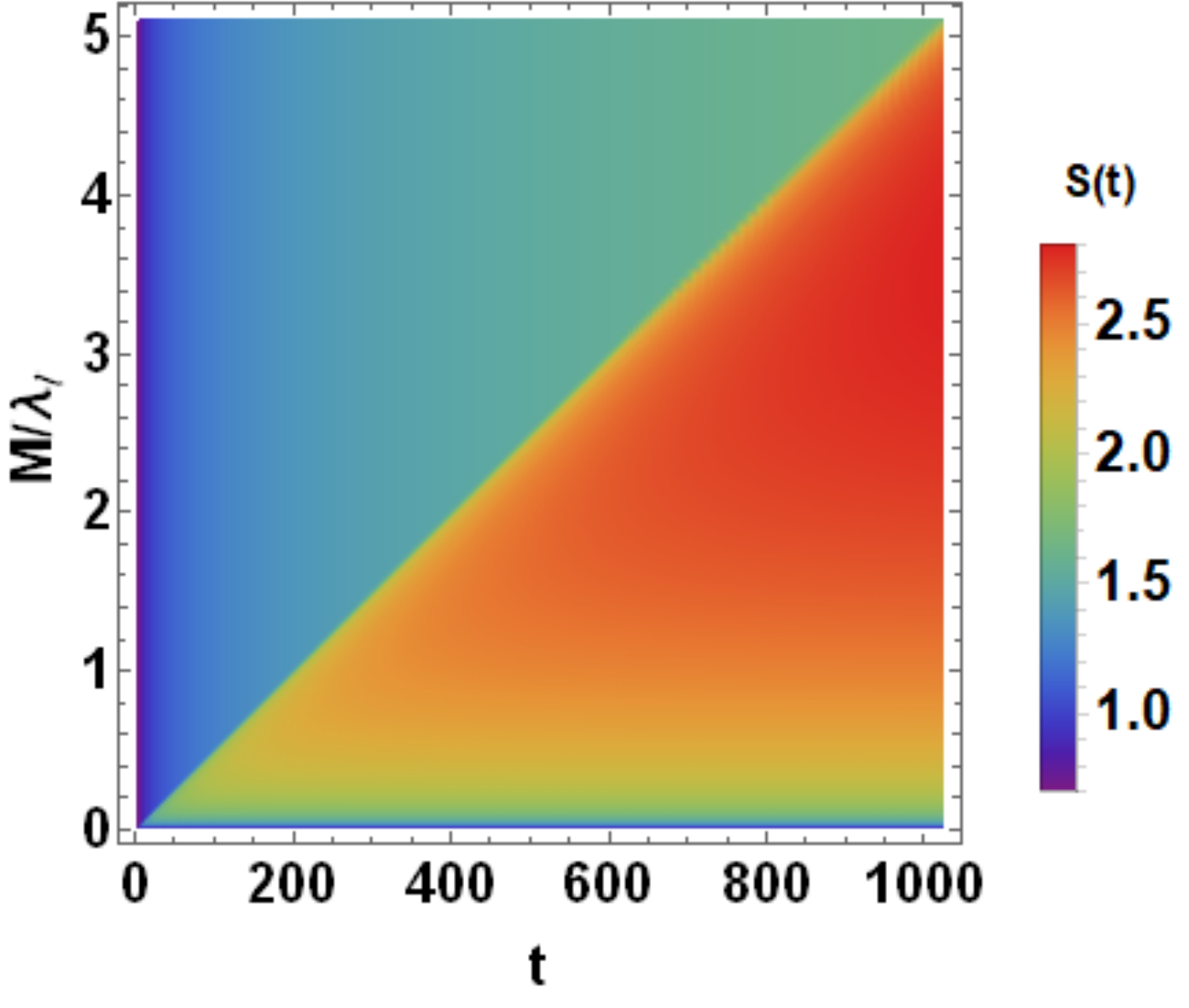}
	\caption{
	Spatio-temporal scale of the enhancement phenomenon. Left panel:- Half chain entanglement entropy (S(t)) as a function of subsystem size (M) (scaled by the maximum Thouless localization length of the system $\lambda_l$) and time (t), for a disordered system of size $L=2048$ and $\mu=\delta \mu=0.1$. Averaging has been done over $80$ disorder realizations. Right:- Same as left panel but for a translationally invariant system. }
				
\label{Scales}
\end{figure*}
%%%%%%%%%%%%%%%%%%%%%%%%%%%%%%%%%%%%%%%%%%%%%%%%%%%%%%%%%%%%%%%%%%%%%%%%%%%%%%%%

\subsection{The Mechanism}
Entanglement between a system and its environment is generated not only when a particle 
enters into the environment from the system, but also when a particle returns/bounces back from the
environment to the system. The mechanism manifests itself most clearly in spreading of particles on a 
finite, TI system with reflecting boundaries. Let us consider a TI lattice, a  part of which 
(consecutive lattice sites) entirely filled (system) with particle, and the rest completely empty (environment).
Now for a quench with such a domain wall like initial states, 
as is already known, the entanglement between the system and the environment (measured by the entanglement
entropy $S$ of any of them) grows logarithmically with time. However, if the system is finite then 
the particles, after exiting the system, traverse the environment, gets reflected from the lattice 
boundary and re-enter the system. This produces initial jumps in $S$ at regular time intervals proportional 
to the average time taken by the particle to leave the system and re-enter. Fig.~\ref{StepMech} precisely shows this.
 Here we consider a setup where the initial state is a half-filled
domain wall state. We refer to the initially filled half of the lattice of size $L$ as ``system", and 
the initially empty half as the environment. The first hike in $S$ is close to $t=0$, due to the entry 
of the particles to the environment right at the onset of the dynamics. 
The next step/jump (the first visible one in the Fig.~\ref{StepMech}) appears around  $t \sim L = 256$, 
which is roughly the time elapsed between the exiting of a particle from the system and 
re-entering it after getting reflected from the lattice boundary once, after traversing the environment. 
The subsequent early jumps appear around $t = n L$ where $n$ is an integer. But with the melting 
of the domain wall the interval between the jump changes, and also, since the system is finite, the steps 
flatten out eventually approaching the saturation.
%%%%%%%%%%%%%%%%%%%%%%%%%%%%%%%%%%%%%%%%%%%%%%%%%%%%%%%%%%%%%%%%%%%%%%%%%%%%%%%%

When TI is broken extensively by disorder, then particles scatter back incoherently to the system 
from all parts of the environment, and that results in a steady increase in $S$ -- the steps
due to contribution from different scatters superimposes, resulting in a smooth 
power-law(approximately linear as discussed in previous section) growth of $S,$
as shown in Fig.~\ref{StepMech}.
It is worth noting that the breaking of TI does not produce a net enhancement in the EE in cases
of the local or global quenches in system with relatively sparse and uniform particle distribution \cite{calabrese4,Igl_i_2013,gobert,Eisler_2007}
Though there is a contribution of back-scattering due to the breaking of TI 
leading to an enhancement in the EE also in those cases, it does not
supersede the cut in the EE growth produced by the effect of localization there. This is because
in those cases the growth of EE relies mainly on the transport of extensive number of 
particles/quasi-particles from all over the system, and localization limits that 
in a severe manner. This extensive cut cannot be compensated by the enhancement due to the
back scattering, which is proportional to the area of the domain boundary in the localized case (only comes 
from the states within a localization length around the domain boundary). 
%Sec. \ref{secrobust} which discusses the robustness of our results when different initial states are considered sheds some more light on this matter.}

%%%%%%%%%%%%%%%%55
%%%%%%%%%%%%%%%%55
\subsection{The Relevant Length and Time Scales}
%%%%%%%%%%%%%%%%%%%%%%%%%%%%%%%%%%%%%%%%%%%%%%%%%%%%%%%%%%%%%%%%%%%%%%%%%%%%%%%%
Clearly, the intriguing feature here is the dynamics of entanglement within a length-scale (subsystem size)
comparable to the localization length. Needless to say, far beyond the localization length, there will be 
practically no spreading of particles due to localization in the disordered cases, and hence the entanglement
growth will saturate to a value which is proportional to the localization length Fig.~\ref{saturation} shows the different saturation values corresponding to different localization length. For a low to intermediate disorder strength where the localization length is large but smaller compared to the length of the system, we see a power law fit $S^{sat}\sim \frac{1}{\delta \mu^{1.82}}$, which clearly indicates the rapid rise in saturation entanglement entropy with larger localization lengths.  In contrast, in the TI case, the
particles will eventually spread everywhere, hence, in spite of the logarithmically slow growth of 
entanglement in this case, the half-chain entanglement for the TI case will overtake the values for the 
disordered cases after sufficiently long time. However, this timescale, in which the TI value 
will match the saturation value of the disordered case, can be enormously large, especially for a weakly
disordered system, since the growth is logarithmically slow in the TI case (see Supplementary material).

The relevant time-scales and length-scales can be studied by taking the subsystem to be of length $M,$ and
putting one end of it at the domain wall boundary and studying the growth of the entanglement entropy $S$
of the subsystem as function of the evolution time $t$ and subsystem size $M$, as shown in Fig.~\ref{Scales}. 
The left panel of Fig. \ref{Scales} shows the value of sub-system entanglement entropy $S$ as a function of the evolution 
time $t$ and the sub-system size $M$ (scaled by the Thouless localization length $\lambda_{l}$). The right frame shows the TI case. 
%%%%%%%%%%%%%%%%%%%%%%%%%%%%%%%%%%%%%%%%%%%%%%%%%%%%%%%%%%%%%%%%%%%%%%%%%%%%%%%%
%%                                   Fig - 4  
%%%%%%%%%%%%%%%%%%%%%%%%%%%%%%%%%%%%%%%%%%%%%%%%%%%%%%%%%%%%%%%%%%%%%%%%%%%%%%%%
\begin{figure}[h!]
\includegraphics[width=0.85\linewidth, height=0.6 \linewidth]{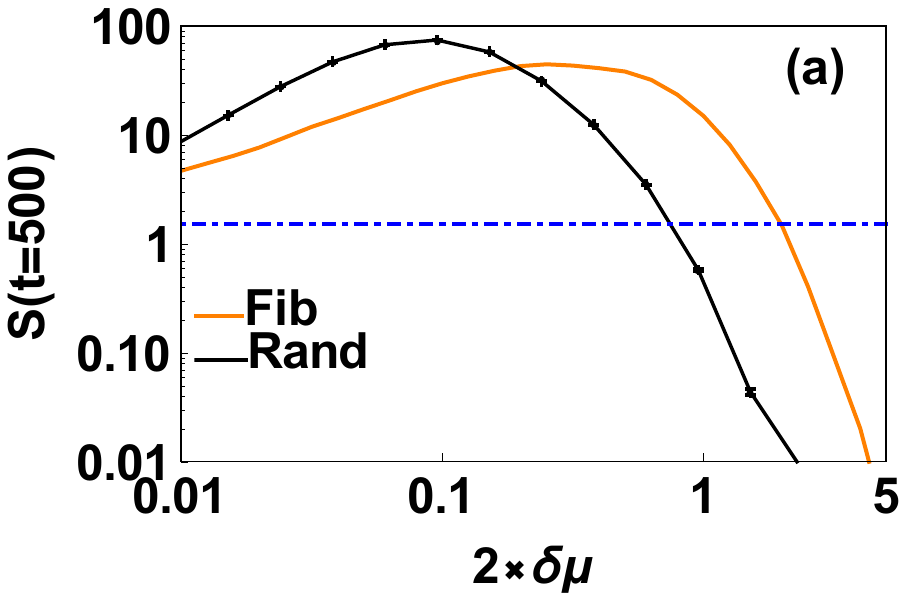}
\includegraphics[width=0.85\linewidth]{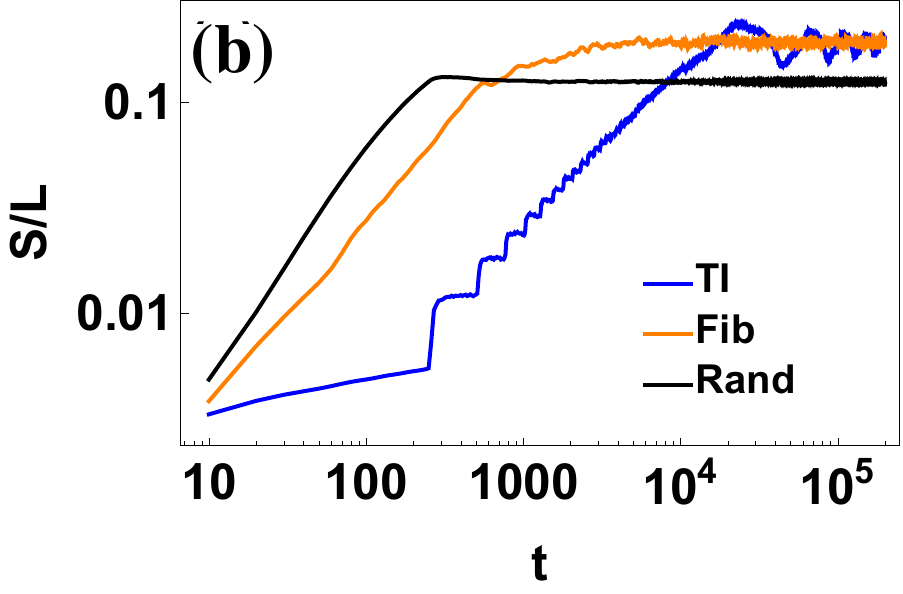}\\
\includegraphics[width=0.85\linewidth]{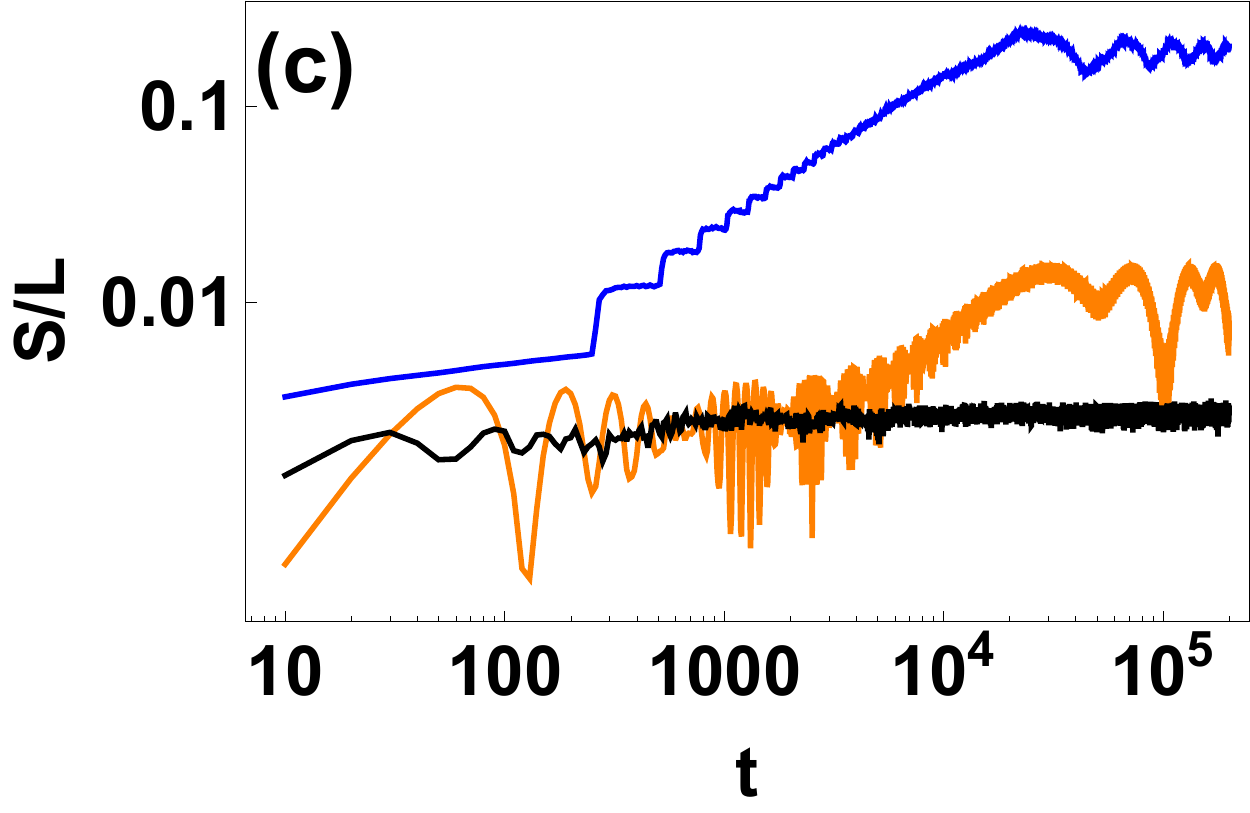}
	\caption{
	The figure demonstrates
	the competing effects produced by the disorder on enhancement of the entanglement growth
	of a subsystem of one-dimensional lattices 
	(see text for the details of the geometry), compared to the TI case of same geometry. 
	Frame {\bf (a)} shows the consequent non-monotonic behaviour 
	of the entanglement $S(t=t^*)$ as a function of the disorder
	strength $\mu$ for a system size $L=2048$. Here we have taken $t^{*}=500.$ 
	The horizontal dash-dot (blue) line marks the 
	value of $S(t=500)$ for the TI case. Frames {\bf (b)} and {\bf (c)} 
	shows  the dynamics of $S$ for $\mu=\delta\mu = 0.1 $ and $\mu=\delta\mu=5.0$
	respectively. We have taken $L=377$ for the Fibonacci lattice and $L=256$ for the random case.
		}
\label{deltamu}
\end{figure}
%%%%%%%%%%%%%%%%%%%%%%%%%%%%%%%%%%%%%%%%%%%%%%%%%%%%%%%%%%%%%%%%%%%%%%%%%%%%%%%%

From the left frame we see, for a given $M,$ the entanglement $S$ saturates rapidly with time. 
The saturation time naturally increases with $M$ (roughly linearly). As $M$ continues to increase,
the saturation time  eventually becomes independent of $M$, i.e., saturates as a function of $M,$
close to $M \sim \lambda_{l}.$
This is because it is actually the underlying localization length that dictates the final extent of 
spreading. To visualize this clearly, we fix $t$ to a sufficiently large value 
(so that saturation is reached for all $M$), and $M$ is increased. 
Then the value to which $S$ saturates, increases with $M,$ meaning 
the saturation value of $S$ scales with $M$ (volume law). However, as $M$ crosses 
the localization length $\lambda_{l}$ ($M/\lambda_{l} \approx 1$), then saturation value of $S$ 
(as a function of $M$) stops increasing any further and settles to a maximum value. 
This is because the spreading of the particles beyond the localization length is 
negligible, hence increasing the subsystem beyond that does not contribute to further enhancement of 
the saturation value of $S.$  The right panel shows the TI case.
Here there is of course no typical length-scale beyond which the saturation time for a fixed $M$ is
independent of $M$ - it is always proportional to $M,$(albeit logarithmically, see Supplementary Material) as can be seen clearly from right panel of Fig. \ref{Scales}.  The linear
contour for saturation of $S$ is formed diagonally. This is expected, since in a TI system the 
entanglement grows till the particles spread over the whole of $M$ (there is no smaller cutoff, e.g., 
due to the localization). The enhancement of spreading of $S$ in the disordered case compared to the TI case
is limited to within the length-scale of order $\lambda_{l}.$ Beyond that scale, the growth in the TI case
continues, while that disordered case comes to a saturation.

%%%%%%%%%%%%%%%%55
\subsection{Tuning the Disorder Strength}
%%%%%%%%%%%%%%%%%%%%%%%%%%%%%%%%%%%%%%%%%%%%%%%%%%%%%%%%%%%%%%%%%%%%%%%%%%%%%%%%%%%%%%%%%%%%%%%%%%%%%%%%%

%%%%%%%%%%%%%%%%%%%%%%%%%%%%%%%%%%%%%%%%%%%%%%%%%%%%%%%%%%%%%%%%%%%%%%%%%%%%%%%%
%%                           Fig - 5
%%%%%%%%%%%%%%%%%%%%%%%%%%%%%%%%%%%%%%%%%%%%%%%%%%%%%%%%%%%%%%%%%%%%%%%%%%%%%%%%
\begin{figure}[h]
\centering{
\includegraphics[width=0.75\columnwidth]{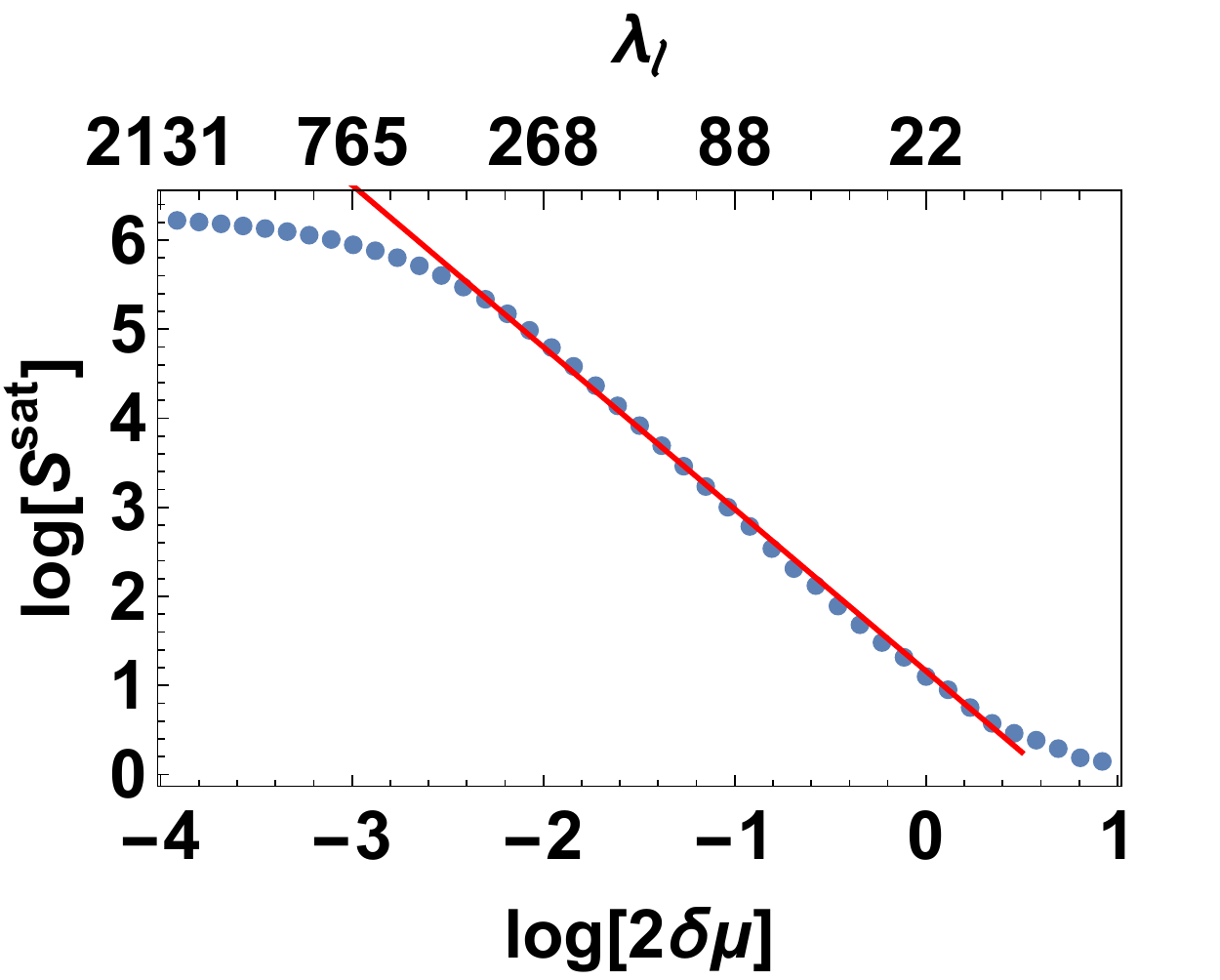}}
\caption{Plot of $\delta \mu$ vs $S^{sat}$ which denotes the saturation value of half chain entanglement entropy after time $t\sim 10^{10}$ for $L=4096$. The x-axis is also labelled by the maximum Thouless localization length $\lambda_{\mathit{l}}$ corresponding to $\delta \mu$. The red line denotes fit to $1.16-1.82 \log(2 \delta \mu)$ }
\label{saturation}
\end{figure}
%%%%%%%%%%%%%%%%%%%%%%%%%%%%%%%%%%%%%%%%%%%%%%%%%%%%%%%%%%%%%%%%%%%%%%%%%%%%%%%%%%%%%%%%%%%%%%%%%%%%%%%%%
The enhancement is most prominent for an intermediate disorder strength. The disorder has two 
competing effects -- on one hand it enhances the back-scattering rate which aids the growth of $S,$
on the other hand, it strengthens localization which arrests the spreading of the particles and
suppresses the growth. This results in a non-monotonic behaviour of $S(t^*)$ measured after a given 
evolution time $t^{*},$ as a function of the strength $\delta\mu$ of the disordered
part of the potential in the respective Hamiltonians for the Fibonacci and random cases.
Initially, $S^{*}$ increases with $\delta\mu$ compared to the value of $S^{*}$ for the 
TI case due to enhanced back scatterings, and reaches a peak. 
Then, with further increase in $\delta\mu$ it reduces sharply and falls below the TI value.
This is demonstrated in Fig.~\ref{deltamu}. Here we consider an initial domain wall state of
width $L/2$ for a lattice of system-size $L= 2048$
 and a subsystem $AB$ of length $M = L/2 - 20,$ with its end $A$ at $i=L/2 + 21,$
i.e., $21$ lattice sites away from the domain wall boundary (see Fig.~1). The entanglement entropy $S$ of $AB$
is measured.(See also App.\ref{appC}) The competing effect of the disorder strength is reflected in the non-monotonic behaviour of 
$S(t=500)$ as a function of $\delta\mu.$ Dynamics for different values of $\delta\mu$ are also shown in the
Fig. \ref{deltamu}~(b) and (c). While for $\delta\mu=0.1,$ the disorder clearly results in a pronounced enhancement compared to the
$TI$ case(though the random disorder shows a faster saturation at late times), for $\delta\mu=5.0,$ the growth is suppressed below that in the TI case even at small timescales. This competing effect is also visible in Fig.~\ref{masterfig} (b) where the mutual information between the two subsystems in the random disordered case shows a lower value than the Fibonacci potential due to onset of localization.
%%%%%%%%%%%%%%%%55
%%%%%%%%%%%%%%%%55

%{\color{green} Some more lines here if possible. RESPONSE: I suppressed Fig. 5, since it is not really reflecting the
%above scenario. That would probably need going far beyond the localization length. Here, it had to show that if you
%increase both $M$ and $t$, then in the TI case entanglement entropy can increase indefinitely, but in the disorder
%case it will hit a saturation beyond $M \sim \lambda$, and increasing $t$ further would not help.}

%fig6a,fig6b,fig6c_ii.
%\iffalse
%%%%%%%%%%%%%%%%%%%%%%%%%%%%%%%%%%%%%%%%%%%%%%%%%%%%%%%%%%%%%%%%%%%%%%%%%%%%%%%%%%%%%%%%%%%%%%%%%%%%%%%%%
\subsection{Randomness in the Initial State}
%%%%%%%%%%%%%%%%%%%%%%%%%%%%%%%%%%%%%%%%%%%%%%%%%%%%%%%%%%%%%%%%%%%%%%%%%%%%%%%%%%%%%%%%%%%%%%%%%%%%%%%%%
\label{secrobust}
%%%%%%%%%%%%%%%%%%%%%%%%%%%%%%%%%%%%%%%%%%%%%%%%%%%%%%%%%%%%%%%%%%%%%%%%%%%%%%%%%%%%%%%%%%%%%%%%%%%%%%%%%
%\iffalse
\begin{figure}[h!]
\centering{
\includegraphics[width=0.94\columnwidth]{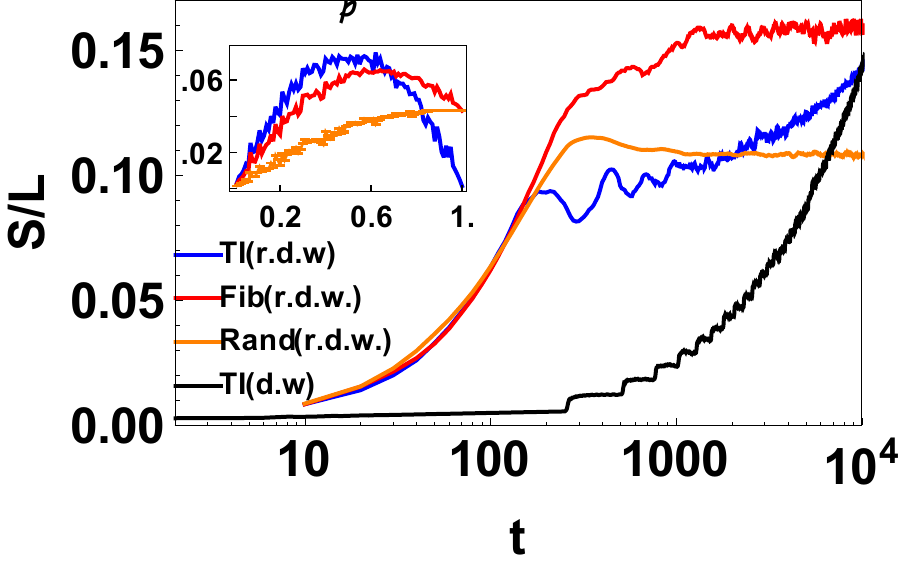}
}
%\includegraphics[width=0.48\columnwidth]{fig7b.pdf}
%\includegraphics[width=0.48\columnwidth]{fig7c.pdf}
%\includegraphics[width=0.48\columnwidth]{fig7d.pdf}\\
%\includegraphics[width=0.48\columnwidth]{fig7e.pdf}
%\includegraphics[width=0.48\columnwidth]{fig7f.pdf}
%\includegraphics[width=0.48\columnwidth]{fig7g.pdf}
%\includegraphics[width=0.48\columnwidth]{fig7h.pdf}}
%\iffalse
	\caption{The figure compares growth of entanglement $S$
	with time for the random domain-wall (r.d.w) initial state (for $\mathcal{P}=0.6$ ) on a TI lattice 
	with those for the domain-wall (d.w.) state in a TI lattice, and r.d.w states on a Fibonacci lattice, and 
	a random lattice respectively. The result shows that the small randomness even just in the
	initial state (r.d.w. case) is sufficient to switch the logarithmic entanglement growth observed in the 
	TI lattice case with the d.w. initial state to a linear growth (see growth till $t \sim 200$) 
	generic to the TI broken lattice cases (also shown for comparison). 
	Results are for ($L =256$, $\delta\mu=0.2$). 
	The inset shows behaviour of $S$ at $t=500$ for $L=1024$ as a function of $\mathcal{P}$ (see text for details) 
	for an r.d.w state .
	}
%	\fi  
\label{robust}
\end{figure}
%\fi
%%%%%%%%%%%%%%%%%%%%%%%%%%%%%%%%%%%%%%%%%%%%%%%%%%%%%%%%%%%%%%%%%%%%%%%%%%%%%%%%%%%%%%%%%%%%%%%%%%%%%%%%%
We have also studied the effect of randomness put solely in the initial state keeping the TI of the lattice intact.
To this end we construct the random domain-wall state (r.d.w.) as follows.
Let $r$ be a pseudorandom number between $0$ and $1$, then the initial state is 
chosen such that, $\langle c_i^{\dagger}c_j\rangle=\delta_{ij}$ for $i<L/2$ and $r>\mathcal{P}$, and $0$ otherwise. 
$\mathcal{P}$ thus defines the probability a site is occupied in the left half of the system. The right half is
kept completely empty. We only use one representative initial state for each probability value to show the scenario, 
and for the disordered Hamiltonian we average over $80$ disorder realizations. 
Our result shows that the logarithmically slow growth of entanglement in the TI case with a single 
domain-wall (d.w.) initial state~\cite{alba, Mossel_2010,XXmodel,calabrese3} is a fine-tuned result. 
We find, the small randomness present in an r.d.w initial state is enough to make 
the growth behaviour switch from logarithmic to a linear time dependence in a TI lattice. 

The figure also shows, a further enhancement of entanglement growth 
(over the TI case with r.d.w) can be achieved in a quantitative level if disorder 
is added to the lattice. This increase however is visible for $\mathcal{P}>\mathcal{P}_c$,dependent of type of disorder as is evident from the inset of the plot. This is expected as both being similar effects, the effect due to disorder in initial state must be low for the effect due to disorder in the potential to be seen. In this sense, the phenomenon of enhancement of entanglement 
due to breaking of lattice TI is not fine-tuned to a particular initial state, 
and is quite robust to the random variation over the r.d.w. initial states we have considered.

\section{The periodic potential case}
\label{periodic}
\begin{figure}[h!]
\centering{
\includegraphics[width=0.48 \columnwidth]{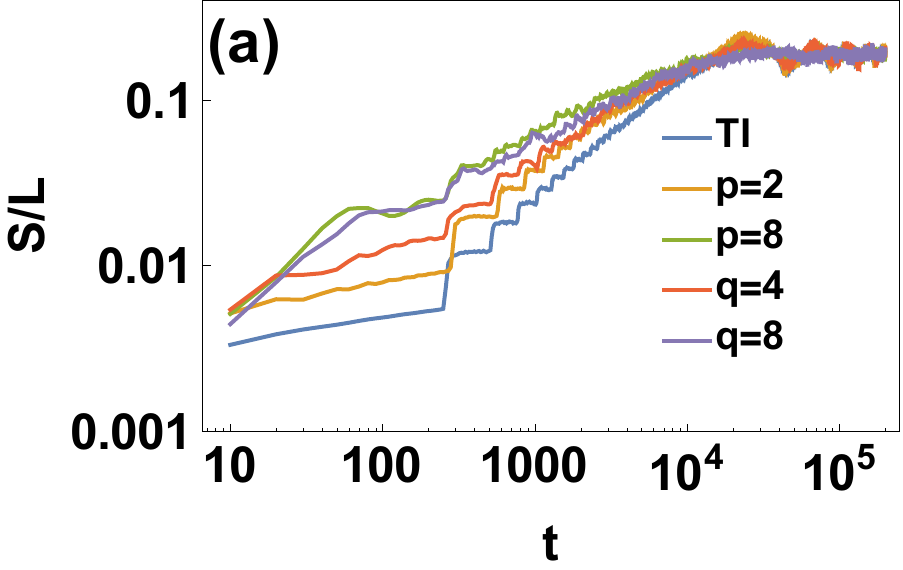}
\includegraphics[width=0.48 \columnwidth]{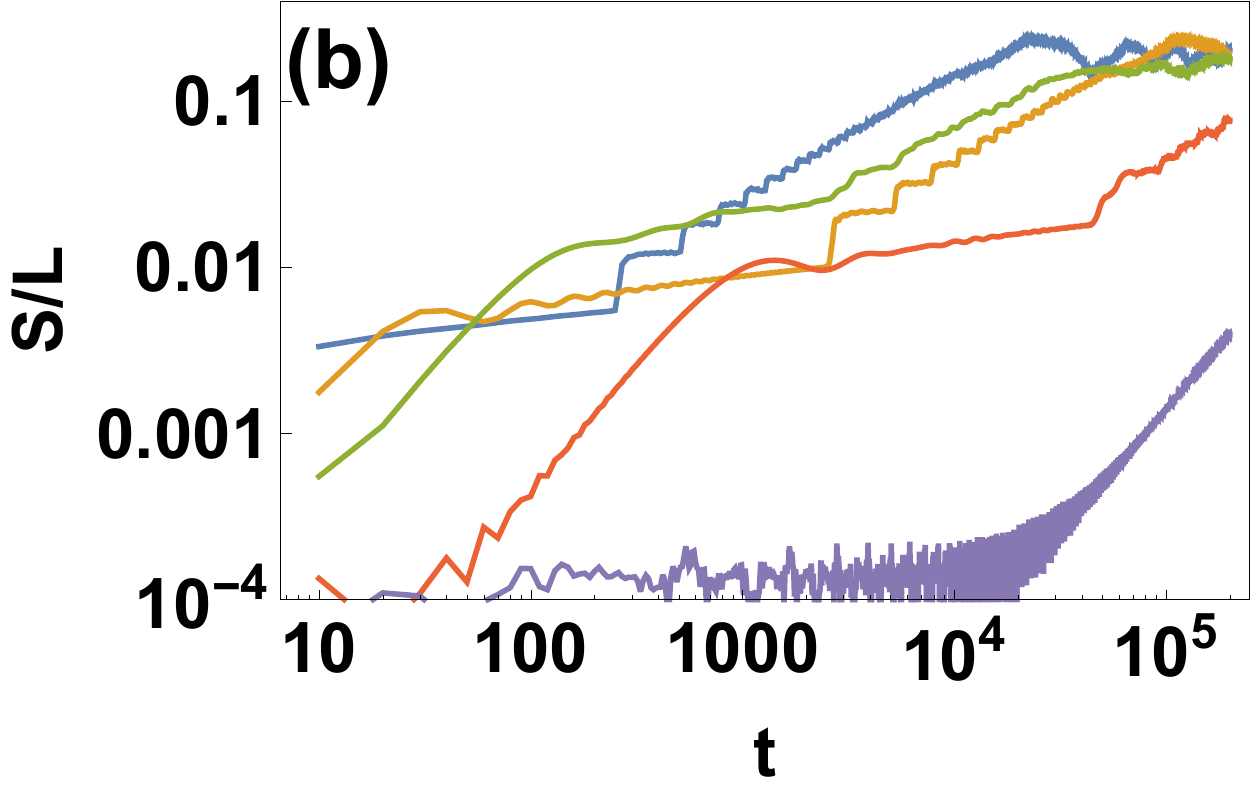}}
\caption{ Plot of a comparison of half chain entanglement entropy growth of a system of length $L=256$ for translationally invariant system (TI)and periodic systems of both kinds with periodicity $p=2,8$ and $q=4,8$ for (a) $\mu=\delta\mu=0.1$ and (b) $\mu=\delta\mu=5$, showing the increase of entanglement entropy with higher periodicity for both $p$ and $q$ type of periodicity for lower strength of potentials, and also showing the difference between them for a large potential strength. Note that $p=2$ and $q=2$ defines the same potential}\label{longperiodic}
\end{figure}

\begin{figure}[h!]
\includegraphics[width=0.48 \columnwidth]{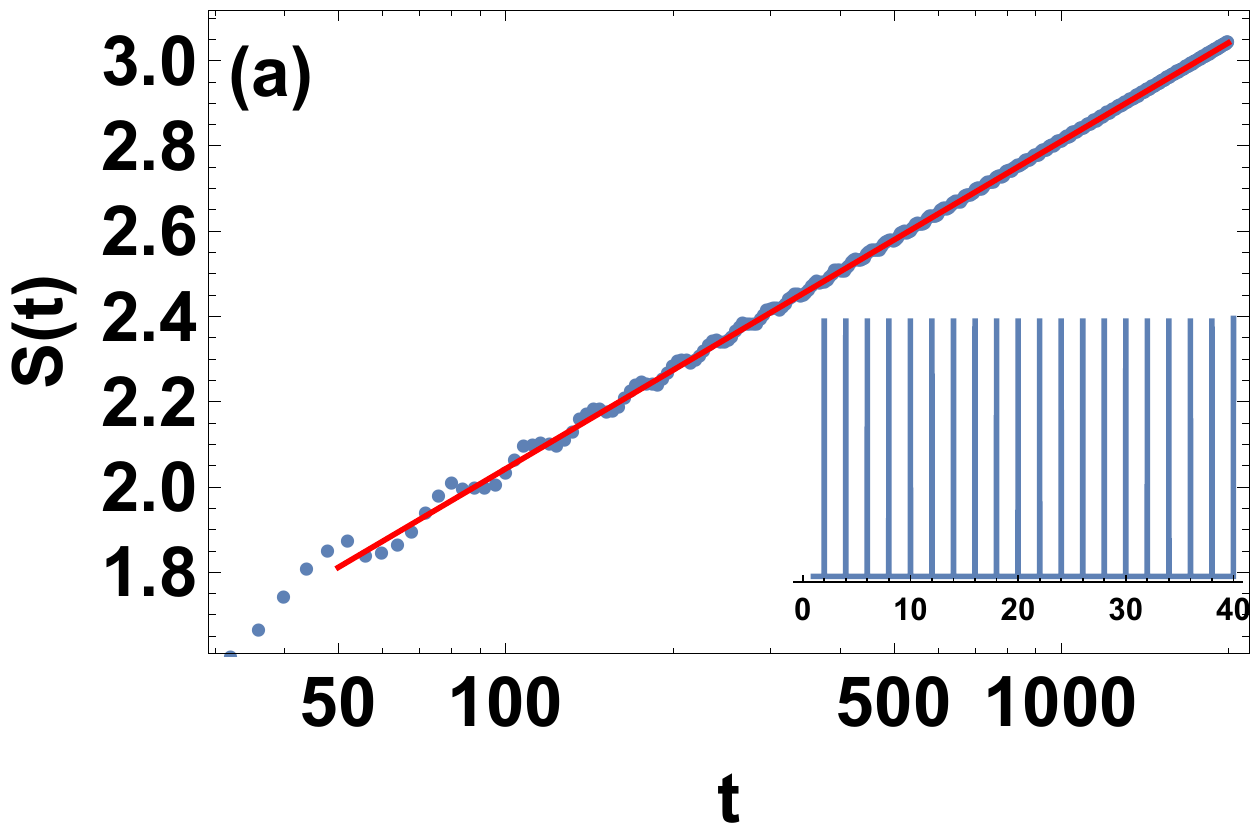}
\includegraphics[width=0.47 \columnwidth]{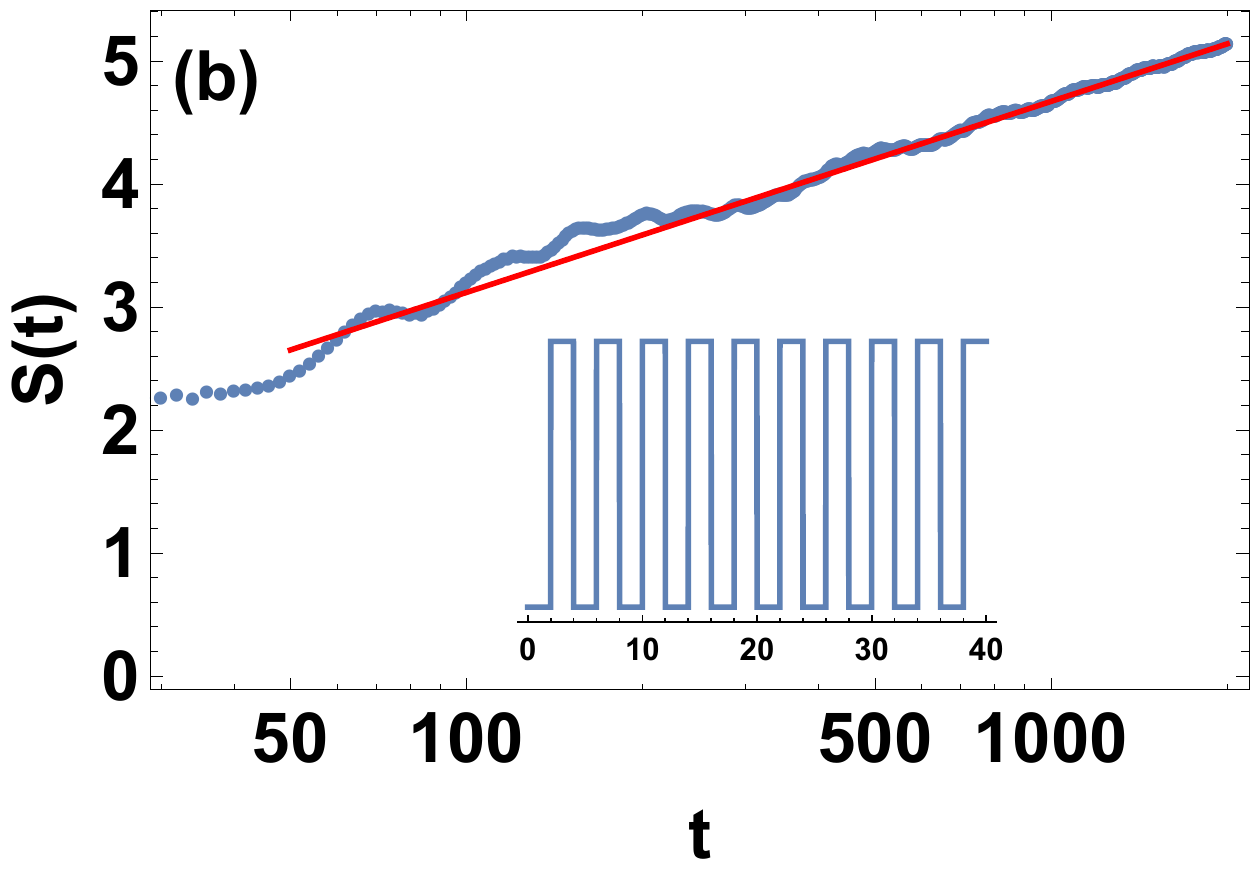} \\
\includegraphics[width=0.48 \columnwidth]{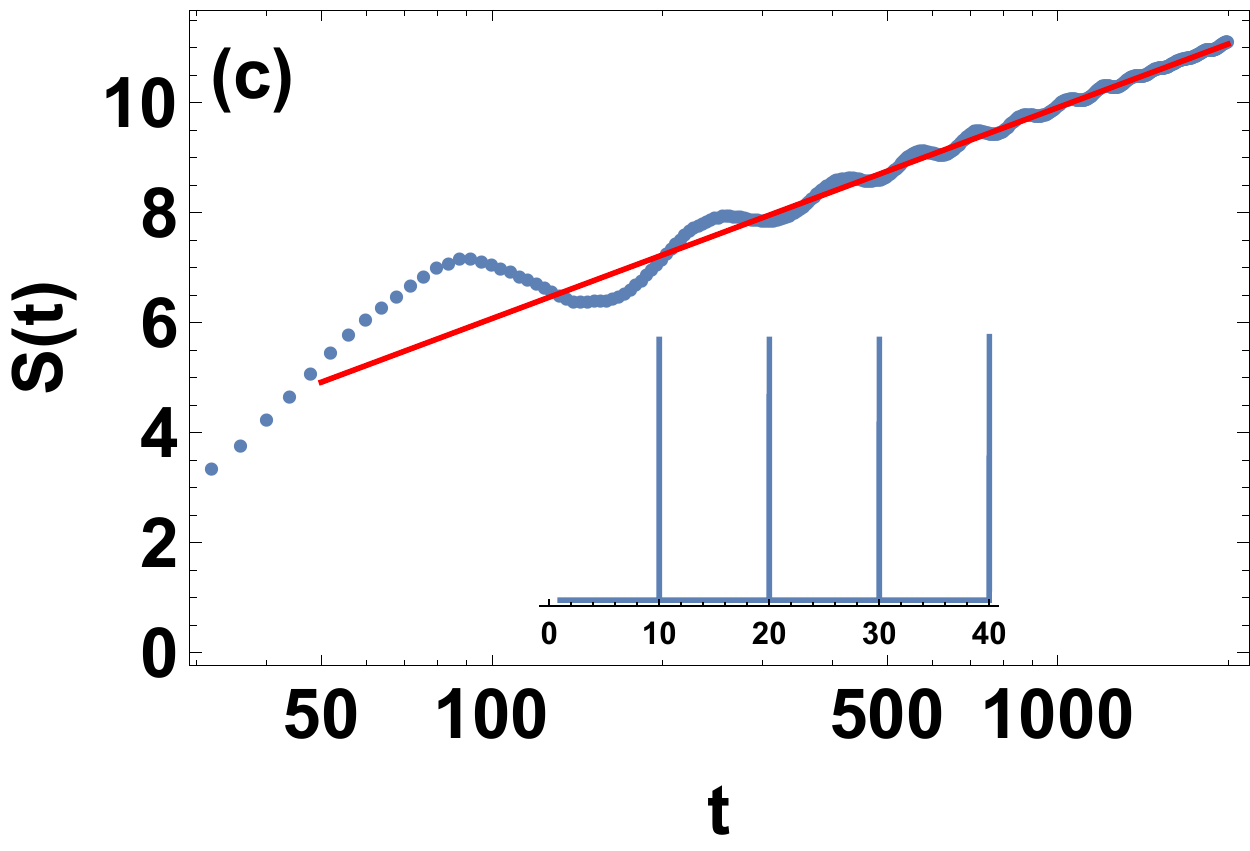}
\includegraphics[width=0.48 \columnwidth]{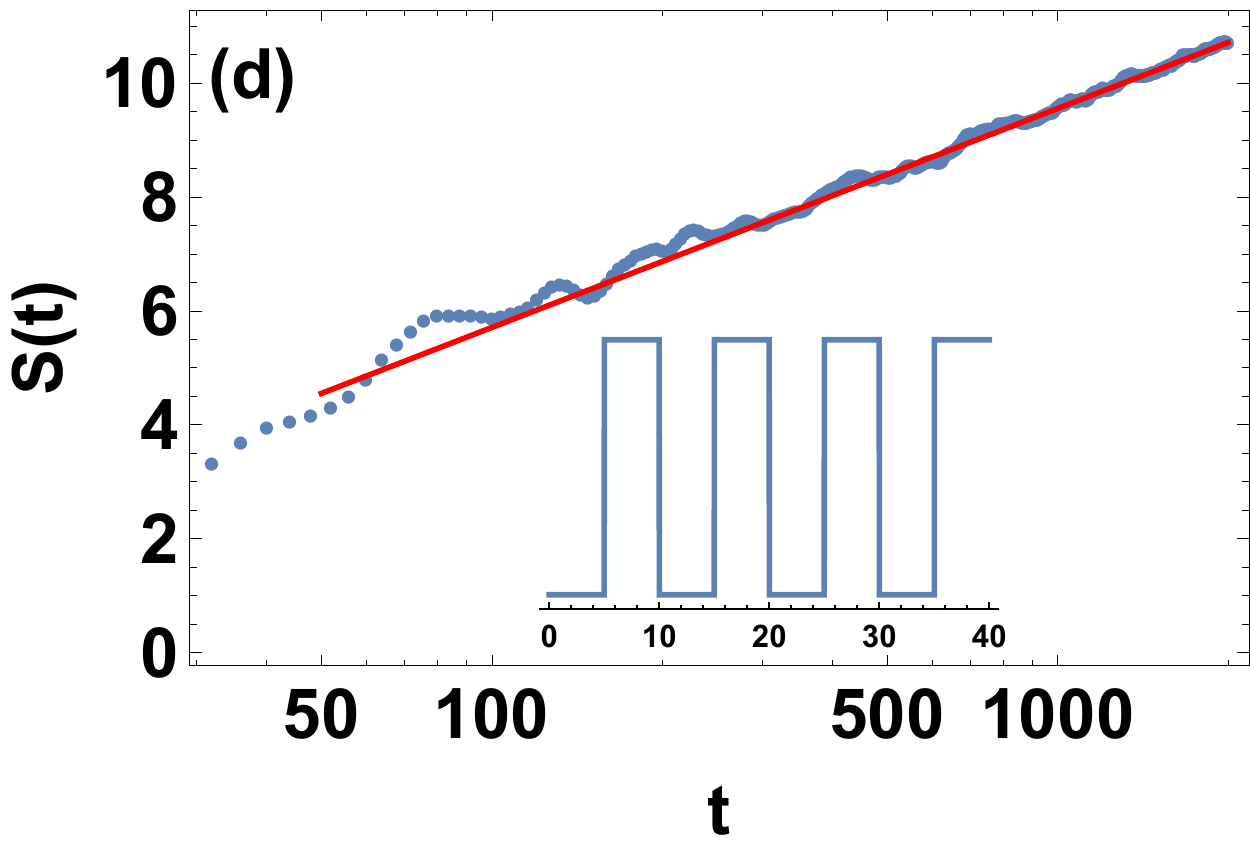}
\caption{(a) Plot showing the logarithmic fit of half chain entanglement $S(t)$ for $p(q)=2$ for a system size of $L=4096$ , the red line is given by the equation $S(t)=0.5+0.333\log(t)$ (b) Similar plot but for $q=4$ and the red line is given by the equation $S(t)=0.01+0.67\log(x)$.(c) Same as (a) with $p=10$. The red line indicates the fit $S(t)=-1.6047+1.66\log t$(d) Same as (c) but with $q=10$, the red line denotes a fit $S(t)=-1.98+1.67 \log t$. The inset in each figure shows the periodicity of the on-site potential used to simulate each plot.   }
\label{logfitt}
\end{figure}

In this section we will skim over the scenario where we break the translational invariance of the system by introducing a spatially periodic potential. Clearly such a system still possesses a translational symmetry of $Z_n$ class and thus is different from the systems in the previous section which had no such symmetry. Figs. \ref{longperiodic} show results for the dynamics of the system when a periodic potential is applied, for the same parameter values as Fig. \ref{deltamu}(b) and (c). Here we clearly see that for low $\delta \mu$, entanglement entropy value increases with increasing periodicity $p$ and $q$. This is a counter-intuitive result in the first glance since one would expect higher number of scattering centres present for lower values of $p$ and $q$. However, due to the periodicity of the lattice the scattered wavefronts interfere coherently unlike in the previous section which results in a counter intuitive scenario. For a concrete understanding, one requires a thorough analytical analysis of this phenomenon, which is involved and left for a future work. In this present work, we focus on the various interesting features which occur in this setup.\\
As can be seen from Fig. \ref{longperiodic}(b) for large values of $\delta \mu$ entanglement entropy for the $q$ type periodic potential is heavily suppressed and $q=8$ show lower entanglement than $q=4$ contrary to the case with $p$ and the results of Fig. \ref{longperiodic}(a). This is simply due to the fact $q$ type potential suppresses particle tunneling more than $p$ type potential due to its nature of being a series of square barrier potentials than a delta function one(shown in inset of Fig. \ref{logfitt}). Additionally, the suppression is larger for larger $q$ because the width of the barrier increases with increasing $q$. Since, at high potential strength particle diffusion plays the most important role in determining entanglement entropy, this factor heavily reduces entanglement growth and we get the current result. \\
 In the rest of this section we would take $\mu=\delta \mu=0.1$ unless otherwise specified. Fig. \ref{logfitt} shows the fit of $S(t)$ for $t<L$ for periodic potentials of different kinds and periodicities for a low potential strength $\mu=\delta \mu=0.1$. It is seen that the growth in $S_{vN}$ follows $a+b \log(t)$ behaviour if $t\sim L$ is sufficiently large for periodic potential systems. This is reminiscent of a similar behaviour seen in conformally invariant TI systems under local/geometric quench \cite{calabrese4, alba}. However, since once we break TI the system is gapped and no longer expected to be conformally invariant, so we do not in general expect this to hold.
 It can be seen after oscillatory behaviour at initial times the fit is quite good. The plot of fit coefficient $b$ vs $p$ is provided in Fig. \ref{deviation}a as well. It is a straight line with slope close to $.167$. For TI systems, it is known that the entanglement growth for inhomogeneous quenches\cite{alba} can be fit at time $t<L$ for large $L$ to $c/6 \log t$ where $c$ , the central charge is of value $1$ for open boundary conditions, which is the configuration we study. For periodic potentials labelled by periodicity $p(q)$ the $c=p(q)$ as evident from Fig. \ref{logfitt}.  As can be seen from the inset of Fig. \ref{logfitt}(a) and (b), the potential type $p(q)$ makes the lattice posses $Z_p(Z_q)$ symmetry . It seems $c=n$ where $Z_n$ is the symmetry of the lattice.\\
\begin{figure}[h!]
\includegraphics[width=0.48 \columnwidth]{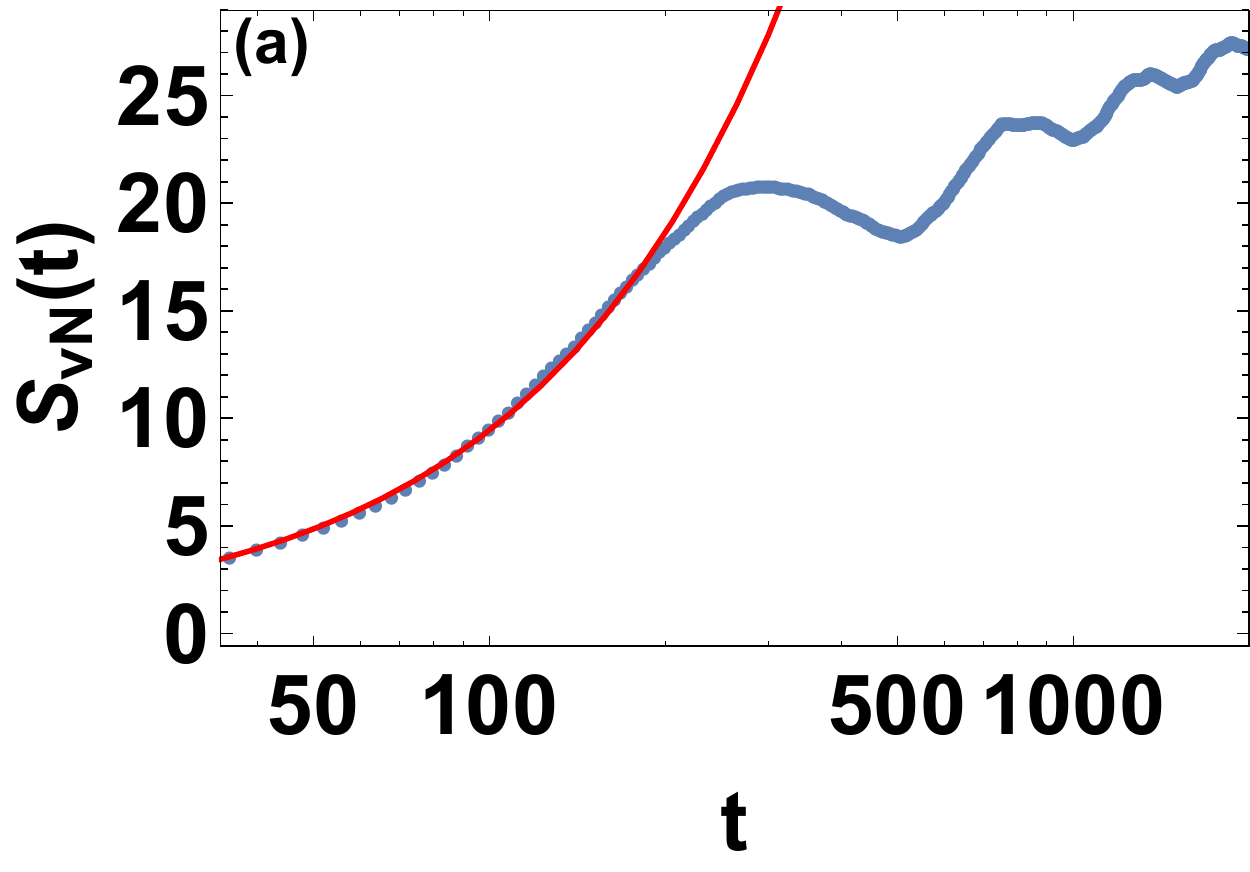}
\includegraphics[width=0.48 \columnwidth]{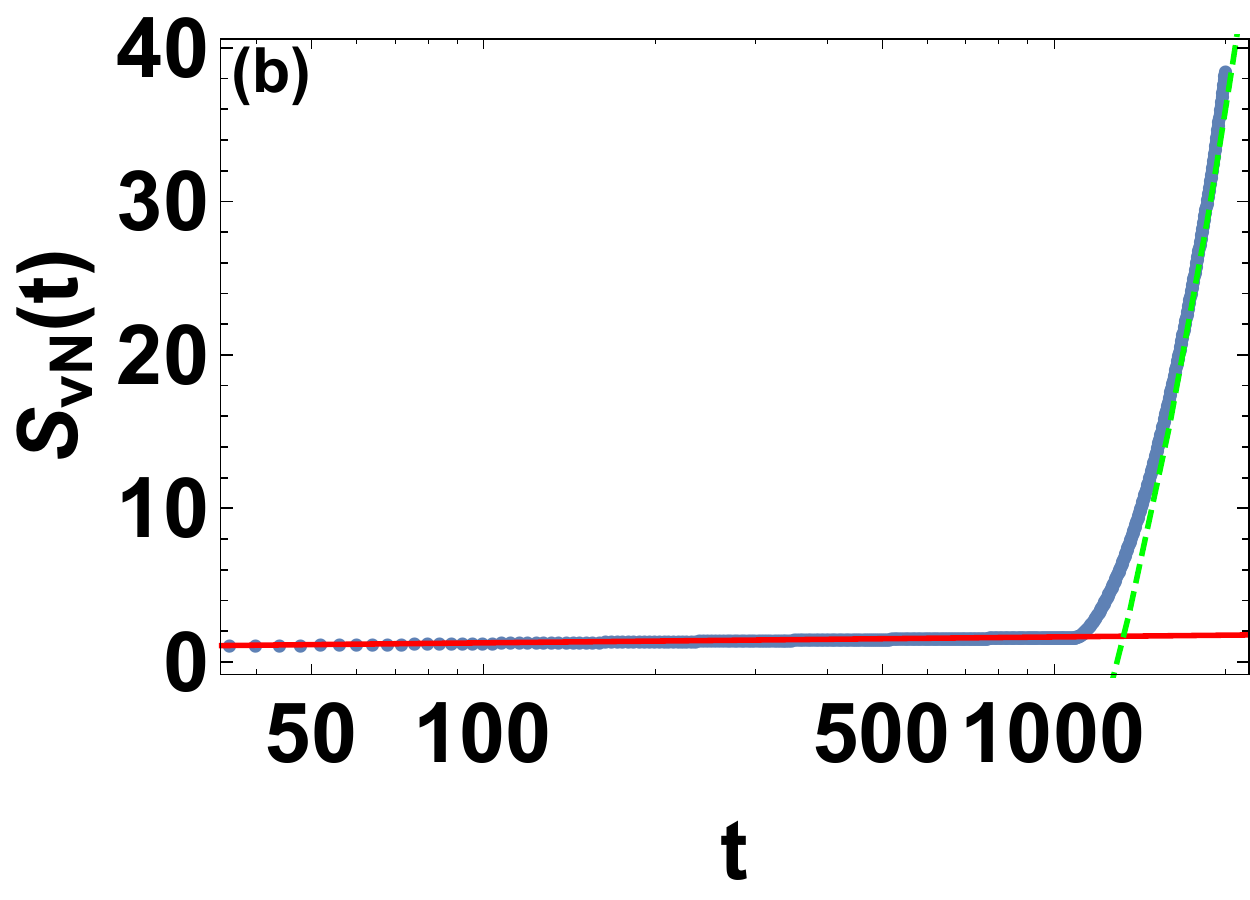}
\caption{(a) Plot showing the linear fit regime of half chain entanglement $S(t)$ for $p=32$ for a system size of $L=4096$ , the red line is given by the equation $S=0.272+0.0917 x$ (b) Plot showing the scenario for $p=1500$ showing the two different fit regimes. The red line shows the fit to $0.467+.167\log(x)$, and the green dashed line shows fit to$-66+.051 x$.  }
\label{linearfit}
\end{figure}

We end this section by showing the limit in which our results correspond to the result of linear growth of EE from a defect site under inhomogeneous quench seen in Ref. \onlinecite{Eisler_2012} and bring in a new perspective. In our case, when $p\ge L/2$ we retrieve their results, since then our system identical to theirs except for position of defect. An example of this can be seen in Fig. \ref{linearfit}(b). In this plot the scattering centre is positioned at $i=3000$ in a lattice of size $L=4096$. Thus it is seen until the wavefront reaches the scattering centre, entanglement follows the $1/6 \log t $ behaviour for TI systems and then we see a linear rise for times larger than it. But in other cases discussed, contrary to their setup, we have multiple scattering centres, hence the wavefront undergoes multiple reflections and the incident and reflected wavefronts interfere in a non-trivial manner to yield $\log(t)$ rise in entanglement. Fig.\ref{deviation} (b) gives an approximate timescale when our results show strong deviations from the linear increase in EE expected from a single defect. In this figure we plot the deviation relative deviation of entanglement entropy($S$) with the linear fit at initial times($t<=2p$), labelled  $S^{fit}$ and plot it against time ($t$) scaled by the periodicity($p$) of the lattice considered. The rescaled time gives us an approximate idea of how many scattering sites the particle has crossed assuming ballistic propagation of particles. (A reasonable assumption for $\delta \mu=\mu=0.1$ for periodic potential. See Supplmentary Material). It shows the linear rise is completely lost due to the interference of wavefunctions after the wavefront crosses $6-8$ scattering sites. Fig. \ref{linearfit}(a) gives the dynamics for a $p=32$ potential which shows how the fit changes from linear to logarithmic.  Thus, one can put a bound of validity of the logarithmic fit in the last paragraph weakly at $p(q)<L/16$ from Fig. \ref{deviation}(b), as at periods lower than $L/16$ for $p$ type potential, the wavefront can cross more than $8$ scattering centres and show the non-linear rise in entanglement.\\
\begin{figure}[h!]
\includegraphics[width=0.48 \columnwidth]{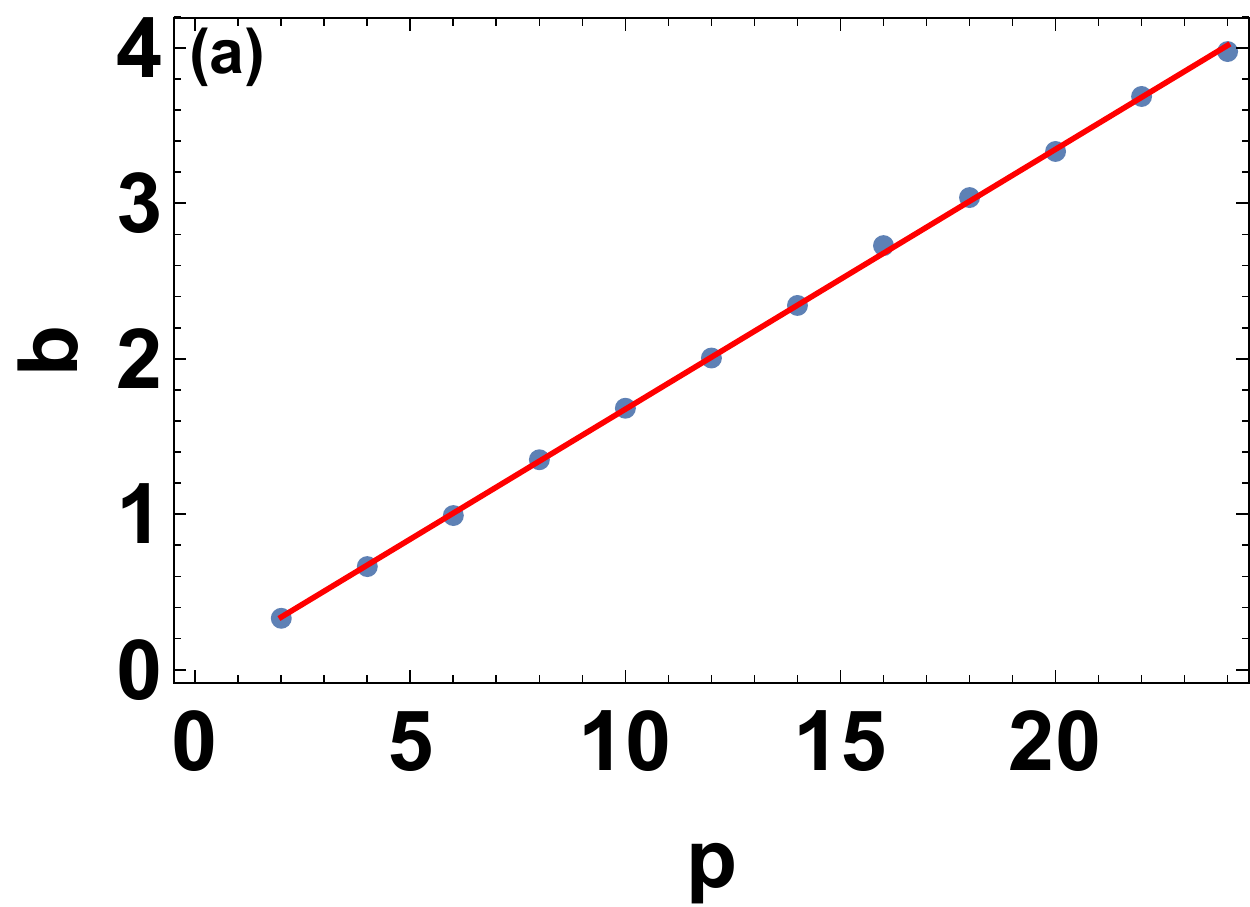}
\includegraphics[width=0.48 \columnwidth]{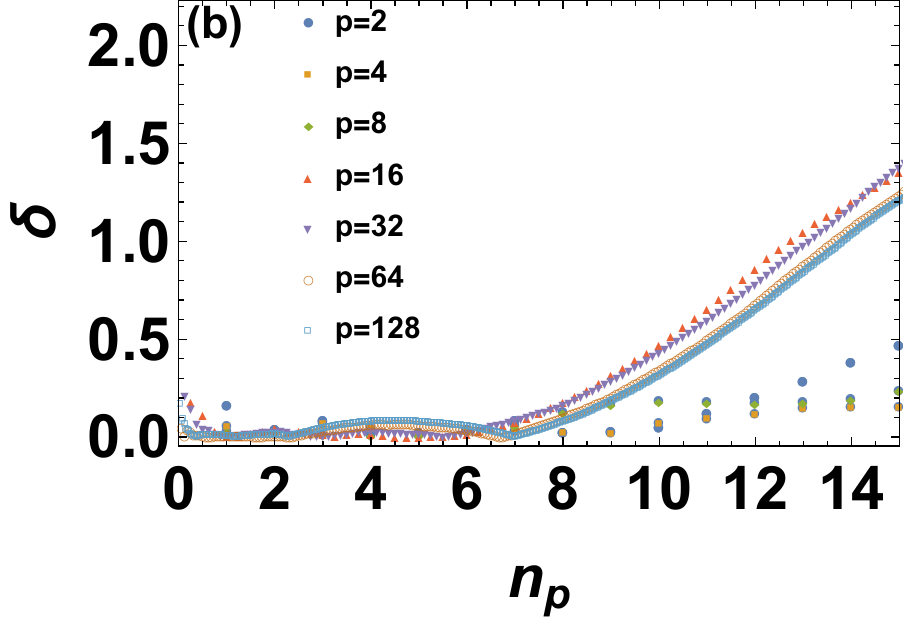}
\caption{(a) Plot of $b$,the coefficient of $\log(t)$ in the fit of $S$ vs $p$, the period of the potential, the red line is given by the equation $S=0.004+0.167 x$ (b) Plot of $\delta=|S-S^{fit}|/S$ vs $n_p=t/p$ showing when the values start deviating from linear fit. See text for details. }
\label{deviation}
\end{figure}
 Fig. \ref{eigenzeta} shows how the eigenvalues $\zeta_l$ of the correlation matrix of the subsystem under consideration behave as the wavefront travels through the lattice. It can be seen that if the timescales are small and the wavefront has not crossed more than one scattering centre, the newer non zero eigenvalues entering into the correlation matrix are $<1$. The analysis of Ref. \onlinecite{Eisler_2012} points out that in such a case, there will be a linear rise in entanglement due to a continued influx of $\zeta_l \neq 0,1$ generated by the propagating wavefront. But once $t>2p$ this analysis is no longer valid as as the wavefront travels through more scattering centers, the scattered and incident wavefronts undergo interference, and there can be new $\zeta_l$ which are $0$ or $1$ thus preventing the linear rise in entanglement entropy. A comparison of Fig. \ref{eigenzeta}(b) and (d) reveals exactly this difference. This happens due to the coherent interference of scattered wavefronts which was absent in the setup considered by Ref. \onlinecite{Eisler_2012}, however a thorough understanding via analytical calculations is left for a future work. It is interesting to note that for disordered systems as Fig. \ref{eigenzeta}(e) and (f) shows, the interference lacking the coherence of periodic systems, this effect is suppressed and hence the rise in entanglement entropy becomes faster and closer to single defect systems. 

\begin{figure}[h!]
\centering{
\includegraphics[width=0.48 \columnwidth]{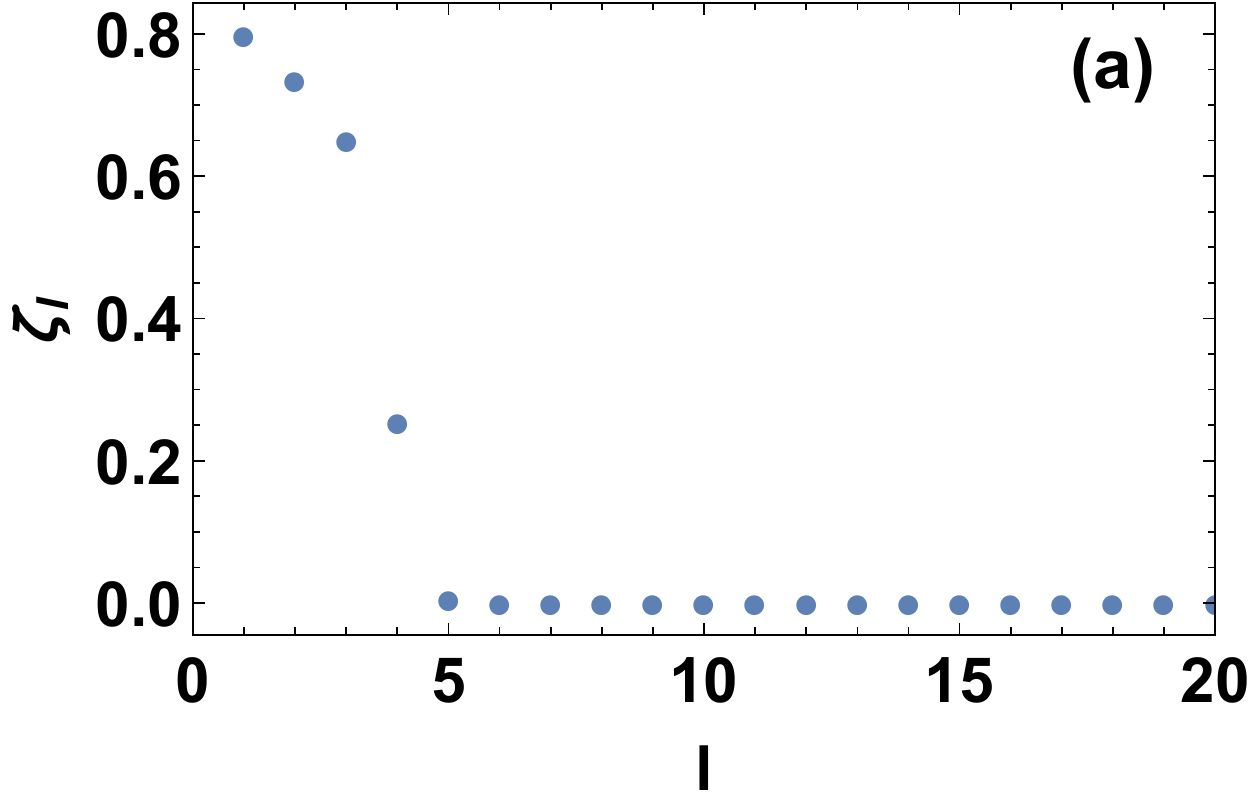}
\includegraphics[width=0.48 \columnwidth]{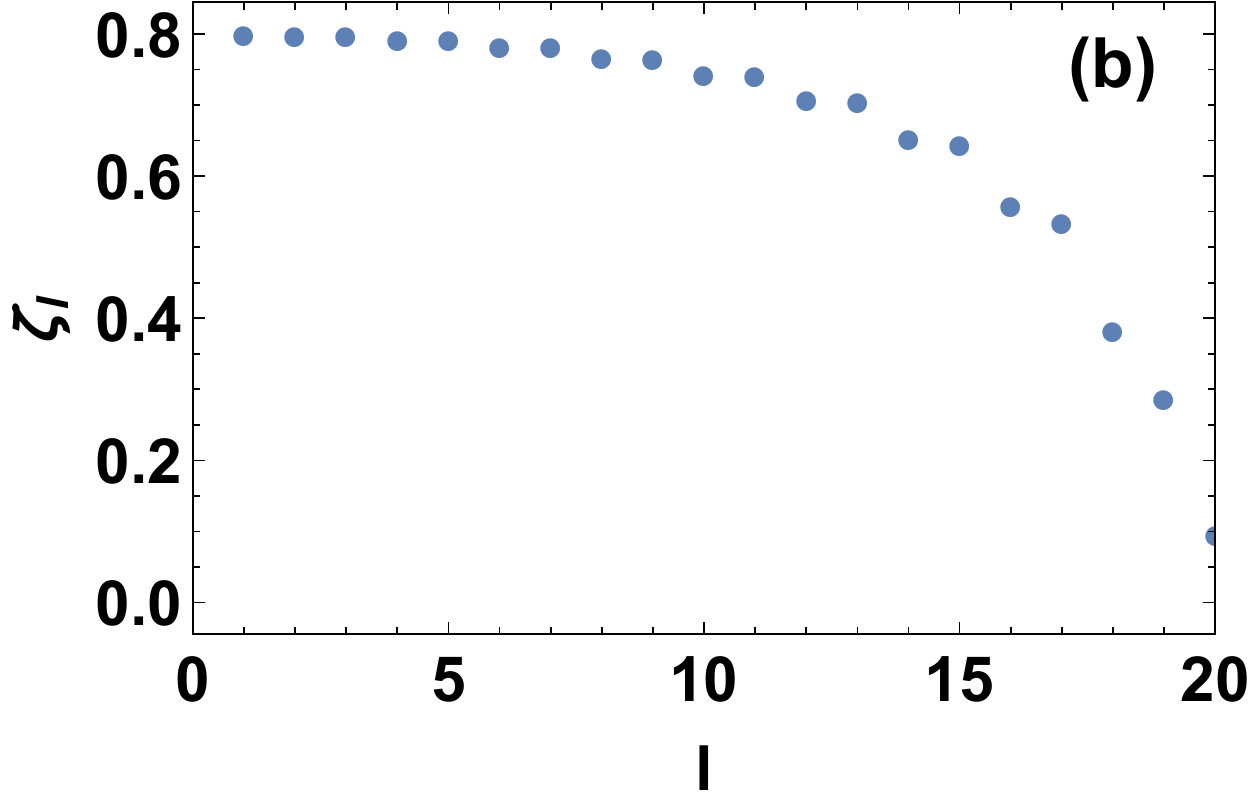}\\
\includegraphics[width=0.48 \columnwidth]{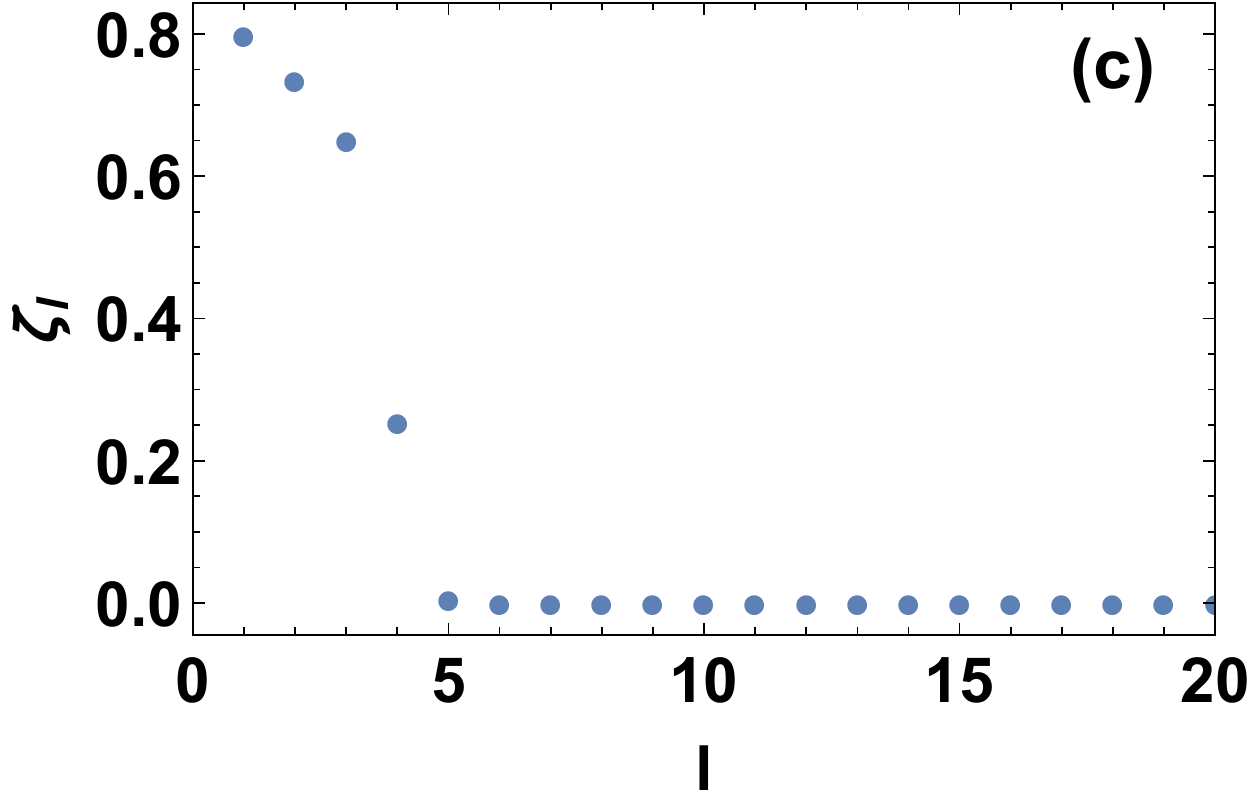}
\includegraphics[width=0.48 \columnwidth]{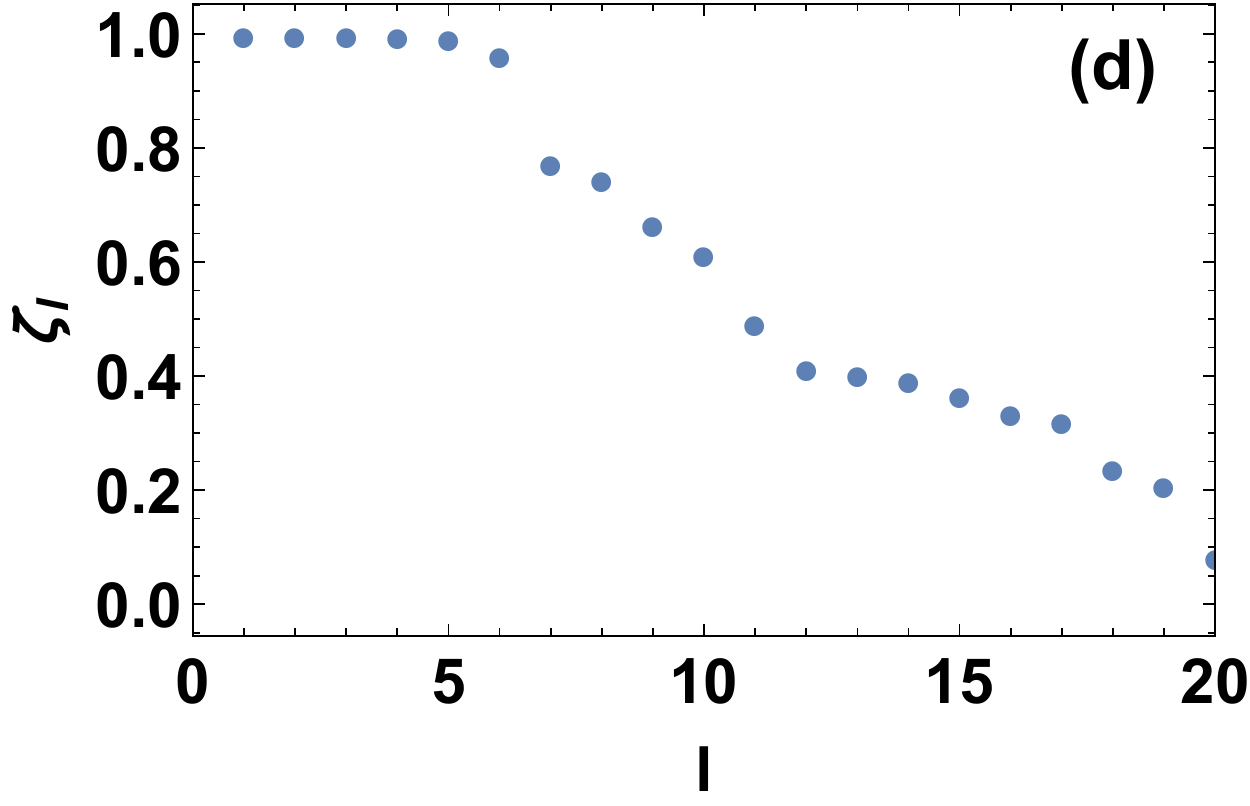}\\
\includegraphics[width=0.48 \columnwidth]{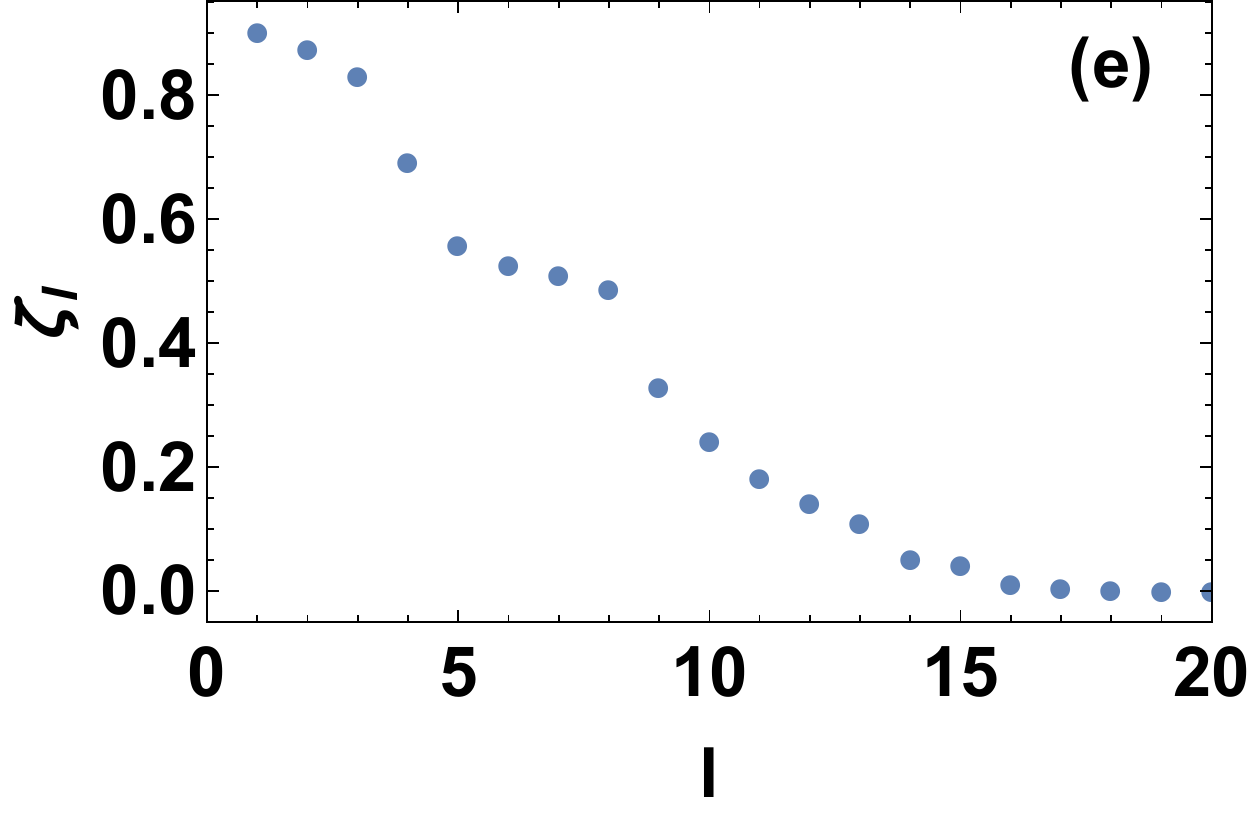}
\includegraphics[width=0.48 \columnwidth]{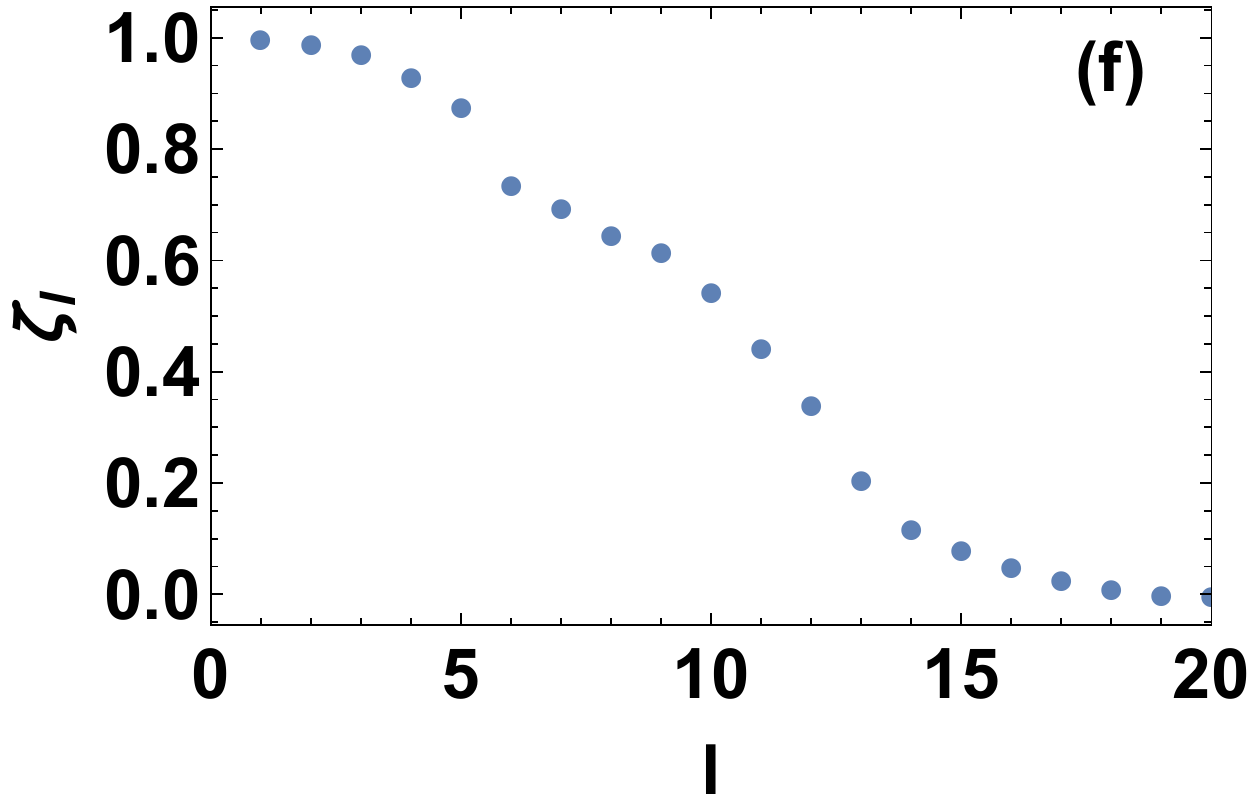}}
\caption{(a) Plot of $\zeta_l$,the eigenvalues of $C_{ij}$ , correlation matrix of the subsystem $M=L/2=128$. Only first 20 values are plotted. (a)In this system $\mu=\delta \mu=0.25$ and $p=L/2$ i.e. only one scattering centre is present at the subsystem environmenment interface. The eigenvalues are plotted at $t=10$. (b) Same system as (a) except $t=60$. (c) Here the eigenvalue at $t=10$ is plotted for a system with $p=16$ showing no quantitative difference with (a). (d) Same as (c) but at $t=60$,. Here we see a completely different result from (b) showing how an increasing number of eigenvalues are of value $1$ and thus do not participate in entangling the subsystem and environment, thus suppressing linear growth of entanglement for the single scattering centre case.(e) Same parameters as (b) and (d) but in a system with random potential. (f) Same as (e) with the system having Fibonacci potential and size $L=377$. Notice how in the last two cases the eigenvalue structure contains less eigenvalues closer to $1$ than plot $(d)$ }
\label{eigenzeta}
\end{figure}
\section{Discussion}
\label{discussion}
In this work we have shown that the breaking 
of translation invariance in a 1D free fermion system 
can result in an enhanced entanglement growth/information 
leakage compared to the translationally invariant case following an
inhomogeneous quench. In disordered systems, this enhancement 
is observed in a sub-system of the size of the localization length.  
We identified back scattering of particles to the sub-system from its surrounding as
the main mechanism behind the enhancement. \\
We also briefly touched upon the special case of breaking of TI by periodic potentials.
There the coherent interference of the back-scattered wave-fronts produces certain peculiarities. For 
example, a larger spatial period $p$ implies lesser number of back-scatterers per unit length (locally closer 
to the TI case) but actually produces stronger enhancement in entanglement growth. 
We found for low periods there exists a $ c/6 \log t$ scaling of entanglement entropy very similar 
to the translationally invariant problem, with $c=n$ where $Z_n$ is the lattice symmetry. 
This is a rather interesting result in the sense these systems do not posses any conformal symmetry and we have similar behaviour obtained as conformally invariant systems. We also recovered the linear rise regime at large periods and numerically analyzed when each regime sets in. We gave a weak bound on the value of period when one can expect to see the $\log t$ scaling behaviour. Then we discussed how the eigenvalues of the correlation matrix of the system change for lower periods compared to larger periods which changes the linear behaviour to logarithmic. Interestingly, the pattern of new eigenvalues for the disordered lattice case follows the single scattering centre scenario when one is within localization length. This points to a requirement of thorough analytical analysis of such systems as coherent scattering changes the physics completely. In principle if one could analytically find closed form expression of the correlation functions in a periodic lattice from the single particle eigenfunctions one can hope to analyze them to figure out this phenomenon. Since the one-particle eigenfunctions will pick up the symmetry of the lattice there might be a possibility of extracting the coefficient of $\log t$ finding how they combine to give the result. This is left for a future work.  

The main result, namely the enhancement of entanglement growth due to introduction of disorder betrays the 
non-trivial character of disordered quantum matter within the scale of localization length. This, to our 
knowledge, is a largely unexplored area. For weak disorder, this length scale can be considerable and the 
understanding the physics might play a key role in designing quantum devices.
An interesting open direction is to study the fate of this enhancement in presence of weak interactions 
and external drive.

\begin{acknowledgements}
RG thanks  A. Sen for discussions and introducing him to problems with domain wall initial states, K. Sengupta, S. Nandy, B. Mukherjee and M. Sarkar for discussions, and CSIR SPM fellowship for financial support.
\end{acknowledgements}
\bibliography{enhoppv3}
\bibliographystyle{unsrt}
\appendix

\section{Calculation of $\la c_m^{\dagger}(t) c_n(t) \ra$}
\label{appA}
Since this is a free fermion model, we can exploit the fact that all the information about the system is contained in the one particle sector of the Hamiltonian. We can thus reduce the $2^N \times 2^ N$ problem to the $N \times N$ problem in theory, but in practice, if we were to work in the Schr\"odinger picture we would have to still deal with $ {N \choose m}$ eigenfunctions for a $m$ particle sector in a N site model which are constructed as Slater determinants of the 1-particle sector. But we are not interested in wave functions here, and hence we can switch to the Heisenberg picture and deal with $ N \times N $ matrices and find the required result. The entire procedure for this is as follows,
\begin{eqnarray*}
\mathcal{H}&=&\bm{c^{\dagger}} \bm{C c} \\
&=& \bm{c^{\dagger}} R^{\dagger} R \bm{C} R^{\dagger}R \bm{c} \\
&=& \bm{b^{\dagger}} \bm{B} \bm{b}
\end{eqnarray*}
where $\bm{c}(\bm{c^{\dagger}})$ denotes a column(row) vector consisting of one-particle Fermionic annihilation(creation) operators on each site. $\bm{C}$ denotes the matrix elements of the Hamiltonian in one-particle occupation basis.  $\bm{B}= R \bm{C} R^{\dagger}$ and $\bm{b^{\dagger}}=\bm{c^{\dagger}} R^{\dagger}$ or $ b^{\dagger}_i=c^{\dagger}_j R^{\dagger} _{ji}=R^* _{ij} c^{\dagger}_j$ and $b_k=R_{kj} c_j$ , $ \bm{B} $ denotes the diagonalized matrix and $\bm{b}$ denotes the diagonal basis. In this basis we know the Hamiltonian is $\mathcal{H}=-\sum_k E_{k} b_k^{\dagger} b_k$($E_k=B_{kk}$), and hence , 
\begin{equation}
b_k(t)=b_k(0)e^{i t E_k}, \hspace{0.2 in } b_k^{\dagger}(t)=b_k^{\dagger}(0) e^{-i t E_k}
\end{equation} 
Using these expressions we can write the time dependent correlation matrix. Also we recall that $\bm{c}=R^{-1}\bm{b}=R^{\dagger} \bm{b}$ or $c_i=R^{\dagger}_{ij} b_j=b_j R^*_{ji}$. Using this and remembering that for the systems considered in this paper the $R$ matrices are all real.
\begin{eqnarray}
\la c_m^{\dagger}(t) c_n(t) \ra&=&\sum_{k,l}R_{km} R_{ln} \la b^{\dagger}_k (t) b_l(t) \ra\label{master} \\
&=&\sum_{k,l}R_{km}R_{ln}e^{-i E_l t}e^{i E_k t} \la b^{\dagger}_k(0) b_l(0) \ra \nonumber\\
&=& \sum_{k,l,i,j}R_{km} R_{ln} R_{ki} R_{lj}e^{i(E_k t-E_l t)} \times \nonumber \\&& \la c^{\dagger}_i (0) c_j(0) \ra \nonumber
\end{eqnarray}
Using the known initial conditions $\langle c_i^{\dagger}(0) c_j(0)\rangle$ we can calculate the evolution of the correlators to any instant of time .
%\begin{figure}
%\centering{
%\includegraphics[width=0.95 \columnwidth]{nifull.pdf}}
%\caption{(Colour online) Plot showing $n_i=\la c_i^{\dagger} c_i 
%$a(T)$ vs $x=(i-L/2)/L$ at $T=500$ for various systems considered in the paper. Notice how, as we go towards chaotic sequences the wave-front gets localized near $x=0$. Also notice the $p=16$ case shows an interesting structure in $n_i$ . This is due to the periodicty of lattice , where there is a smaller density of particles on every $p^{th}$ site for $x>0$ and vice-versa for $x<0$, this effect is more prominent with increasing p, upto a value which is same as the $p$ for maximum entropy.}
%\label{nifull}
%\end{figure}
\section{Thouless localization length}
\label{appB}
This section we give a brief idea about the calculation of Thouless localization length as defined by Ref. \onlinecite{Thouless_1972}. It gives the localization length of the single particle eigenfunctions of a non-interacting Anderson localized system. For a system with a Hamiltonian as given by Eq.\ref{hammaster}, the eigenstate equation is of the form ,
\begin{equation}
\mu_i a_i^{\alpha}-\frac{J}{2}(a_{i+1}^{\alpha}+a_{i-1}^{\alpha})=E_{\alpha}a_i^{\alpha}
\end{equation}
where $i$ runs from $1$ to $L$ and $E^{\alpha}$ is an eigenvalue of the system labelled by $\alpha$, and $a_i^{\alpha}$ are the amplitudes of eigenstates. The Green's function of such a system can be written as,
\begin{equation}
(E-\mu_i) G_{ij}(E)+\frac{J}{2}(G_{i+1,j}(E)+G_{i-1,j}(E))=\delta_{ij}
\end{equation}
where $G_{ij}=(E \mathbf{I}-\mathbf{H}^{-1})$ , $\mathbf{H}$ being the Hamiltonian of the system. 
From this one can find $G_{1N}$ as, 
\begin{equation}
G_{1N}=(\frac{J}{2})^{N-1}/\prod_{\alpha=1}^N (E-E_{\alpha})
\end{equation}
Identifying $G_{1N}$ has a pole of residue $a_1^{\beta}a_N^{\beta}$, we can rewrite the equation as,
\begin{equation}
\log|a_1^{\beta}a_N^{\beta}|=(N-1) \log|\frac{J}{2}|-\sum_{\alpha \neq \beta}\log|E_{\beta}-E_{\alpha}|
\end{equation}
The key thing to note is now, for a localized eigenstate exponentially decayse from its peak. hence if the peak is at site $x$ one would expect $a_1=Exp[-i \kappa (1-x)]$ and $a_N=Exp[-i \kappa (x-N)]$, where $\lambda=1/\kappa$ is the localization length.
Hence, in the limit of large $L$, one can define ,
\begin{equation}
\lambda=-[(N-1) \log|\frac{J}{2}|-\sum_{\alpha \neq \beta}\log|E_{\beta}-E_{\alpha}|]^{-1}
\end{equation}
Clearly for different $E_{\beta}$ the $\lambda$ would be different. In our work we consider $\lambda_l=\lambda_{max}$ i.e. the largest Thouless localization length in the problem to provide an idea of the lengthscales.

\section{A peculiarity of the entanglement at high $\delta \mu$}
\label{appC}
\begin{figure}
\centering
\includegraphics[width=0.95\columnwidth]{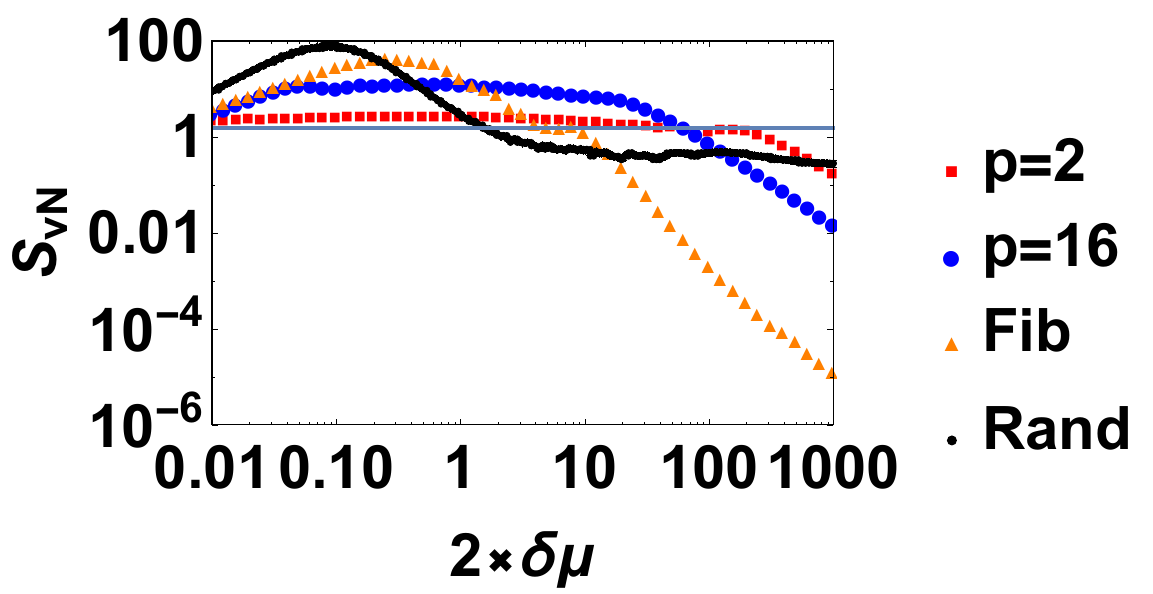}
\caption{(Colour online)Left: Plot showing Von Neumann entropy vs strength of inhomogeneity denoted by $\delta\mu$ at $t=500$ for different representative cases. The subsystem is of length $M=L/2$, and the position of $A$ is at $i=L/2+1$. The rest of the parameters are same as Fig \ref{deltamu}(a). }
\label{mucompare2}
\end{figure}

In Fig. \ref{deltamu}(a) we showed the behaviour EE with increasing $\delta \mu$ for different potentials when the subsystem $AB$ is chosen from $i=L/2+21$ to $i=L$. As expected, for most of the cases the entanglement drops off as a power law with $\delta\mu$. It is also important to note different potentials follow a different power law which occurs due to the nature of scattering events encountered. It is to be expected correlated and uncorrelated scatterings result in different behaviour. Even in correlated scattering the scattering from a periodic potential is different from a quasi-periodic one. The understanding of the various power laws observed is left for a future work. 

Fig. \ref{mucompare2} shows the half chain entanglement and in this case $A$ is taken at site $i=L/2+1$. Here as expected, when we are in the low disorder limit, the entanglement increases with disorder until a maximum after which it starts decreasing for all the cases of inhomogeneity. The random case shows the steepest ascent and descent compared to the other two cases. However at very high disorder entanglement still stays at a finite value. This peculiarity is due to the fact that, for certain disorder realizations(which are significant in number) a very small particle density can be present in the $x>0$ sector. This effectively ensures a finite number of wave-fronts reaching the subsystem and carrying information even with a low particle density. This raises the entropy to a significant non-zero value even if the value is smaller compared to all other potentials. This events which cause such anomaly in EE become exponentially smaller in number the further we move from $i=L/2$, hence our choice of subsystem in the main-text was such that these events do not interfere with our general analysis.  However,in cases where we have considered a periodic potential or a Fibonacci potential, correlated scattering events show no such peculiarity. 

\section*{Supplementary Material}
\renewcommand{\theequation}{S.\arabic{equation}}
\renewcommand\thefigure{S\arabic{figure}}
\subsection{Constant potentials}
\label{appB}
\begin{figure}
\centering{
\includegraphics[width=.47 \columnwidth]{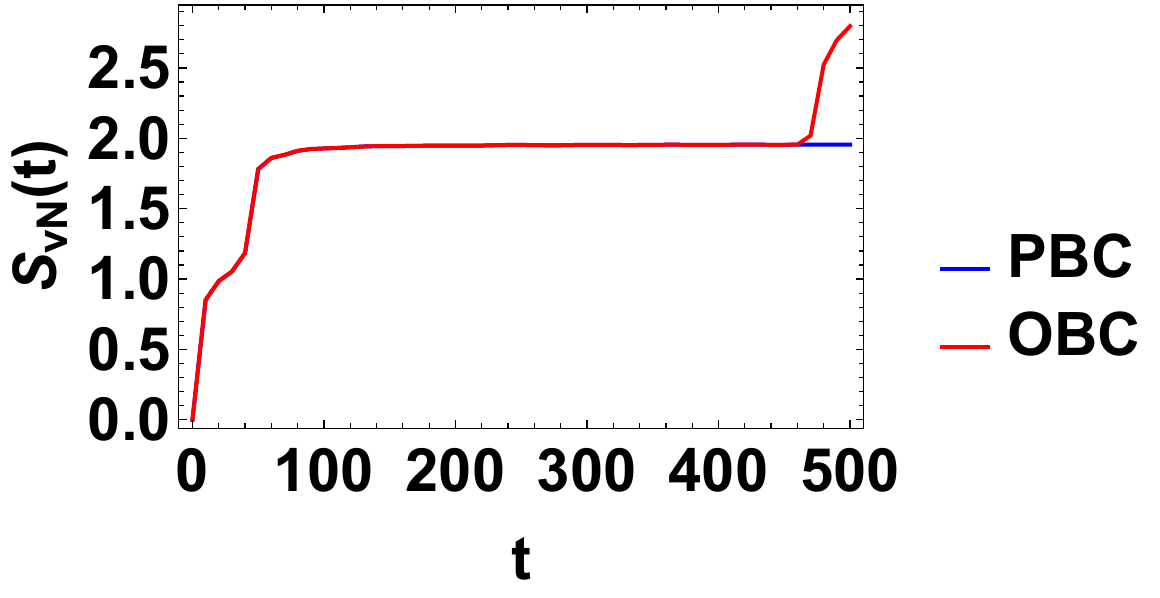}
\includegraphics[width=.47 \columnwidth]{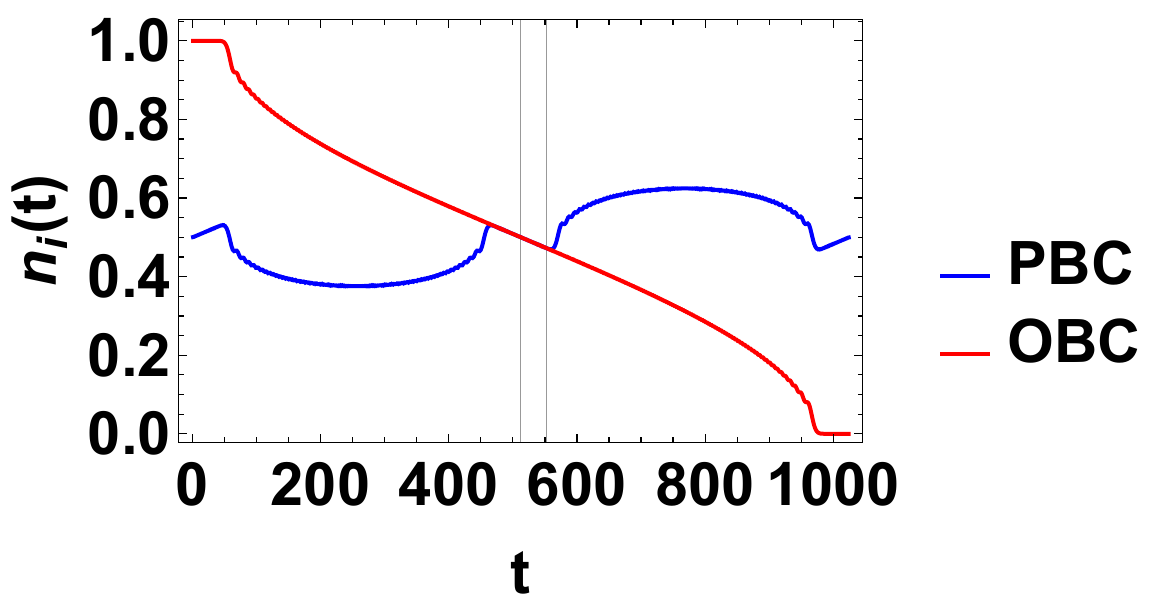}}
\caption{(Colour online)Left: Plot of $S$ vs $t$ for a subsystem of size $M=40$ for a system with Periodic Boundary Conditions(PBC) compared to one with Open Boundary Conditions(OBC) showing exact same results till thermodynamic limit. Right: Plot of $n_i=\la c_i^{\dagger} c_i 
\ra$ vs $i$ at $t=460$ in units of $J^{-1}$, for PBC and OBC conditions, showing the region where they overlap. The thermodynamic limit for the $M=40$ subsystem would be broken at $t=472$ considering ballistic propagation. It is seen at $t=460$ the overlap is still perfect in the region of the subsystem denoted by the two gridlines, and hence either analysis gives equivalent results.}
\label{compare}
\end{figure}

For periodic boundary conditions , if $\mu_p=\mu$ then, this system can be diagonalized in momentum space to give,
\begin{equation}
\mathcal{H}=- \sum_k ( \cos (k)-\mu) b_k^{\dagger} b_k  \label{master1}
\end{equation}
where $k=\frac{2 \pi n}{L}$ and $n=0,1, \hdots, L-1$ is the momentum index of the system. Now using Heisenberg's equation of motion, we can figure out the time evolution of $b_k$ as, 
\begin{equation}
\dot{b_{k^{\prime}}}=-i [b_{k^{\prime}}, \mathcal{H}] =i(\cos(k)-\mu) b_{k^{\prime}}
\end{equation}
which gives us,
\begin{equation}
b_k(t)=b_k(0)e^{i ( \cos(k)-\mu)t}, \hspace{0.2 in } b_k^{\dagger}(t)=b_k^{\dagger}(0) e^{-i ( \cos(k)-\mu) t}
\end{equation}
Now our aim is to find $\la c_m^{\dagger} c_n \ra(t)$, which written in Heisenberg picture looks like $\la c_m^{\dagger}(t) c_n(t) \ra$. The set of steps to do this are as follows,
\begin{align*}
c_n(t)&=  \sum _k e^{i n k} b_k(t) \\
&=e^{-i \mu t}\sum_k e^{i n k} b_k(0) e^{i t \cos(k)} \\
&=e^{-i \mu t}\sum_{k,j}e^{i n k} e^{ i t \cos k} e^{- i k j} c_j(0)\\
&=e^{-i \mu t}\sum_{k,j} \sum_{\alpha=-\infty}^{\infty}e^{i[n-j]k} i^{\alpha} J_{\alpha}(t) e^{i \alpha k} c_j(0)\\
&=e^{-i \mu t}\sum_j i^{n-j} J_{n-j}(t) c_j(0)
\end{align*}
Hence,
\begin{eqnarray}
\la c_m^{\dagger} (t) c_n(t) \ra &=&\sum_{j,k}i^{n-j}(- i)^{m-k}J_{n-j}(t) J_{m-k}(t)\nonumber \\ \times  \la c_k^{\dagger}(0) c_j(0) \ra  \label{rougheqn}
\end{eqnarray}
From the initial condition of the system,
\begin{align}
\la c_p^{\dagger} c_q \ra&= \delta_{pq} \hspace{0.2 in} p \le N/2 \nonumber\\
	&=0 ~ {\rm otherwise} \label{init2}
\end{align}
after several straightforward but tedious lines of algebra using recursion relation of Bessel functions, we obtain a more tractable form as follows, 
\begin{eqnarray}
C_{mm}(t)&=&\la c_m^{\dagger} c_m(t) \ra=(\frac{1}{2}[1- J_0^2(t)]-\sum_{l=1}^{m-1}J_l^2(t)) \nonumber \\
C_{mn}^{m \neq n} (t)&=& \frac{ i^{n-m}t}{2(m-n)}[J_{m-1}(t) J_n(t)-J_{n-1}(t) J_m(t)] \nonumber
\end{eqnarray}
For open boundary conditions, one can also find the eigenvalues as,
\begin{equation*}
\mathcal{H}=- \sum_k ( \cos (k)-\mu) b_k^{\dagger} b_k 
\end{equation*}
where $k=\frac{ \pi n}{L+1}$ and $n=0,1, \hdots, L-1$. However, it is easier to follow the numerical prescription given in the main text to find out the time evolution. For a sanity check, in Fig. \ref{compare} we show the comparison of the results in the periodic BC and open BC. As expected for a finite subsystem size if we are in the thermodynamic limit, (which is defined by the position of the fastest particle at a particular instant of time, as, if the system size is smaller than that, reflected particles will cause interference and results from larger system sizes will not match it for the same instant of time) the results show no difference. But on breaking the limit, for PBC the transmitted wave-front going towards left from $i=L/2$ will interfere with the wave-front going towards the right at the boundary of the system at $i=L$. This happens for OBC at a time $O(L/2)$ later than PBC as wave-fronts get reflected there and hence results start to differ. For the translational invariant system we mostly consider results in the thermodynamic limit and hence thw two boundary conditions yield similar results.
\begin{figure*}
\centering{
\includegraphics[width=0.70\columnwidth]{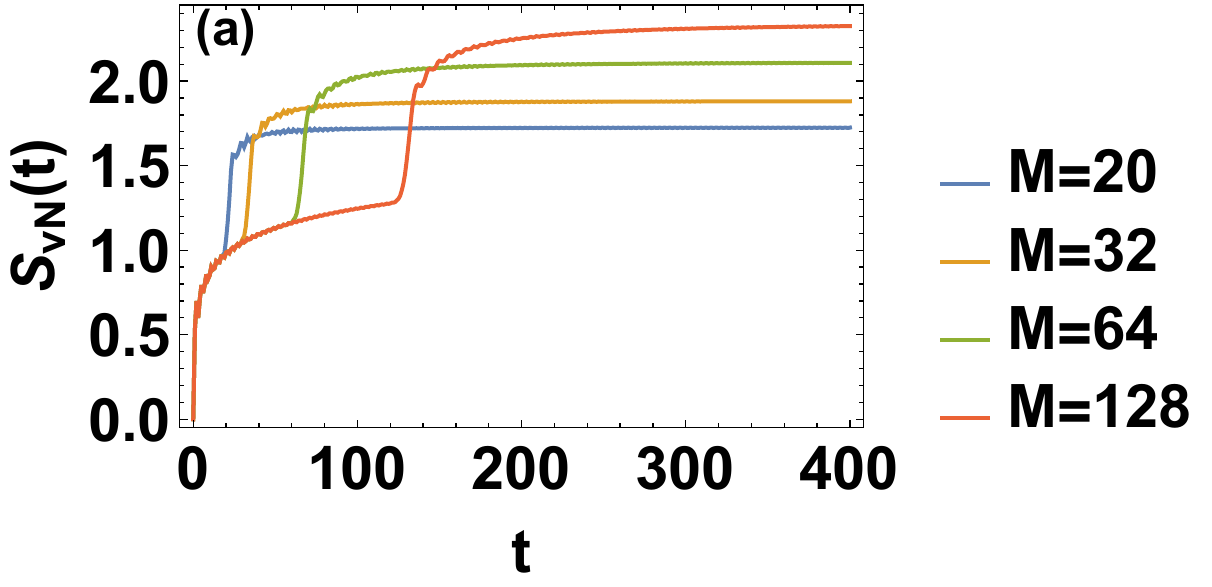}
\includegraphics[width=0.66\columnwidth]{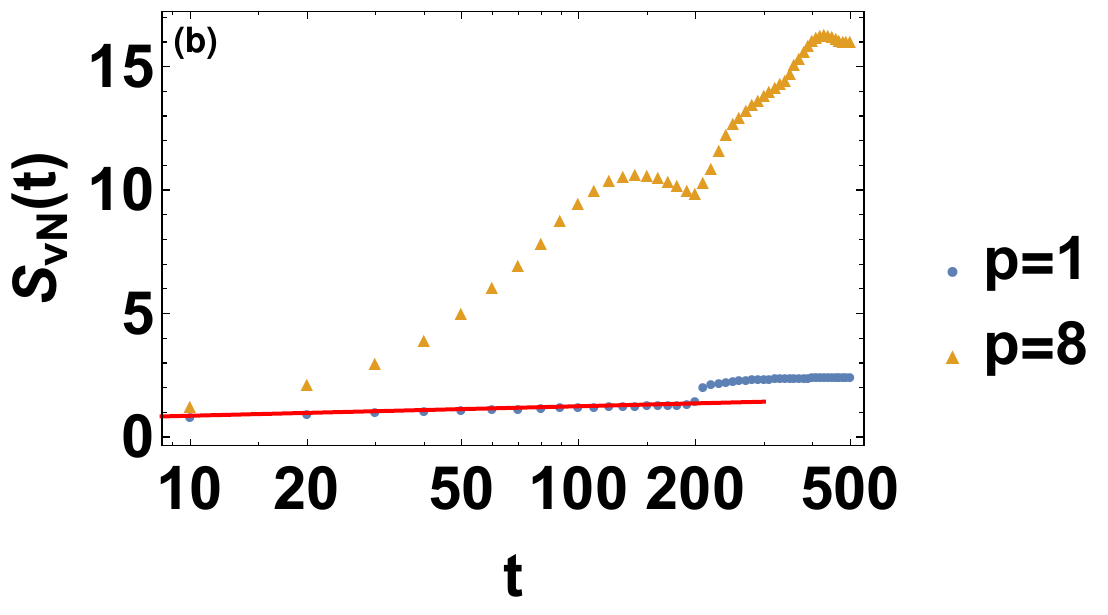}
\includegraphics[width=0.60\columnwidth]{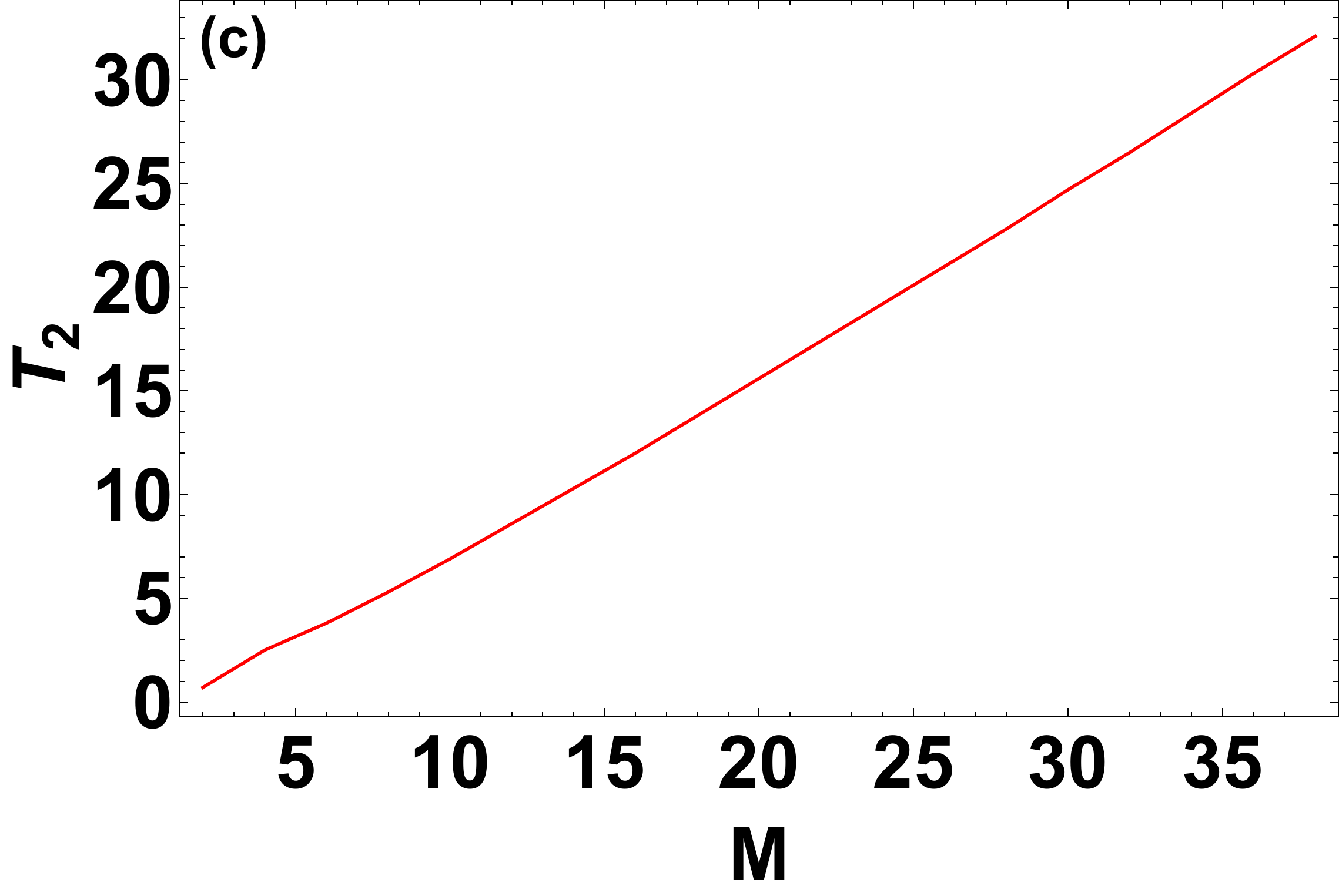}}
\caption{(a) Plot showing growth of von Neumann entropy with time for a susbsystem of $M$ sites to the right of $L/2$ with time. (b) Plot showing the fit of $\log t$ with the data for a subsystem of size $M=200$. $p$ denotes the periodicity of the potential. Hence $p=1$ denotes the constant $\mu$ case. The red line shows the fit of $S= 0.476-0.167 \log (t)$ with the data.(c) The figure shows a comparison of the point of time($T_2$) at which the entanglement entropy shows a sudden jump for different subsystem sizes $M$. See text for details}
\label{figconst1}
\end{figure*}
\begin{figure}
\centering{
\includegraphics[width=0.45\columnwidth]{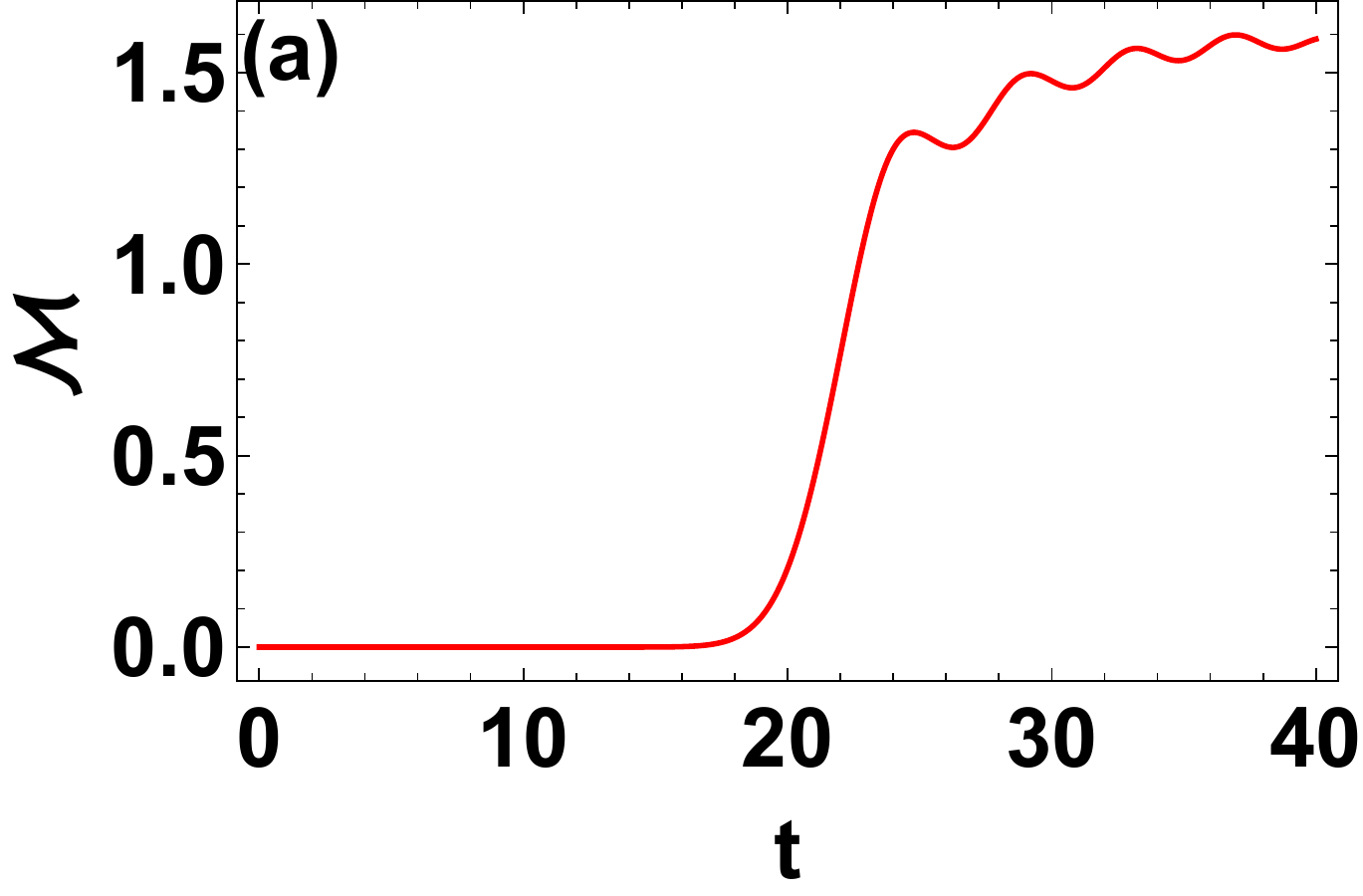}
\includegraphics[width=0.50\columnwidth, height=0.30\columnwidth ]{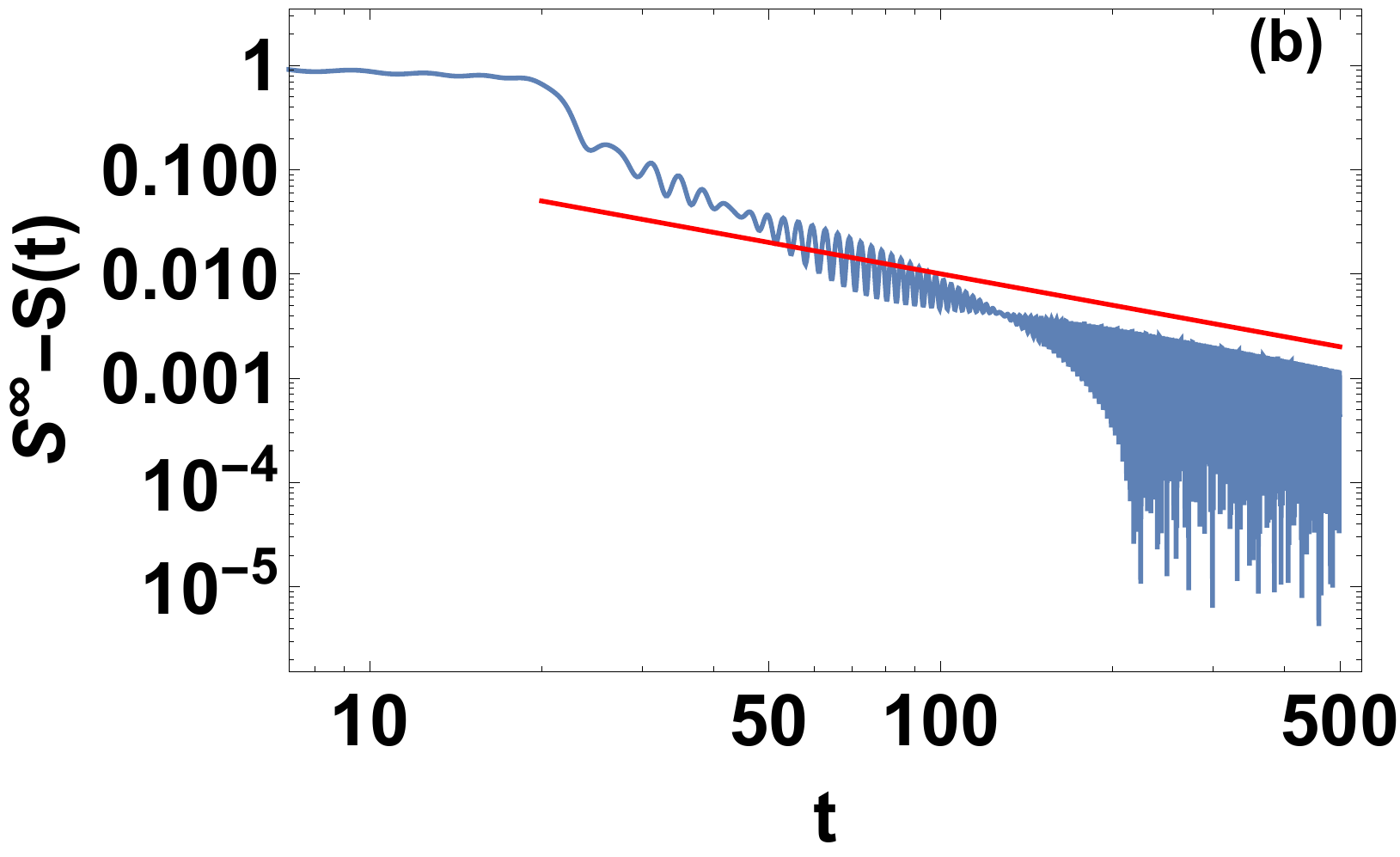}\\
\includegraphics[width=0.43\columnwidth]{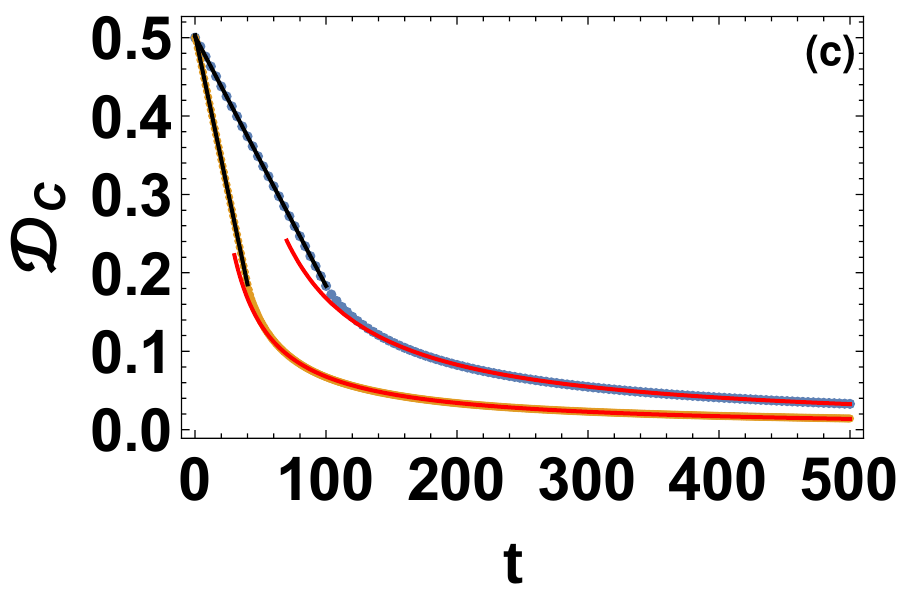}
\includegraphics[width=0.54\columnwidth]{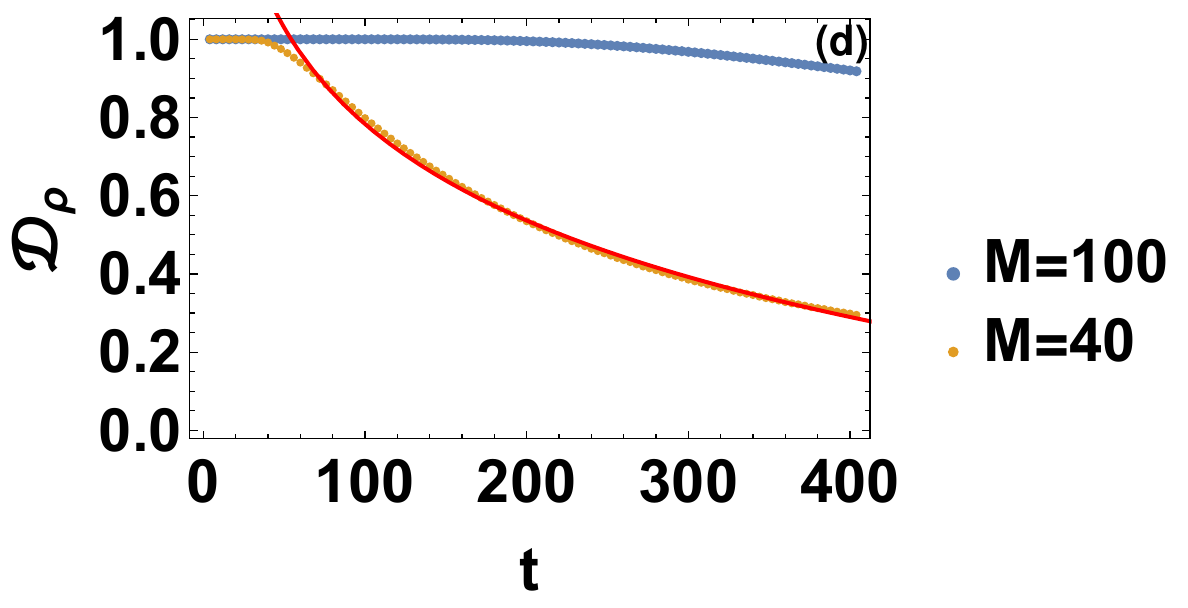}}
\caption{(Colour online)(a) Plot showing growth of mutual information $\mathcal{M}$ with time for a susbsystem of $M$ sites to the right of $N/2$ and a subsystem of $M$ sites to the right of $N/2+M$ with time. M is chosen to be $20$ . (b) Plot showing approach of $S$ to the diagonal ensemble of a subsystem of size $M=100$, showing presence of two distinct timescales. For $t \gg T_2$ it is clear the entanglement goes to the diagonal value as $t^{-1}$. (c) Plot showing the evolution of the trace distance between the diagonal ensemble correlation matrix and $C_{ij}(t)$ for two different system sizes. The black line is the fit for the section of data at $t<T_2$ as $\mathcal{D}_c\propto t$. The red line is the fit of the data at times $t > T_2$ as $\mathcal{D}_c\propto 1/t^{1.017}$ for $M=100$ and $\mathcal{D}_c\propto 1/t^{0.99}$ for $M=40$. (d) Plot showing the evolution of the trace distance between the diagonal ensemble reduced density matrix and $\rho(t)$ for the same two system sizes. The red line shows the fit of the data to $\log(t)$. See text for details}
\label{figconst3}
\end{figure}
We now show the numerical results obtained for the system described in this section. 
\\
Since we have derived exact analytical formulas for the correlation function, we would be using those expressions to calculate entanglement measures in this system. Fig. \ref{figconst1}(a) shows the growth of Von Neumann entropy with time for different subsystem sizes denoted by $M$. As expected,\cite{cheong1} the entanglement entropy at large times is higher for higher susbsystem sizes. Due to the fact that the correlation functions at large times equal that to the correlations calculated in the ground state of this Hamiltonian with an irrelevant overall phase factor, it can be shown the entanglement goes as $\log M$.\cite{cheonghabis}. In fact it was shown that the data can be fitted exactly to
\begin{eqnarray}
0.3374 \log M+1.4052 \hspace{0.3 in} \text{even M} \nonumber \\
0.3346 \log M+0.72613 \hspace{0.3 in} \text{odd M}
\label{fit}
\end{eqnarray} 
A peculiar feature of this figure however is that the $S$  vs t shows two sharp increases, one at $t=0$ and another at $t=T_2$ before it saturates. This is unlike the case of global quenches in systems where the initial state can be written in terms of pairs of quasi-particles, \cite{calabrese2,PhysRevLett.96.136801,Alba_2017} where it has been shown via CFT arguments that there would be a linear increase in entropy at first after which it tends to saturate, and there is no second point of increase. Additionally, the increase till $t=T_2$ can be nicely fit in a $\log(t)$ behaviour. This is reminiscent of the behaviour of entanglement entropy dynamics in two halves after a local quench as predicted analytically \cite{calabrese4}, and was verified numerically \cite {gobert,Eisler_2007}. Hence this quench shows features in between a global and a local quench for the TI system. The argument is that, even if the initial state considered here has an extensively different energy than the ground state of the system, the Pauli exclusion principle prevents particles/wave-fronts to be created in an extensive manner throughout the system at $t=0$. Another thing to note is that the transport of particles and the local correlations are ballistic but the non local correlations travel slower than that as presented by the trace distance of the density matrix(Fig \ref{figconst3}(d)) and the entanglement evolution.(Fig \ref{figconst1}(b)) In fact one can see there are two distinct time scales of the system and until the wave reaches the end of the subsystem, non local correlators evolve exponentially slowly. This can again be attributed to the fact that a quench from such an initial state behaves more like a local quench than a global quench, as there is only one developing wave-front. However note that, in local quenches where only an intensive number of quasi-particles are pumped into the subsystem, the initial $\log(t)$ behaviour will be replaced by a decrease once the particle has left the subsystem or a saturation value as the quasi-particle goes deeper inside the subsystem. Here as long as particles are being pumped into the subsystem from on defect site the $\log(t)$ behaviour will continue, that is unless the particles see a `wall'. Several things can qualify as a `wall'. It can be the end of a finite sized subsystem under consideration, labeled by $B$, where the $S$ shows another jump at $t=T_2$, or it can be the fact the particles get reflected(transmitted) from a finite sized system with open(periodic) boundary conditions and reach $x=0(i=L/2)$ which was the position of $A$. In either of these cases $S$ shows a jump since a wave-front entering or leaving a system generates entanglement with the surrounding. How $S$ approaches equilibrium for a finite sized translationally invariant system after multiple reflections from the boundary has been shown in Alba \cite{alba} and plays a major role in the explanation of the observed results for TI broken case as described in the main text. Our findings is completely consistent with the light cone spreading of correlation function and the Lieb-Robinson bound for the group velocity of quasi-particles.\cite{Lieb}.\\

To elaborate a bit more about the sudden rise in entanglement at the time $t=T_2$, initially, all correlations to the right of $x=0$ point was 0, so there could not be any entanglement between any subsystem and the surroundings. As soon as we switch on the Hamiltonian, correlations between neighbouring sites start developing near the origin and spread towards left and right with a group velocity $v$. One can think of this as carried by two wave-fronts (consisting of particles) travelling left and right from the domain-wall at the origin entangling the subsystem with the other half of the lattice(environment). Hence there is an initial rise in entanglement. \\

With time the wave-front travelling to the right reaches the end point of the subsystem and then crosses over to the part of the environment not accessible initially. This allows correlations to form between the part which was inaccessible earlier and the subsystem under consideration. It is at this moment the second jump in entanglement is observed. This theory is further strengthened by the plot of $T_2$ vs $M$ where $T_2$ is the time in which the second jump starts and $M$ is the subsystem size, plotted in Fig. \ref{figconst1}(c). The tolerance level for calculation of $T_2$ was chosen to be $10^{-4}$. The time $T_2$ exactly matches with the time at which mutual information between this subsystem and a subsystem of $20$ sites chosen just to the right starts increasing from a $0$ value which is shown in the top left panel of Fig. \ref{figconst3}(a). This also shows a nice picture of how entanglement is generated between two systems as particles leave the system.  \\

Thus, the two timescales in this problem are before and after $t=T_2$. The rise of entanglement after $t=T_2$ does not follow a $\log(t)$ behaviour here as can be seen from  Fig. \ref{figconst1}(b). In fact after $t=T_2$ the $S$ goes to its diagonal ensemble value as $t^{-1}$ shown in \ref{figconst3}(b).   The intriguing thing is, even so, in this quench diagonal ensemble $S$ exactly matches the $S$ of the ground state of the Hamiltonian and thus it is not an extensive quantity unlike what is expected of a global quench, within the thermodynamic limit. This is due to the fact, the correlation functions, at $t \rightarrow \infty$ have an absolute value same as the ground state correlators with a position-dependent phase in the off-diagonal terms. Thus even though state goes to an eigenstate of the Hamiltonian which is not the ground state,\cite{XXmodel} the entropy has the same value. This happens mainly due to the nature of the system considered. Mathematically, the Eigenvalues of $C_{mn}$, compared to the eigenvalues of the correlation matrix of the ground state of the system, has no difference, as these two Toeplitz matrices are similar matrices. In the wave picture, we believe that when we choose a finite sized subsystem $AB$, the wave from the domain-wall at $x=0$ would propagate through the subsystem, but on seeing site $B$ it breaks up into a reflected and transmitted wave. The reflected and the incoming wave undergo interference to set up standing waves which in turn are eigenfunctions of the Hamiltonian, and thus the DE behaves as it does.  This anomalous behaviour carries over to other non local correlators like the trace distance measure between $\rho_D$ and $\rho(t)$ which we plot in Fig \ref{figconst3}(d). In Fig \ref{figconst3}(c) , we plot the trace distance between just the local correlation functions.($\mathcal{D}_C=C^{\infty}-C(t)$). As one can see for $t<T_2$, it goes towards the DE in an approximately linear manner because it can be fit almost perfectly to $t$ behaviour. For $t > T_2$ the behaviour changes to $\sim t^{-1}$ which again points out $t=T_2$ is a special point. It is worthy to note that even though the entanglement falls off as $t^{-1}$ for $t > T_2$ , the trace distance of the density matrix $\mathcal{D}_{\rho}$ shows a $\log(t)$ sort of a fall. And for $t<T_2$ the density matrix goes exponentially slowly towards the diagonal ensemble. Hence we state that for this system the local correlations reach towards the DE faster than non local quantities. Physically, when the wave-front travels through the subsystem, until $t=T_2$ it does not see the right boundary or $B$. At this time the correlators go towards the DE value of the subsystem linearly and entanglement grows as $\log(t)$. Once it reaches the point $B$ at $t=T_2$, the reflected wave-front which gets formed interferes with the incoming wave. This changes how the local and non-local quantities go to their DE values. The local correlators as well as $S$ are seen to follow $t^{-1}$ law and the Trace distance of the density matrix follows $\log(t)$. \\
\subsection{Thue Morse and Rudin Shapiro sequences}
\label{appE}
Thue Morse sequence or more formaly, Prohuet-Thue-Morse sequence \cite{Prohuet} is another automatic sequence defined on the set of alphabets $\{0,1\}$. There are several equivalent definitions for this sequence, here we state the most commonly used one. If $T_n$ is the $n^{th}$ word of the Thue Morse sequence and $T_n^i$ is $i^{th}$ letter(digit) in the word, then,\cite{allouche}
\begin{equation}
T_n^i=s_2(i)\hspace{0.1in} mod \hspace{0.1in}2
\end{equation} 
where $s_2(i)$ is the sum of binary digits of the decimal number $i$. The length of the $n^{th}$ word is given by $2^n$. The first few words in the sequence are,
\begin{eqnarray*}
T_0 &=& 0 \\
T_1 &=& 01 \\
T_2 &=& 0110 \\
T_3 &=& 01101001 \\
T_4 &=& 0110100110010110 \\
\vdots
\end{eqnarray*}
A similar sequence can be created by taking $1's$ compliment of $T_n$ and it can be shown to have same properties as the original TM sequence.
Similar to the case of Fibonacci words, we generate the on site potential using,
\begin{eqnarray}
\mu_i=\mu-\delta\mu , \hspace{0.3 in} T_L^i=0 \nonumber \\
\mu_i=\mu+\delta\mu , \hspace{0.3 in} T_L^i=1
\end{eqnarray}
The Thue Morse sequence is said to exist at the margin between quasi periodicity and randomness. This feature will be important for our analysis later. \\

The Golay-Rudin-Shapiro sequence is another automatic sequence defined on the set of alphabets $\{0,1\}$. If $R_n$ is a word sequence of length $n$ , then the $i^{th}$ letter is given by the parity of the number of $11$(including overlaps) occurring in the binary representation of the number $i$\cite{allouche}.the length of the $n^{th}$ word is the same as in the case for TM sequence, which is $2^n$. Mathematically,
\begin{eqnarray}
a_i=\sum_{k=1}^i\epsilon_k \epsilon_{k+1} \nonumber \\
R_n^i=1+(-1)^{a_i}
\end{eqnarray}
where $\epsilon_k$ is the $k^{th}$ digit in the binary representation of $i$.
The first few terms of this sequence are,
\begin{eqnarray*}
R_1&=& 11 \\
R_2 &=& 1110 \\
R_3 &=& 11101101 \\
R_4 &=& 1110110111100010\\
\vdots
\end{eqnarray*}
The potential is arranged in the same way as it was done for a TM sequence. It is interesting to note, if the potential arranged in Rudin Shapiro sequence shows the same characteristics in eigenvalues and eigenvectors of a free fermion system as a completely random potential. Indeed, Allouche \cite{allouche2} defined a quantity called Inconstancy of a curve to characterize the nature of a sequence, showed that the RS sequence and the random sequence have exactly the same value of this parameter that can be used to characterize the sequence. The TM sequence also had a value which was close to the value obtained for the Random sequence, while the Periodic sequences had completely different values determined by the periodicity of the sequence considered. It can also be shown that the second moments of the RS sequence match with that of a random sequence while higher moments differ. \\

\begin{figure}
\centering{
\includegraphics[width=0.96 \columnwidth]{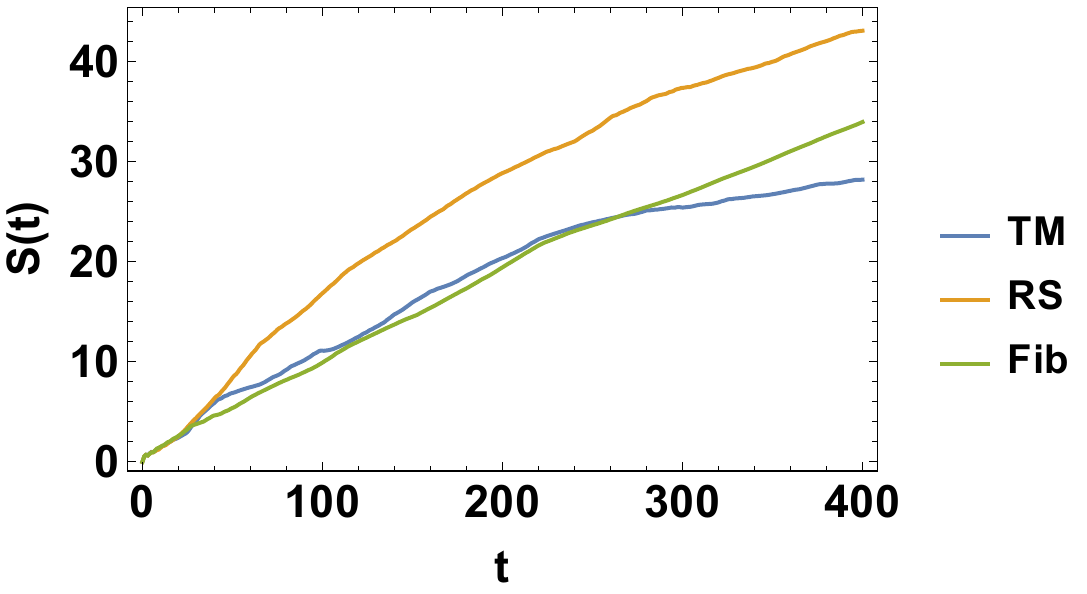}}
\caption{Plot of half-chain Von Neumann entropy vs Time in a system where on site potential has values $\mu+ \delta\mu$ and $\mu -\delta \mu$ arranged in a Thue-Morse word sequence, Rudin-Shapuro sequence and Fibonacci sequence. The system size is $L=2048$ for TM and RS sequence and $L=2584$ for Fibonacci sequence. $\mu=\delta \mu=0.1$}
\label{figquas2}
\end{figure}
A comparison plot for the half-chain Von Neumann entropy vs time  EE when the on-site potential is arranged in a Thue Morse sequence, Rudin Shapiro and Fibonacci is given in Fig \ref{figquas2}. It seems while the growth of EE for RS sequence is the fastest, TM sequence shows slower growth tha Fibonacci sequence after the initial faster rise. This feature of TM sequence is unexpected because TM sequence is considered a border between a quasi-periodicity and randomness, and we have shows a random sequence results in a much higher growth rate of entropy with the same parameters. \\
\begin{figure*}
\centering{
\includegraphics[width=0.96 \columnwidth]{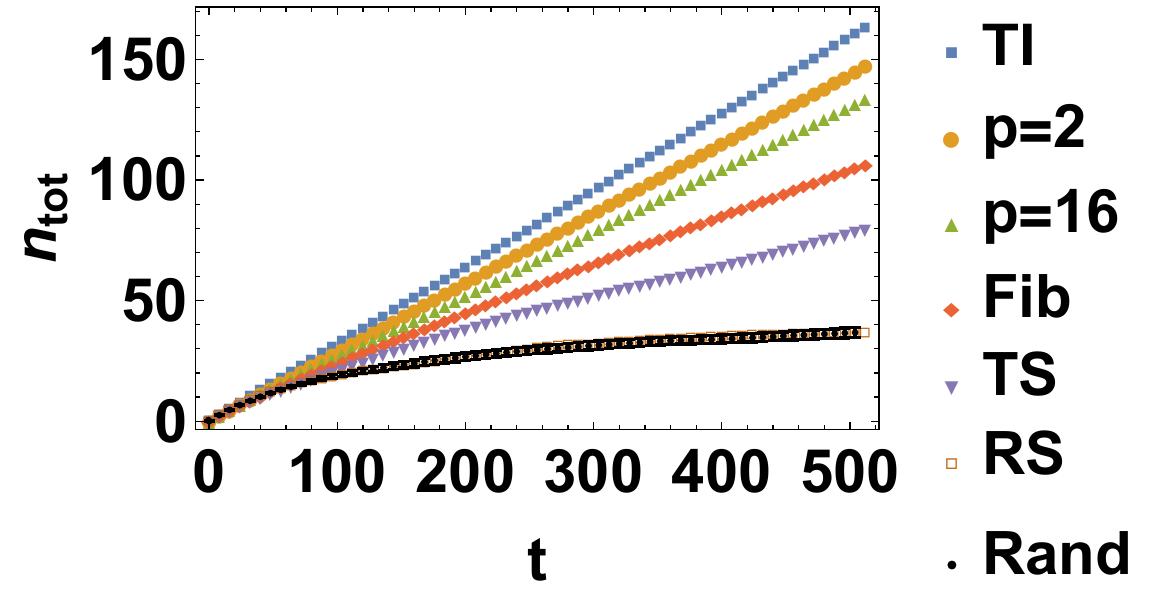}
\includegraphics[width=0.96 \columnwidth]{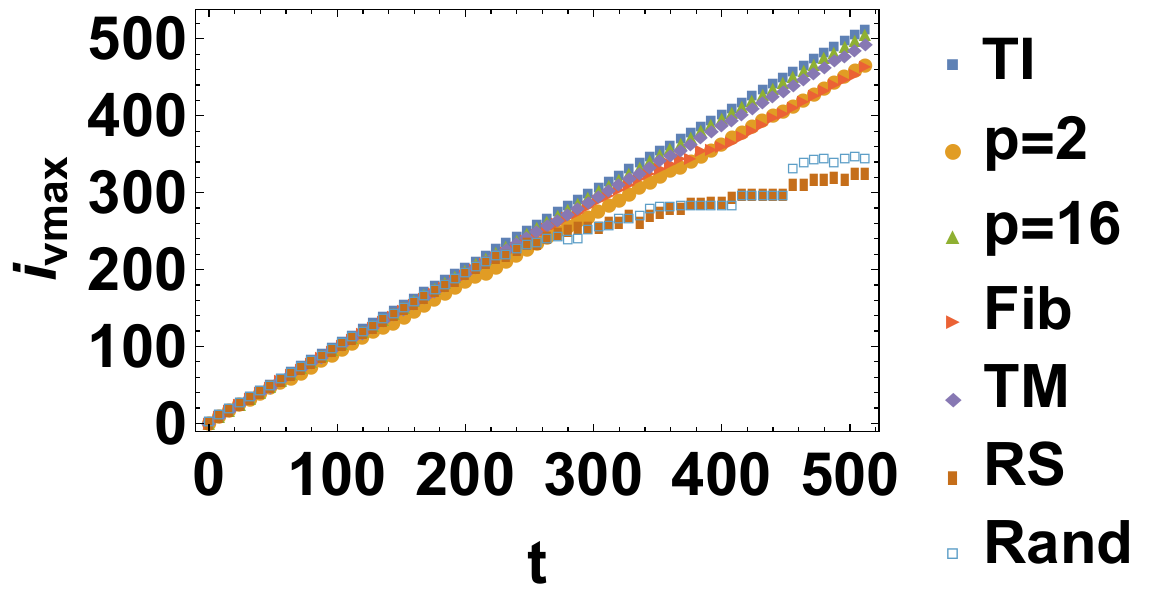}}
\caption{(Colour online) Left:Plot of $n_t=\sum_{i=L/2+1}^L \la c_i^{\dagger} c_i\ra(t) $ vs t in each of the distributions discussed in the paper.For the random sequence errorbars are shown as we have averaged over several realizations Right: Plot showing the position of the fastest particle in the system at time t. In this case only one realization of the random sequence was simulated.}%Right: Plot of Mutual Information ($\mathcal{M}$) between two subsystems $\alpha$ and $\beta$  where $\alpha$ consists of sites $L/2+1$to $L/2+40$ and $\beta$ consists of sites $L/2+41$ to $L/2+80$. For the random sequence case, the error bars are denoted by the thickness of the line. See text for details. }
\label{ni}
\end{figure*}
Qualitatively, this feature can be attributed to the specific distribution of $n_i(t)$, the form of the quasi-periodic sequence allows. As shown in the left panel of Fig \ref{ni}, the number density of particles present in the $x>0$ part of the system is actually greater in the Fibonacci case than in the Thue Morse case. This suggests that there are less incidents of scattering in the Fibonacci case than the Thue Morse case, which is logical considering Thue-Morse sequence is closer to a random sequence. However, in our geometry, $S$ is actually controlled by the number of wave-fronts crossing $x=0$. In the right panel of the same plot we show the first instance where $\la c_i^{\dagger} c_i 
\ra$ goes below a tolerance level of $10^{-2}$ which we have used to characterize the end of wave-front. This plot gives an idea of the position of the fastest particle in the system. As a sanity check we see for the TI system the particle is ballistic.  We also see that the fastest particle travels further in the case of TM sequence than Fibonacci sequence. Furthermore, a plot of $n_i$ vs $i$ for $t=500$ as shown in the left panel of Fig. \ref{compare2}, where the x axis is scaled by the system length, shows the exact distribution, from where it is easy to see the spread in $n_i$ for the TM case is much more pronounced than the Fibonacci sequence. This indicates even if the particle density on the right half of the system for Fibonacci sequence is greater, the particles tend to be closer to the system-environment bounday and thus can transfer more information between them then in the TM sequence case.
\begin{figure}
\centering{
\includegraphics[width=0.45\columnwidth]{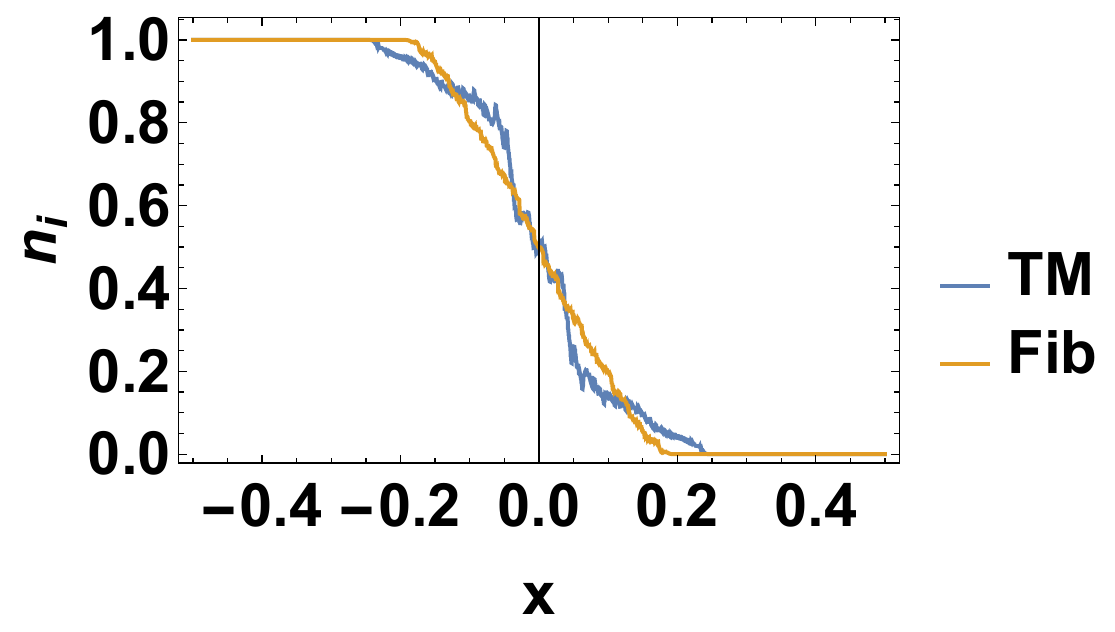}
\includegraphics[width=0.47\columnwidth]{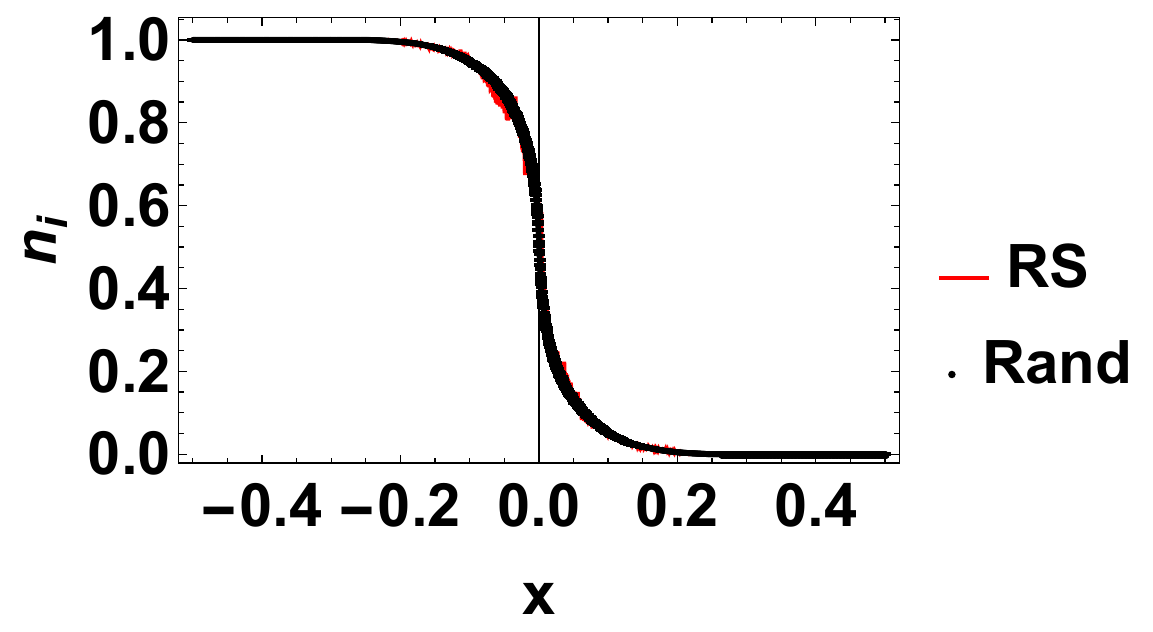}}
\caption{(Colour online) A comparison of $n_i=\la c_i^{\dagger} c_i 
\ra$ vs $x=(i-L/2)/L$ for on site potentials arranged in (left) Thue Morse and Fibonacci sequence and (right) Rudin Shapiro sequence and randomly at time $t=500$. Rest of the parameters are same as Fig \ref{figquas2}. See text for details.}
\label{compare2}
\end{figure}

Fig \ref{figquas2} also shows the growth of Von Neumann entropy vs time for a system in which the on-site potential is arranged in a Rudin Shapiro sequence. It seems,  scatterings under this type of potential are also larger in number to correlated scatterings of the Fibonacci sequence and thus the entanglement entropy grows faster to any of the cases discussed above. A distribution of $n_i$ at time $t=500$ in the right panel of Fig. \ref{compare2}, shows that it propagates exponentially slowly, and strikingly similar to random distribution values, a feature further corroborated by the extremely low value of $n_{tot}$ compared to the other distributions and almost matching the case of random distribution. Also the speed of the fastest particle significantly slows down after some time and the wave does not reach the end of the system for the system size considered here , even for large time. This essentially means the system is going to a localized state in spite of being under a quasi-periodic potential. This behaviour can be attributed to  second moments of the RS sequence showing exactly similar behaviour as second moments of a random sequence. Since we have a free-fermion system higher moments are not relevant to our calculation and thus we see a localization for this system. \\ 
Nevertheless this shows, that there are large number of scattering events happening near $x=0$ which causes a large number of wave-fronts to cross $A$ of the subsystem causing in the rise in Entropy. However, similar to the case of the random disordered sequence, there is a saturation of entanglement due to particle localization and thus if we see a very long time result, systems with periodic or other quasi periodic potential show a higher entanglement. 
\subsection{Inhomogeneous quench with TI breaking only in environment} 
\begin{figure*}
\centering{
\includegraphics[width=0.64\columnwidth]{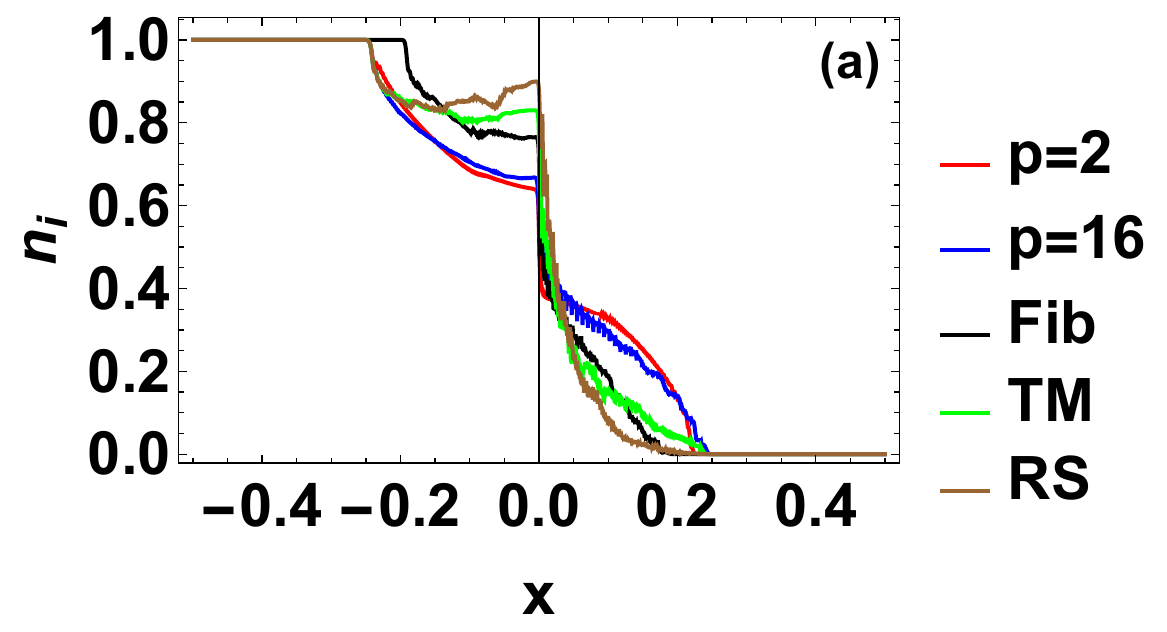}
\includegraphics[width=0.64\columnwidth]{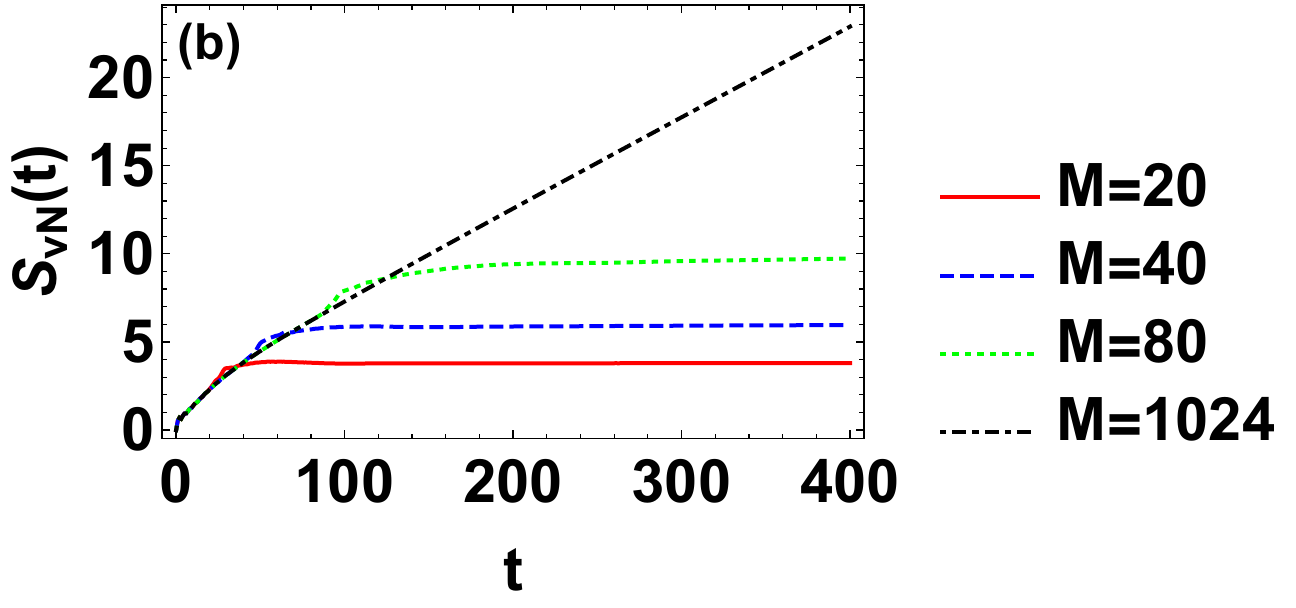}
\includegraphics[width=0.64\columnwidth]{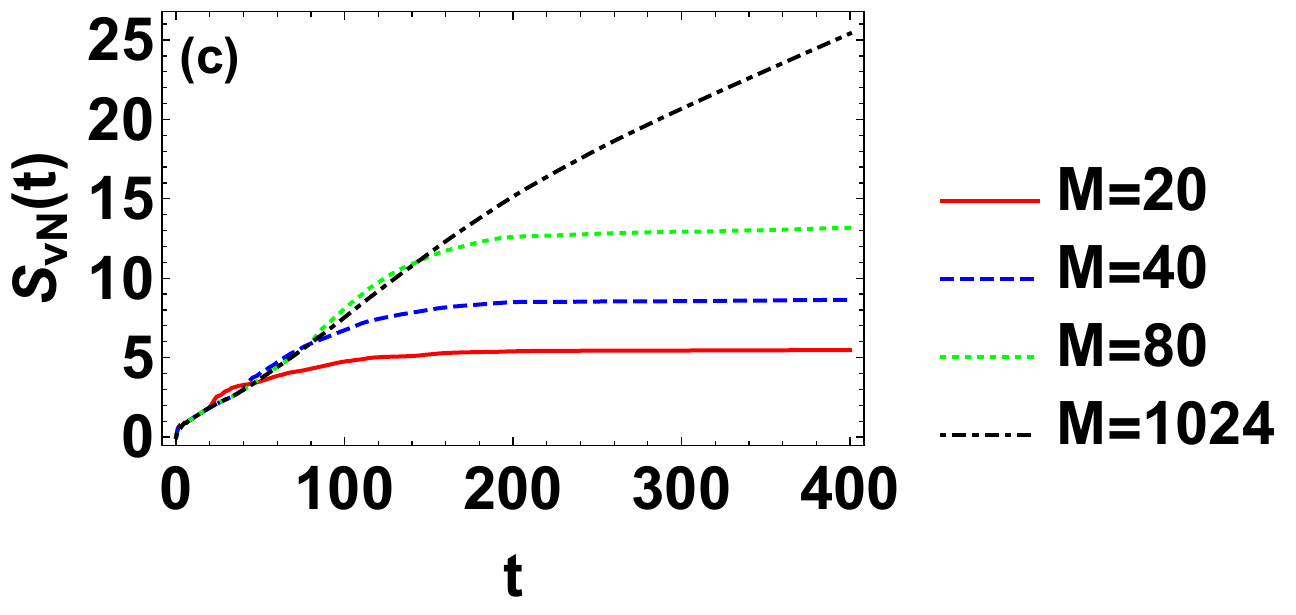}\\
\includegraphics[width=0.64\columnwidth]{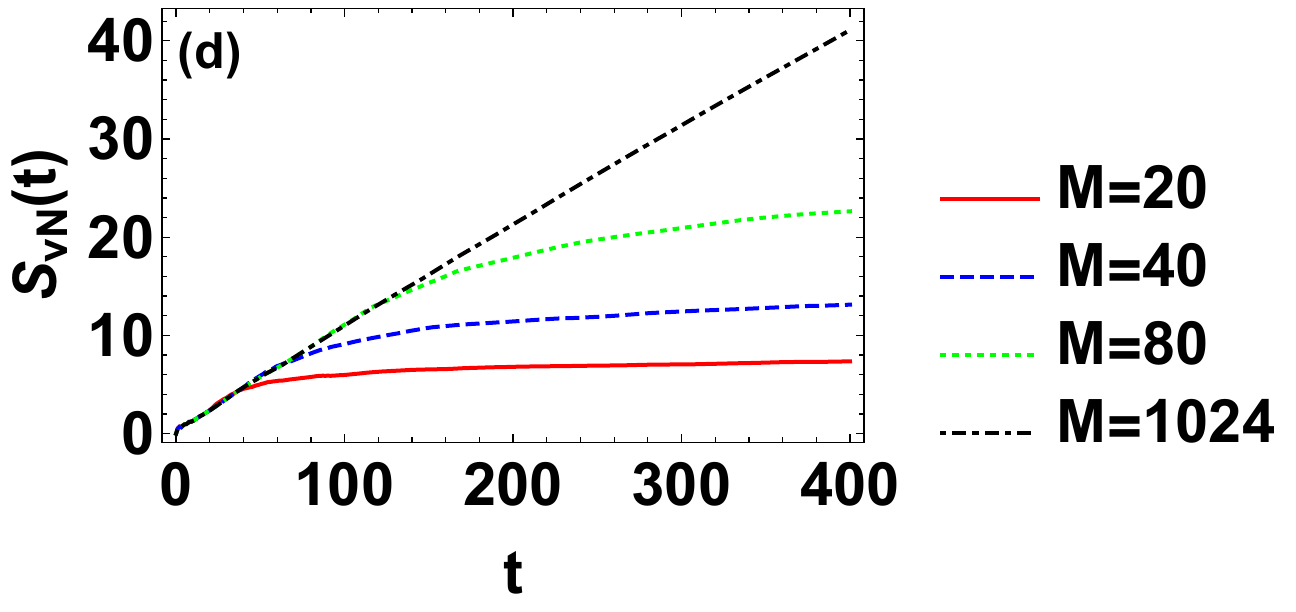}
\includegraphics[width=0.64\columnwidth]{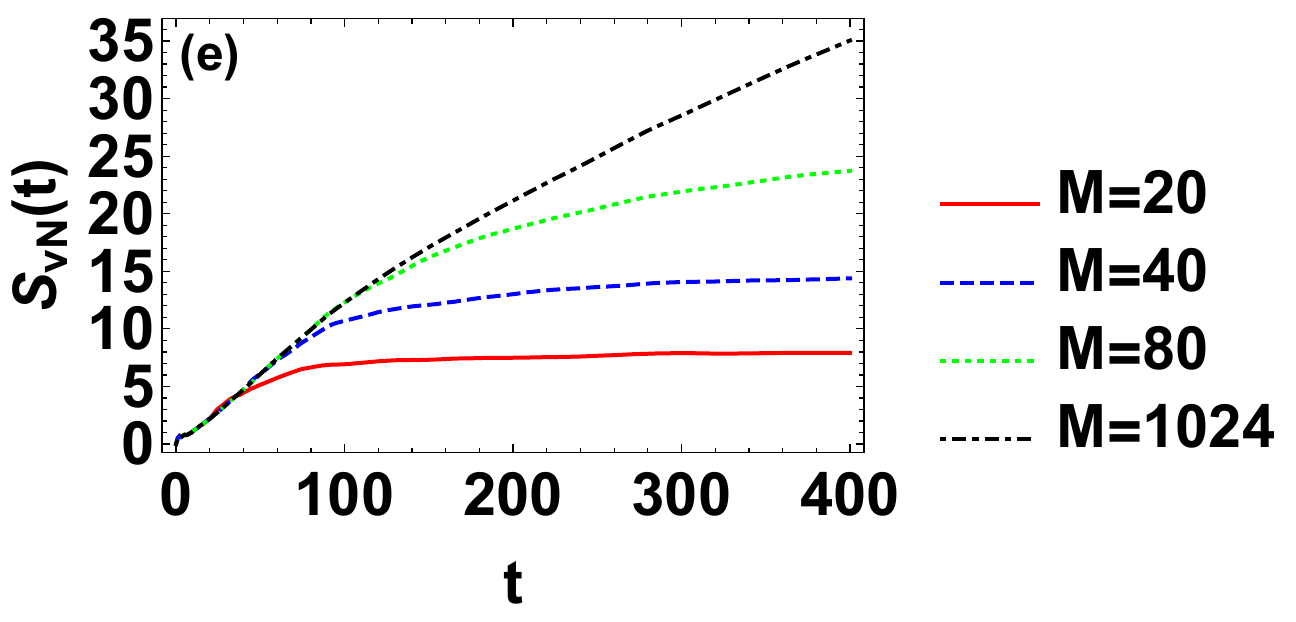}
\includegraphics[width=0.64\columnwidth]{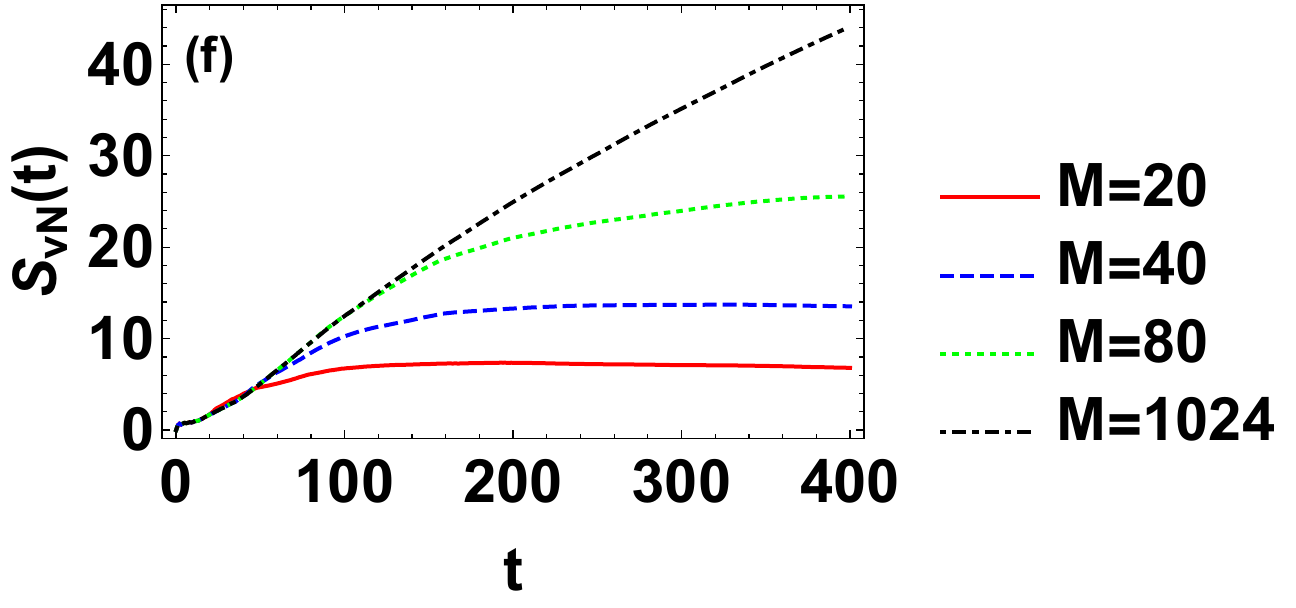}}
\caption{(Colour Online)(a) Plot of $n_i=\la c_i^{\dagger} c_i \ra$ vs $x=(i-L/2)/L$ for t=500, 
	for the system with Hamiltonian given by Eqn. \eqref{hammaster3} for different sequences. 
	(b) Plot of $S(t)$ vs t for different system sizes for a periodic sequence potential of period $p=2$. 
	(c) Same as (b) but with a period $p=16$. (d), (e), (f): Same as (b) but the potential is arranged in 
	Fibonacci, Thue Morse and Rudin Shapiro sequence respectively}
\label{quench2}
\end{figure*}
In this section we will describe an inhomogeneous quench in which the system is chosen to be Translationally Invariant but the environment has TI breaking. The Hamiltonian is chosen as follows,
\begin{equation}
\mathcal{H}=-t\sum_{p=1}^L(c_p^{\dagger} c_{p+1}+c_{p+1}^{\dagger} c_p)+\mu \sum_{p=1}^{L/2} c_p^{\dagger} c_p+ \sum_{p=L/2+1}^{L} \mu_p c_p^{\dagger} c_p
\label{hammaster3}
\end{equation}
with the same initial condition, 
\begin{align}
\la c_p^{\dagger} c_q \ra &=\delta_{pq} \hspace{0.2 in} p \le L/2 \nonumber\\
	&=0 ~ {\rm otherwise} \label{init3}
\end{align}
Thus we take two systems of length $L/2$ with open boundary conditions. One of them is translationally invariant, and the other with broken translation symmetry by on site potential arranged in the sequences described in the paper. In some cases this becomes a better approximation to the system-environment setup, where one can ensure the system has no inhomogeneities but the environment does. Our aim is to find if such a change produces changes in the results we obtain.\\
Since the system on the left is translationally invariant, one would expect no scattering events there and the quasi-particles travelling ballistically towards the right. However on crossing $i=L/2$ it encounters the environment where scattering processes occur and thus has a chance of going back to the side $i<L/2$, however, there due to lack of scattering the particle will again try to travel to the right and then get impeded by scattering events. This suggests an accumulation of particles near $i=L/2-\epsilon$ where $\epsilon$ is a small integer. This in turn is expected to increase entanglement as there is a very high number of wave-fronts crossing the $i=L/2$ boundary due to accumulation of particles scattered back from the surrounding.
In Fig. \ref{quench2}, we show the numerical results for the case considered corroborates our analysis. For each of the cases  Fig. \ref{quench2}(a) shows an accumulation of particle density near $x=0$ in the left half system which was not present in the case considered in the paper. Consequently one sees the entanglement for the cases $p=2$ and $p=16$ shown in (b) and (c) shows a much higher value than what was seen in the Periodic potential section of the main text. Thus it shows accumulation of particles near $x=0$ increases the chances of a particle oscillations between the subsystem and the environment and thus increases its entropy. It is expected if the point $A$ of the subsystem is moved away from $x=0$  to a positive value of x, this effect would gradually become non existent. \\
In the cases of the automatic sequence, the increase due to accumulation of particles is also present but less pronounced. In fact for the RS sequence it is negligibly small. This is because in these cases a localization had already set in in the setup considered in the rest of the work, thus effectively having a large number of particles which oscillate between the system and the subsystem. The further accumulation of particles in the $x<0$ region thus has negligible effects since the increase in entropy due to multiple oscillations of the particle was already present in the case considered in the paper. Hence the increment in entropy due to the change in the quench is much less pronounced.

\end{document}